\documentclass[11pt,a4paper]{article} 
 
        \usepackage[utf8]{inputenc}
        \usepackage{fontenc}
	\usepackage[italian,english]{babel}
	\usepackage{amsmath}
	\usepackage{geometry}
	\usepackage{amsthm}
	\usepackage{mathrsfs}
	\usepackage{bbold}
	\usepackage{mathtools}
	\usepackage{amsfonts}
	\usepackage{graphicx}
	\usepackage{setspace}
	\usepackage{physics}
	 \usepackage{comment}
	  \usepackage{layout}
	\usepackage{xcolor}
     \usepackage[big]{layaureo}
	 \usepackage{dsfont}	
	 \usepackage{slashed}
 	\usepackage{amssymb}
	 \usepackage{tensor,multirow}
	\usepackage{fancyhdr} 
 	\usepackage{tikz-cd}
 	\usepackage{braket}
 	\usepackage{tikz}
 	\usepackage[compat=1.1.0]{tikz-feynman}
 	\usepackage{subfigure}
	\usepackage{fancyhdr} 
	\usepackage{booktabs}
	\usepackage{braket,tikz-cd,iftex}
	
\usepackage{jheppub} 

\usepackage{natbib}
\preprint{CERN-TH-2022-146}
\title{Constant primary operators and where to find them: The strange case of BPS defects in ABJ(M) theory}

\author[a]{Nicola Gorini,} 
\author[b]{Luca Griguolo,} 
\author[c]{Luigi Guerrini,} 
\author[a]{Silvia Penati,}
\author[d]{Domenico Seminara,}
\author[b,e]{Paolo Soresina}

\affiliation[a]{ Dipartimento di Fisica, Universit\`a degli studi di Milano--Bicocca, and INFN, Sezione di Milano--Bicocca, Piazza della Scienza 3, I-20126 Milano, Italy }
\affiliation[b]{Dipartimento SMFI, Universit\`a di Parma and INFN Gruppo Collegato di Parma, Viale G.P. Usberti 7/A, 43100 Parma, Italy}
\affiliation[c]{Department of Mathematical Sciences, Durham University, Durham DH1 3LE, United Kingdom}
\affiliation[d]{Dipartimento di Fisica, Universit\`a di Firenze and INFN Sezione di Firenze, via G. Sansone 1, 50019 Sesto Fiorentino, Italy}
\affiliation[e]{Theoretical Physics Department, CERN, 1211 Geneva 23, Switzerland\\}

\emailAdd{nicola.gorini93@gmail.com}  
\emailAdd{luca.griguolo@pr.infn.it} 
\emailAdd{luigi.guerrini@durham.ac.uk} 
\emailAdd{silvia.penati@mib.infn.it} 
\emailAdd{seminara@fi.infn.it}
\emailAdd{paolo.soresina@cern.ch}

\abstract{We investigate the one-dimensional defect SCFT defined on the $1/2$ BPS Wilson line/loop in ABJ(M) theory. We show that the supermatrix structure of the defect imposes a covariant supermatrix representation of the supercharges. Exploiting this covariant formulation, we prove the existence of a long multiplet whose highest weight state is a constant supermatrix operator. At weak coupling, we study this operator in perturbation theory and confirm that it acquires a non-trivial anomalous dimension. At strong coupling, we conjecture that this operator is dual to the lowest bound state of fluctuations of the fundamental open string in AdS$_4\times \mathbb{CP}_3$ around the classical $1/2$ BPS solution. Quite unexpectedly, this operator also arises in the cohomological equivalence between bosonic and fermionic Wilson loops. We also discuss some regularization subtleties  arising in perturbative calculations on the infinite Wilson line.
}

	\newcommand{\beq}{\begin{equation}}
	\newcommand{\bea}{\begin{eqnarray}}
	\newcommand{\eea}{\end{eqnarray}}
	\newcommand{\eeq}{\end{equation}}
	\newcommand{\non}{\nonumber}

	\newcommand{\bY}{\bar{Y}}
	\newcommand{\bZ}{\bar{Z}}

	\newcommand{\de}{\partial}
	
	\renewcommand{\a}{\alpha}
	\renewcommand{\b}{\beta}
	\newcommand{\g}{\gamma}

	\newcommand{\STr}{\text{STr}}
	\newcommand{\La}{\mathcal L}

	\makeatletter
	\newcommand{\aextp}{\@ifnextchar^\@aextp{\@aextp^{\,}}}
	\def\@aextp^#1{\mathop{\bigwedge\nolimits^{\!#1}}}
	\makeatother
	
	\makeatletter
	\newcommand{\extp}{\@ifnextchar_\@extp{\@extp_{\,}}}
	\def\@extp_#1{\mathop{\aextp\nolimits_{\!#1}}}
	\makeatother

	\theoremstyle{definition}

\begin{document}
\maketitle

\section{Introduction}

Defects play a central role in Quantum Field Theory. They are valuable tools to study physical configurations far from ideal, e.g., finite size effects, presence of boundaries, interfaces, or impurities, and give insight into how a theory responds to the insertion of a probe. 
\\
When restricting to a $n$ dimensional Conformal Field Theory (CFT), a codimension-$m$ conformal defect is a lower dimensional operator which breaks the conformal group to the conformal group in $(n-m)$ dimensions on the defect, plus transformations in the directions orthogonal to the defect. 
The presence of the defect enlarges the spectrum of observables of the CFT. On the one hand, the set of correlation functions of operators localized on the defect defines a \emph{Defect Conformal Field Theory} (dCFT) that can be analyzed through standard methods of CFTs. On the other hand, local bulk operators acquire a non-trivial VEV near the defect, leading to a new set of CFT data \cite{Billo:2016cpy}. These coefficients are not independent but constrained by crossing symmetry and unitarity in the defect bootstrap equations \cite{Liendo:2012hy,McAvity:1995zd}.

In particular, one-dimensional defects have been extensively considered. Physically, they represent worldlines of physical particles and provide a universal language to describe phenomena in condensed matter physics, like the Kondo problem (see \cite{Affleck:1995ge} for a review) or in high energy physics. In the latter case, the preeminent example of a line operator is the Wilson loop operator in gauge theories \cite{Wilson:1974sk}. It serves as an order parameter for phase transitions, e.g., the confinement/deconfinement phase transition. Moreover, the Wilson line spectrum characterizes the form of the gauge group, a piece of information that is not accessible by using only local observables \cite{Aharony:2013hda,Gaiotto:2014kfa}.

In supersymmetric theories, it is natural to consider extensions of the Wilson loops, preserving a fraction of the supercharges (BPS). They are constructed by adding extra couplings with the matter fields to the Wilson connection.  Linear and circular paths can often support BPS Wilson loops that keep maximal superconformal symmetry, namely half of the total supercharges \cite{Zarembo:2002an}. The first example is in 4d $\mathcal{N}=4$ SYM where, in the context of the AdS/CFT correspondence, the BPS Wilson loop was introduced as the CFT dual of the fundamental string \cite{Maldacena:1998im, Rey:1998ik}. 
BPS Wilson loops in 4d \(\mathcal{N}=4\) Super Yang-Mills provide a natural example of dCFT \cite{Giombi:2017cqn, Cooke:2017qgm, Giombi:2018qox}.
Generally, the study of Wilson loops in supersymmetric theories has a broad horizon. In particular setups, we can combine supersymmetric localization and perturbation theory to compute exactly physical observables, such as the Bremsstrahlung function \cite{Pestun:2007rz, Correa:2012at}. 
This possibility establishes an exciting bridge with integrability, as the same quantity is accessible using integrability-based methods \cite{Correa:2012hh}.

Having in mind to investigate if similar features are present in superconformal field theories (SCFTs) defined in different spacetime dimensions, it is natural to explore BPS Wilson operators in the three-dimensional analog of $\mathcal{N}=4$ SYM, namely ABJ(M) theory. This is a class of three dimensional $\mathcal{N}=6$ Super Chern-Simons-matter theories with gauge group $U(N_1)_k\times U(N_2)_{-k}$ \cite{Aharony:2008ug, Aharony:2008gk}, dual to type IIA string theory in AdS$_4\times \mathbb{CP}_3$ or M-theory in AdS$_4\times S^7/Z_k$. 

In ABJ(M) theory, the structure of BPS Wilson loops is much richer than in four dimensions due to the possibility of using not only scalar matter but also fermions to build up a supersymmetric loop connection. We will limit the discussion to the maximally supersymmetric (1/2 BPS) Wilson operator. This observable is naturally defined in terms of a $U(N_1|N_2)$ supermatrix connection, which involves gauge fields and scalars in the diagonal terms, and matter fermions in the anti-diagonal ones \cite{Drukker:2009hy}. The fermionic couplings constitute a defect marginal deformation \cite{Ouyang:2015iza, Ouyang:2015bmy, Mauri:2017whf}, which connects the fermionic loops to the less supersymmetric bosonic Wilson loops that do not involve any fermions \cite{Drukker:2008zx, Berenstein:2008dc, Chen:2008bp}. The exact quantum value of these Wilson loops is accessible via localization, 
either using the cohomological equivalence between fermionic and bosonic Wilson loops or exploiting a more recent representation involving the background connection \cite{Kapustin:2009kz, Drukker:2019bev, Drukker:2020opf}. In the AdS/CFT context, the 1/2 BPS Wilson loop is dual to the fundamental open string living in AdS$_4\times \mathbb{CP}_3$, with appropriate boundary conditions on the contour at the boundary. 

The maximally supersymmetric Wilson line provides an example of a conformal line defect with non-trivial boundary conditions induced by the fermions. In this paper, we will focus on the related defect CFT \cite{ Bianchi:2017ozk, Bianchi:2020hsz}\footnote{For the dCFT on the bosonic Wilson loop see \cite{Bianchi:2018scb}. }. This dCFT admits a Lagrangian formulation and a weak coupling limit and is thus amenable to perturbative investigation. Moreover, according to the AdS/CFT dictionary, its strong coupling limit is described by the AdS$_2$ theory for the fluctuations of the fundamental open string in AdS$_4\times \mathbb{CP}_3$. This fact provides a controlled example of non-maximally supersymmetric AdS$_2$/CFT$_1$ correspondence \cite{Giombi:2017cqn}.

The exploration of this theory was initiated in \cite{ Bianchi:2017ozk, Bianchi:2020hsz}, where the main focus was the \emph{displacement multiplet}. The presence of the displacement operator is a universal feature of defects arising from the non-conservation of the stress-energy tensor in the direction orthogonal to the defect. It describes the response of the defect to contour deformations. 
The displacement operator is always the top component of a superconformal multiplet and arises from breaking transverse translations. Its correlation functions encode the information on the energy exchanged between the bulk and the defect, usually called the Bremsstrahlung function. 
Here we continue the investigation of the dCFT defined on the 1/2 BPS fermionic Wilson line, whose set of local operators are naturally represented as supermatrices. We will be primarily interested in constructing the lowest dimensional multiplets and evaluating their correlation functions at the perturbative level. 

At weak coupling, in any Lagrangian theory, a multiplet can be built by acting on its superconformal primary (SCP) operator with the supersymmetry charges preserved by the theory. However, unlike its 4d counterpart, the superconnection of the Wilson line is supersymmetric invariant, only up to a super-gauge transformation. 
Following \cite{Bianchi:2020hsz}, we compensate for this residual transformation by dressing the standard supersymmetry with a further super-gauge transformation. We generalize this idea, building this structure on the space of supermatrices. The resulting algebra is called \emph{covariant} superconformal algebra. We explore its algebraic structure in detail and make contact with the abstract representation theory for the worldline superconformal algebra.

Exploiting this covariant formalism, we find, quite surprisingly, that at weak coupling, the lightest multiplet living on the defect is a long multiplet whose SCP is classically dimensionless. It corresponds to the insertion of a constant supermatrix on the Wilson line. Formally, it is the supermatrix that swaps the supertrace and the trace prescriptions in the definition of the Wilson line. Physically, it can be related to the action of transverse rotations on the defect. Since the superconnection involves fermions, one might think that the Wilson line breaks transverse rotations. However, this is not the case since its variation can always be modded out by a gauge transformation \cite{Agmon:2020pde}. Equivalently, we prove that the defect operator produced under transverse rotations is a descendant (a total line derivative) of the constant operator.

Strong coupling considerations corroborate the physical relevance of the constant operator. As in the 4d case, \cite{Giombi:2017cqn}, in the ABJ(M) theory, the displacement multiplet of the 1/2 BPS Wilson line is naturally mapped into the fluctuations around the classical open string solution \cite{Bianchi:2020hsz}. In the 4d case, the lowest dimensional long multiplet is conjectured to be dual to the lightest bound state of fluctuations in the string theory setup. In the same spirit, we conjecture that our constant operator is the CFT dual of the lowest dimensional bound state built out from the fluctuations corresponding to the displacement multiplet. We find a non-trivial agreement with the weak coupling structure. The constant operator at weak coupling is the only possible operator matching the SCP in the holographic representation.

We begin by studying the quantum properties of the constant multiplet at weak and strong coupling. First, we provide the explicit construction of the multiplet in terms of the ABJ(M) fundamental fields. We study its correlation functions at weak coupling in standard perturbation theory. 
Our main result is the evaluation of the anomalous dimension of the constant operator at one loop. We also borrow the results from the bootstrap analysis of \cite{Bianchi:2020hsz} to get the anomalous dimension at strong coupling. It turns out that the quantum dimension of the constant operator is compatible with an interpolating function between the weak and strong coupling. 
We provide a next-to-leading order computation of the Brehmsstrahlung function at weak coupling from the two-point of the displacement, finding agreement with previous results \cite{Lewkowycz:2013laa, Bianchi:2014laa, Bianchi:2017svd, Bianchi:2018bke, Griguolo:2021rke}.
\\
In performing perturbative calculations, long-distance divergences associated with the infinite length of the Wilson line arise and need to be regularized. Usually, this is done by cutting the line to a segment, but this produces unwanted terms which mix the IR regulator with the parameter of the UV dimensional regularization in a way that renders the dCFT at quantum level sensible to the regularization scheme and therefore ambiguous. To clarify the origin of these terms and find a consistent way to remove them, we compare the results from the dCFT on the line with those of the same dCFT on the maximally supersymmetric circular Wilson loop where the IR divergences are absent. This comparison allows us to identify a definite procedure to avoid these issues, which eventually seems to correspond to putting extra degrees of freedom at the two edges of the cut-off line. 

The paper is organized as follows. In section \ref{sect:WL}, we begin by summarizing the main features of 1/2 BPS Wilson loops in ABJ(M) theory. We then study the symmetries of the defect in section \ref{sect:symmetries}. The section includes the construction of the supermatrix covariant generators of the superconformal algebra. Section \ref{sec:dSCFT} is devoted to introducing the constant operator, constructing its superconformal multiplet, and discussing its properties in connection with the would-be breaking of transverse rotations. We also discuss the role of the constant operator in connection with the cohomological equivalence between bosonic and fermionic Wilson loops. We investigate the new multiplet at weak coupling in section \ref{sect:perturbative}, where we exploit a non-trivial Ward identity that arises from the covariant algebra to read its anomalous dimension at one loop directly from the coefficient of the two-point function of its descendant. The regularization prescription that we adopt to tame IR divergences on the line is checked against the computation of the Brehmsstrahlung function at two loops, which is consistent with previous results in the literature. 
Finally, the realization of the constant operator in terms of the dual bound state is discussed in section \ref{sect:strong}. 
We summarize the main results and collect insights on future directions, in section \ref{sect:conclusions}. 
Six appendices complete the paper. They cover technical details on supermatrices, the ABJ(M) theory, and the superconformal algebra on the defect and its representations. A detailed discussion on the regularization of large distance divergences on the line and its comparison with the theory defined on the circle is presented in appendix \ref{sect:cut-off}. 

\section{$1/2$ BPS defects in ABJ(M) theory}\label{sect:WL}
This section briefly reviews the structure and properties of $1/2$ BPS Wilson operators in ABJ(M) theory \cite{Drukker:2009hy}, primarily to fix notations and conventions. 

Given the ABJ(M) theory associated with the $U(N_1)_k \times U(N_2)_{-k}$ quiver and described by the action in eqs. (\ref{action}, \ref{CS} -- \ref{S4pt}), we consider the fermionic Wilson operator defined as 
\begin{equation}\label{WL}
W[C] =\mathcal P\exp (-i\int_C \mathcal L)
\end{equation}
The super-connection $\mathcal L$ is given by the following {\em even} supermatrix\footnote{See App. \ref{supermatrices} for the basic properties of supermatrices.} in the Lie superalgebra of $U(N_1|N_2)$:
\begin{equation}\label{connec}
\mathcal L=
\begin{pmatrix}
 A_\mu\dot x^\mu- 2\pi i \frac{\ell}{k}|\dot x|{M_J}^IC_I\bar C^J & i\sqrt{2\pi\frac{\ell}{k}}|\dot x|\eta^\alpha_I\bar\psi^I_\alpha\\
-i\sqrt{2\pi \frac{\ell}{k}}|\dot x|\psi_I^\alpha\bar\eta_\alpha^I& \hat A_\mu\dot x^\mu- 2\pi i \frac{\ell}{k}|\dot x|{\hat M_J}^I\bar C^JC_I
\end{pmatrix}
\end{equation}
In \eqref{connec} $C$ denotes a generic smooth contour in ${\mathbb R}^3$ parametrized as $x^\mu=x^\mu(\tau)$ (we work in Euclidean signature with conventions given in appendix \ref{ABJ(M)}). The super-connection depends on the Chern-Simons level $k$, whereas $\ell$ is an arbitrary parameter that can take values $\pm1$. The quantities $M_{J}^{\ \ I}(\tau)$, $\hat M_{J}^{\ \ I}(\tau)$, $\eta_{I}^{\alpha}(\tau)$ and $\bar{\eta}^{I}_{\alpha}(\tau)$ control the possible
local couplings. The latter two, in particular, are considered Grassmann even quantities even though they transform in the spinor representation of the Lorentz group. We shall focus on locally 1/2 BPS operators that possess a local $U(1)\times SU(3)$ $R-$symmetry invariance. Thus the couplings in \eqref{connec} can be taken of the form 
\begin{equation}
\begin{split}
\label{cc}
&\eta_{I}^{\alpha} (\tau)=n_{I} (\tau)\eta^{\alpha} (\tau),\ \ \  \bar\eta^{I}_{\alpha} (\tau)=\bar n^{I} (\tau) \bar\eta_{\alpha} (\tau),\ \ \
M_{J}^{\ \ I} (\tau)=
\widehat M_{J}^{\ \ I} (\tau)=\delta^{I}_{J}-2 n_{J} (\tau) \bar n^{I} (\tau)
\end{split}
\end{equation}
The functional dependence of $n^I(\tau)$ and $\eta_\alpha(\tau)$ on $\tau$ is then determined by requiring that the loop preserves some superconformal transformations (see \cite{Drukker:2009hy,Cardinali:2012ru,Lietti:2017gtc}). For generic closed paths, the net result of this subset of transformations on \eqref{WL} can be represented as a field-dependent super-gauge transformation belonging to $u(N_1|N_2)$. However, this super-gauge transformation is not, in general, periodic when $\tau$ spans the contour. Thus, to obtain a BPS quantity, it is not enough to take the super-trace of the line operator, but we have to introduce a twist matrix $\mathcal{T}$ that takes care of the lacking of periodicity, i.e.
\begin{equation}
\mathcal{W}=\mathrm{Str}\left( W[C] \mathcal{T}\right)
\end{equation}
See \cite{Cardinali:2012ru} for details. Alternatively, this issue can be cured by introducing a classical background connection along the path, which makes the super-gauge transformation periodic \cite{Drukker:2019bev}. Although it is more elegant from a geometric point of view, this latter approach is less suited for performing perturbative computations, and thus we shall not use it in the following.

One can also consider Wilson operators, which are supported on unbounded contours. A typical example is the Wilson line, where the contour is an infinite straight line. In this case, to obtain a BPS operator, we must carefully choose the boundary condition at infinity. This choice is not always unique or unambiguous. In the following, we choose to fix the boundary conditions and consequently the twist matrix by requiring that the unbounded contour is obtained as the decompactification limit of a closed 
path.

Our analysis will focus on two types of (conformally equivalent) operators/defects: the infinite straight line and the great circle in $S^2$. 

\noindent
\paragraph{Linear defect.} This is described by the Wilson operator in \eqref{WL}, with $C$ being the infinite straight line parametrized as $x^\mu = (0,0, s)$, $ -\infty < s < +\infty$. When the matter couplings are chosen as 
\begin{equation}\label{eq:couplings}
{M_J}^I={\hat M_J}^I=
\begin{pmatrix}
-1&0&0&0\\
0&1&0&0\\
0&0&1&0\\
0&0&0&1
\end{pmatrix},
\qquad \eta^\alpha_I=\sqrt{2}
\begin{pmatrix}
1\\
0\\
0\\
0
\end{pmatrix}_I
(1,0)^\alpha,\qquad
\bar\eta_\alpha^I=i\sqrt{2}\ (1,0,0,0)^I
\begin{pmatrix}
1\\
0
\end{pmatrix}_\alpha
\end{equation}
with the fermionic couplings satisfying the conditions
\begin{equation}\label{eq:couplingsprop}
\delta^\beta_\alpha=\frac{1}{2i}(\eta^\beta\bar\eta_\alpha-\eta_\alpha\bar\eta^\beta)\qquad {(\dot x\cdot\gamma)_\alpha}^\beta=\frac{\ell}{2i}|\dot x|(\eta^\beta\bar\eta_\alpha+\eta_\alpha\bar\eta^\beta)
\end{equation}
the operator preserves half of the supersymmetry charges \cite{Cardinali:2012ru}, {\em i.e.} it defines a $1/2$ BPS linear defect.

Using this parametrization and organizing the elementary fields in $SU(3)$ representations (see eq. \eqref{su3breaking} for field redefinitions), we rewrite
\begin{equation}\label{connecsu3}
\begin{aligned}
\mathcal L &= 
\left(
 \begin{matrix}
A_3&0\\
0&\hat A_3
\end{matrix}
\right)
+2\pi i\frac{\ell}{k}
\left(
\begin{matrix}
Z\bar Z-Y_a\bar Y^a&0\\
0&\bar Z Z-\bar Y^a Y_a
\end{matrix}
\right)
+2\sqrt{\pi\frac{\ell}{k}}
\left(
\begin{matrix}
0&i\bar\psi_{(1)}\\
\psi^{(1)}&0
\end{matrix}
\right)\\
&\equiv \mathcal L_A+\mathcal L_B+\mathcal{L}_F
\end{aligned}
\end{equation}  

Since we shall think of the line as the decompactification limit of the circle, we shall select the same twist-matrix of the circle, i.e., $\sigma_3$, (see \cite{Drukker:2009hy,Cardinali:2012ru}). This choice amounts to taking the trace instead of the super-trace of \eqref{WL} and it is also the prescription that leads to an operator that is dual to the 1/2 BPS string configuration in AdS$_3 \times {\mathbb {CP}}^3$ or a 1/2 BPS M2-brane configuration in M-theory \cite{Lietti:2017gtc}.  
With this choice, the vacuum expectation value (VEV) at the tree level is given by
\begin{align}
	\langle \mathcal{W} \rangle \equiv \langle \Tr[W] \rangle=N_1+N_2 \label{eq:w+} 
\end{align}

\paragraph{Circular defect.} A circular defect can be easily obtained by conformally mapping the line onto the circle. 
This requires to first shift the line in the $(2,3)$-plane by half unit along the $x^2$-direction, namely taking $x^\mu=(0,\frac{1}{2},s)$, in order to have a complete invertible mapping $\forall s\in\mathbb R$. Then, taking a special conformal transformation generated by the vector $b^\mu=(0,1,0)$, the line-to-circle map reads \cite{Griguolo:2012iq}
\begin{align}
&x^\mu=\bigg(0,\frac{1}{2},s\bigg)\ \mapsto\ x'^\mu=\bigg(0,\ \frac{\frac{1}{4}-s^2}{\frac{1}{4}+s^2},\ \frac{s}{\frac{1}{4}+s^2}\bigg) \equiv (0,\ \cos \tau,\ \sin \tau)  \\
&\Lambda(\tau)=\frac{1}{4}+s^2=\frac{1}{4\cos^2\frac{\tau}{2}}
\end{align}
where, in the last equality, we have defined $s\equiv \frac{1}{2}\tan \frac{\tau}{2}, \tau\in (-\pi,\pi)$ being the proper time of the circle. 

Scalars couplings are not affected by these transformations, while for the fermionic ones, we obtain
\begin{equation}
\eta'^\alpha_I=\sqrt{2}\delta^1_I\left(\begin{matrix}\cos\frac{\tau}{2} & i\sin\frac{\tau}{2}\end{matrix}\right)^\alpha\qquad \bar{\eta}'^I_\alpha=i\sqrt{2}\delta_1^I\left(\begin{matrix}\cos\frac{\tau}{2} \\ -i\sin\frac{\tau}{2}\end{matrix}\right)_\alpha
\label{sceltacer}
\end{equation}
The super-connection on the circle then reads 
\begin{equation}
\resizebox{\hsize}{!}{\(\La=\left(\begin{matrix}
\left(A_3\cos \tau\!-\!A_2\sin\tau\right)-2\pi i\frac{\ell}{k}\left(Z\bar{Z}\!-\!Y_a\bar{Y}^a\right) & -i\sqrt{2\pi\frac{\ell}{k}}\, \eta\bar{\psi} \\
-i\sqrt{2\pi\frac{\ell}{k}}\, \psi\bar{\eta} & \left(\hat{A}_3\cos\tau\!-\!\hat{A}_2\sin\tau\right)-2\pi i \frac{\ell}{k}\left(\bar{Z}Z \!-\!\bar{Y}^aY_a\right)
\end{matrix}\right)\)}
\label{superconnectioncircle}
\end{equation}
The antiperiodicity of the fermionic couplings immediately suggests that the twist matrix is $\sigma_3$ and thus  $\mathcal{W}=\Tr[W]$. At tree level it evaluates to \eqref{eq:w+}. 

The role of the parameter \(\ell\) will be clarified in a following paper \cite{forthcoming}; we proceed in our analysis setting \(\ell=1\).

\section{Symmetries of the defect}\label{sect:symmetries}
Aimed at studying the one-dimensional SCFTs defined on linear and circular defects, we first discuss in detail the symmetries that the defects inherit from the bulk theory. 
For simplicity, we restrict to the case of a linear defect. Everything can be easily rephrased for circular Wilson loops.

The ABJ(M) theory is invariant under the action of the superconformal algebra \(\mathfrak{osp}(6|4)\). The insertion of the defect breaks this symmetry as 
\[\mathfrak{osp}(6|4)\rightarrow \mathfrak{su}(1,1|3)\oplus \mathfrak{u}(1)_b\]
The \(\mathfrak{su}(1,1|3)\) superalgebra contains as the maximal bosonic subalgebra \(\mathfrak{su}(1,1) \times \mathfrak{su}(3) \times \mathfrak{u}(1)_M\). Here \(\mathfrak{su}(1,1)\) is the conformal algebra in one dimension, \(\mathfrak{su}(3) \) is the R-symmetry algebra on the defect and the $\mathfrak{u}(1)_M$ abelian factor is generated by a linear combination of the generator of rotations transverse to the defect and a broken generator of the bulk R-symmetry algebra (see eq. \eqref{eq:M}). The fermionic sector is generated by twelve odd generators, six Poincar\'e supercharges \(Q^a, \bar{Q}_a\) and 
six superconformal generators \(S^a, \bar{S}_a\), where $a=1,2,3$ is a \(\mathfrak{su}(3) \) fundamental index. The residual \(\mathfrak{u}(1)_b\) is generated by the operator in \eqref{eq:B}. This symmetry plays the role of a flavor symmetry for the defect SCFT. 

For more details on the \(\mathfrak{su}(1,1|3)\) algebra and the classification of its representations, we refer to appendix \ref{sect: su(1,1|3)}. Here, we discuss the covariant realization of the $\mathfrak{su}(1,1|3)$ superconformal algebra induced by the  defect.  

\subsection{Supersymmetry invariance}We begin by studying the behavior of the linear defect introduced in section \ref{sect:WL} under the action of the $\mathfrak{su}(1,1|3)$ supercharges $Q^a, \bar{Q}_a$, and 
$S^a, \bar{S}_a$, $a=1,2,3$. 
Since the Wilson operator is defined in terms of a $U(N_1|N_2)$ superconnection,  SUSY variations are better described in terms of matrix supercharges, defined as
\begin{equation}\label{matrixQ}
Q^a \rightarrow {\mathbb Q}^a \equiv Q^a {\mathbb 1} = \left( \begin{matrix}
Q^a &0\\
0&- Q^a
\end{matrix} \right)  \qquad \qquad \bar{Q}_a \rightarrow \bar{\mathbb Q}_a \equiv \bar{Q}_a {\mathbb 1} = \left( \begin{matrix}
\bar{Q}_a &0\\
0&- \bar{Q}_a
\end{matrix} \right) 
\end{equation}
\begin{equation}\label{matrixS}
\hspace{-1cm} S^a \rightarrow {\mathbb S}^a \equiv {\mathbb S}^a\mathbb 1 = \left( \begin{matrix}
S^a &0\\
0&- S^a
\end{matrix} \right)   \qquad \qquad \quad \bar{S}_a \rightarrow \bar{\mathbb S}_a \equiv \bar{\mathbb S}_a \mathbb 1 = \left( \begin{matrix}
\bar{S}_a &0\\
0&- \bar{S}_a
\end{matrix} \right) 
\end{equation}
where ${\mathbb 1} $ is the identity in the space of $U(N_1|N_2)$ supermatrices.  Here, we have used identity \eqref{left}, taking into account that 
$Q^a, \bar{Q}_a, S^a, \bar{S}_a$ are grade-1 scalars. 

In the case of the Poincar\'e supercharges\footnote{Identical definitions hold for superconformal generators.}, the SUSY variation of a generic supermatrix $T$ is defined as
\beq\label{matrixtransf2}
\delta_Q T = [ \theta_a {\mathbb Q}^a, T\} \; , \qquad \quad \bar\delta_Q T = [\bar\theta^a \bar{\mathbb Q}_a, T\}
\eeq 
where \( \theta_a, \bar{\theta}^a\) are constant odd parameters and the graded commutators are defined in (\ref{supercomm}). The products explicitly read as
\begin{equation}\label{matrixtransf}
 \theta_a {\mathbb Q}^a = \left( \begin{matrix}
\theta_aQ^a &0\\
0& \theta_aQ^a
\end{matrix} \right) \qquad \qquad  \bar\theta^a \bar{\mathbb Q}_a = \left( \begin{matrix}
\bar\theta^a \bar{Q}_a &0\\
0& \bar\theta^a \bar{Q}_a
\end{matrix} \right)
\end{equation}

According to these definitions, the SUSY variation of the superconnection in \eqref{connecsu3} reads\footnote{For simplicity, here we set $\ell =1$.}
\begin{equation}
	[ {\mathbb Q}^a, \mathcal{L}] =\left(\begin{matrix}
		-\frac{4\pi i  }{k}\bar{\psi}_1 \bar{Y}^a & \; \; 0 \\ 2\sqrt{\frac{\pi}{k}}\left(i D_3 \bar{Y}^a+ i\frac{2\pi}{k}\left(\bar{Y}^a l_B-\hat{l}_B\bar{Y}^a\right)\right) & \; \; \frac{4\pi i}{k}\bar{Y}^a\bar{\psi}_1
	\end{matrix}\right)
	\label{}
\end{equation}
\begin{equation}
	[\bar{\mathbb Q}_a , \mathcal{L}] =\left(\begin{matrix}
		-\frac{4\pi}{k} Y_a \psi^1  &  \; \; \;  -2\sqrt{\frac{\pi}{k}}\left(i D_3 Y_a + \frac{2\pi i}{k}\left( Y_a \hat{l}_B- l_B Y_a\right)\right)  \\
		0& \; \; \; \frac{4\pi}{k}\psi^1 Y_a
	\end{matrix}\right)
	\label{}
\end{equation}
These identities can be rewritten as \cite{Drukker:2009hy, Bianchi:2020hsz} 
\begin{equation}
	[ {\mathbb Q}^a, \mathcal{L}] =i\partial_3 {\mathbb G}^a - \left[\mathcal{L},{\mathbb G}^a\right] \equiv i\mathfrak{D}_3 \mathbb{G}^a 
	\qquad [\bar{\mathbb Q}_a , \mathcal{L}] =-i\partial_3\bar{{\mathbb G}}_a +\left[\mathcal{L},\bar{{\mathbb G}}_a\right] \equiv -i\mathfrak{D}_3 \bar{{\mathbb G}}_a
	\label{eq:QtransfL2}
\end{equation}
where we have defined 
\begin{equation}\label{Gmatrices}
{\mathbb G}^a = 2 \sqrt{\frac{\pi}{k}} \begin{pmatrix}
0 & 0\\
\bar{Y}^a  & 0
\end{pmatrix} \qquad \qquad 
\bar {\mathbb G}_a = 2\sqrt{\frac{\pi}{k}} 
\begin{pmatrix}
0 & Y_a \\
0  & 0
\end{pmatrix} 
\end{equation}
and $\mathfrak{D}_3 = \partial_3 + i [ {\mathcal L}, \cdot \}$. 

As a consequence, a generic linear Wilson operator defined on a segment $[s_1,s_2]$ 
\begin{equation}\label{WL2}
W(s_2,s_1) =\mathcal P\exp (-i\int_{s_1}^{s_2} \!\!ds \, \mathcal L(s))   
\end{equation}
is not invariant under the action of SUSY charges \eqref{matrixQ}, rather it transforms as 
\begin{align}\label{GW}
& \delta W = [ \theta_a{\mathbb Q}^a, W] = G(s_2)W(s_2,s_1)-W(s_2,s_1)G(s_1) \nonumber \\
& \bar{\delta} W = [ \bar{\theta}^a\bar{\mathbb Q}_a, W] = \bar G(s_2)W(s_2,s_1)-W(s_2,s_1)\bar G(s_1) 
\end{align}
where $G \equiv \theta_a {\mathbb G}^a$, $\bar G \equiv -\bar\theta^a \bar {\mathbb G}_a$.

However, if we define the {\em covariant} SUSY charges 
\begin{equation}\label{covsusy}
{\mathcal Q}^a \equiv {\mathbb Q}^a - {\mathbb G}^a  = \begin{pmatrix} 
Q^a & 0 \\
-2 \sqrt{\frac{\pi}{k}} \, \bar{Y}^a & - Q^a \\
\end{pmatrix} \quad \quad \bar {\mathcal Q}_a \equiv \bar{\mathbb Q}_a  + \bar {\mathbb G}_a = \begin{pmatrix} 
\bar{Q}_a &  2 \sqrt{\frac{\pi}{k}} \, Y_a \\
0  & - \bar{Q}_a \\
\end{pmatrix}
\end{equation}
from \eqref{eq:QtransfL2} we obtain
\begin{equation} \label{Ltransf}
[ {\mathcal Q}^a, \mathcal{L} ] =  i \partial_3 {\mathbb G}^a \qquad \qquad [\bar {\mathcal Q}_a ,\mathcal{L}] = -i \partial_3 \bar {\mathbb G}_a
\end{equation}
or equivalently
\beq
\delta_{\mathcal Q}  \mathcal{L} \equiv [\theta_a {\mathcal Q}^a, \mathcal{L}] = i \partial_3 G \; , \qquad \quad \bar\delta_{\mathcal Q} \mathcal{L} \equiv [\bar\theta^a {\mathcal Q}_a, \mathcal{L}] = i \partial_3 \bar G
\eeq

Now, if, in addition, we define the non-local covariant variation 
\begin{equation}
\delta_{12}^{\mathcal Q}  \equiv \delta  - G(s_2) (\cdot) + (\cdot)G(s_1)
\end{equation}
from identities \eqref{GW} it follows that the invariance of the Wilson line under covariant transformations reads 
\begin{equation}\label{WQinvariance}
\delta_{12}^{\mathcal Q} W(s_2, s_1) = 0
\end{equation}

Similar results are obtained by applying the superconformal $S^a, \bar S_a$ charges to the superconnection. Defining the variations $\delta^{\text s} \equiv \lambda_a {\mathbb S}^a$ and  $\bar\delta^{\text s} \equiv \bar\lambda^a \bar{\mathbb S}_a$, we obtain that under superconformal transformations the superconnection transforms as 
\begin{equation}
	[ {\mathbb S}^a, \mathcal{L}] =i\partial_3 \left(s{\mathbb G}^a,\mathcal{L}\right] \equiv i\mathfrak{D}_3 \left(s{\mathbb G}^a\right) \qquad 
	[\bar{\mathbb S}_a , \mathcal{L}] =-i\partial_3\left(s\bar{{\mathbb G}}_a\right) -\left[s\bar{{\mathbb G}}_a,\mathcal{L}\right] \equiv -i\mathfrak{D}_3 \left(s\bar{{\mathbb G}}_a\right)
		\label{eq:QtransfL}
\end{equation}
with ${\mathbb G}^a$ and $\bar {\mathbb G}_a$ given in \eqref{Gmatrices}.

Now, defining covariant superconformal charges as 
\begin{equation}\label{covsuperconf}
{\mathcal S}^a \equiv {\mathbb S}^a - s{\mathbb G}^a  = \begin{pmatrix} 
S^a & 0 \\
-2s \sqrt{\frac{\pi}{k}} \bar{Y}^a & - S^a \\
\end{pmatrix} 
\quad \quad 
\bar {\mathcal S}_a \equiv \bar{\mathbb S}_a  + s\bar {\mathbb G}_a = \begin{pmatrix} 
\bar{S}_a &  2s \sqrt{\frac{\pi}{k}} Y_a \\
0  & - \bar{S}_a \\
\end{pmatrix}
\end{equation}
and covariant variations, $\delta_{{\mathcal S}} \equiv \lambda_a {\mathcal S}^a = \delta - sG$ and $\bar\delta_{{\mathcal S}}\equiv \bar\lambda^a \bar {\mathcal S}_a = \bar\delta - s\bar G$, a reasoning similar to the one which led to eq. \eqref{WQinvariance} allows concluding that the defect is invariant under the following {\em covariant} superconformal transformations 
\begin{equation}\label{WSinvariance}
\delta_{12}^{{\mathcal S}} W(s_2, s_1) = 0 \; , \quad {\rm with} \qquad \delta_{12}^{{\mathcal S}}   \equiv \delta  - sG(s_2) (\cdot) + (\cdot)sG(s_1)
\end{equation}

\subsection{The covariant $\mathfrak{su}(1,1|3)$ superconformal algebra}\label{sec:cov_alg}
From the previous analysis it follows that the correct supersymmetry and superconformal charges which leave the defect invariant are the covariant supercharges \eqref{covsusy} and \eqref{covsuperconf}. 

In order to construct the whole one-dimensional superconformal algebra, we start evaluating their anticommutators. As we are going to show, the main novelty is that these anticommutators close on a covariantized version of the $\mathfrak{su}(1,1|3)$ superalgebra, where the differential representation of the spacetime bosonic operators is given in terms of supercovariant derivatives \eqref{eq:QtransfL2} taken along the defect. 

To prove this statement we start acting with the covariantized anticommutator $\acomm*{\mathcal{Q}^b}{\bar{\mathcal{Q}}_c}$ on supermatrix local operators defined on the Wilson line, for instance ${\mathbb G}^a$ or $\bar{{\mathbb G}}_a$ in \eqref{Gmatrices}. We obtain
\begin{equation}
\acomm*{\mathcal{Q}^b}{\bar{\mathcal{Q}}_c} {\mathbb G}^a =-\delta^b_c\left(\partial_3 {\mathbb G}^a+i\comm*{\mathcal{L}}{{\mathbb G}^a}\right) = - \delta_b^c \, \mathfrak{D}_3 {\mathbb G}^a\,,
\end{equation}
and comparing it with the first identity in \eqref{anticomm} we find that $\mathcal{P} = -\mathfrak{D}_3$. 

Proceeding in an analogous way, we derive all the other (anti)commutators, and comparing them with the general structure of the $\mathfrak{su}(1,1|3)$ algebra given in appendix \ref{sect: su(1,1|3)}, we find the explicit realization of all the other generators. Details of the derivation and further examples are reported in Appendix \ref{app:closure}. Here we simply list the final result. We find that the expressions for the spacetime generators are given by 
\begin{equation}\label{covariant_gen}
{\mathcal P}=-\mathfrak{D}_3 \; , \qquad 
{\mathcal D}=-s\mathfrak{D}_3+\Delta\; , \qquad 
{\mathcal K}=-s^2\mathfrak{D}_3+2s\Delta
\end{equation}
where $\Delta$ is the scaling dimension\footnote{When this generator acts on supermatrix operators, it has to be thought as the supermatrix 
$
\begin{pmatrix}
\Delta & 0 \\
0  & \Delta
\end{pmatrix}
$.}. It is easy to prove that they satisfy the correct $\mathfrak{sl}(2)$ algebraic relations
\begin{equation}\label{sl2algebra}
[{\mathcal D}, {\mathcal P}]={\mathcal P}\qquad [{\mathcal D},{\mathcal K}]=-{\mathcal K}\qquad [{\mathcal P},{\mathcal K}]=-2{\mathcal D}
\end{equation}
in agreement with \eqref{su1,1}. Therefore, $\{{\mathcal P}, {\mathcal K}, {\mathcal D} \}$ correctly realise the covariant conformal algebra on the defect. 

These generators, together with the covariant supercharges \eqref{covsusy}, \eqref{covsuperconf}, the R-symmetry generators and the residual $\mathfrak{u}(1)_M$ symmetry generator \eqref{eq:M} suitably promoted to supermatrices, provide a representation of the $\mathfrak{su}(1,1|3)$ superalgebra on the space of supermatrices. 

The covariantization of the generators is required in order to make the superconformal algebra compatible with the gauge invariance on the defect generated by its superconnection. The net effect of the covariantization can be thought of as a modification of the supersymmetry generators obtained by adding a gauge transformation, in analogy with the "gauge-restoring" gauge transformations that modify SUSY transformations in order to preserve the Wess-Zumino gauge.

In section \ref{sec:dSCFT} we are going to use the covariantized supercharges to characterize the supersymmetry properties of the dCFT living on the fermionic Wilson line.

\section{The defect SCFT} \label{sec:dSCFT}
We now study the defect superconformal field theory (dSCFT) generated by local operators ${\mathcal O}$ defined as even/odd $U(N_1|N_2)$ supermatrices localized on the Wilson line and belonging to a given representation of the covariant superconformal algebra $\mathfrak{su}(1,1|3)$. They can be easily constructed by promoting ABJ(M) fields localized on the defect to supermatrices. One example is the ${\mathbb G}^a$ (or the $\bar{{\mathbb G}}_a$) supermatrix \eqref{Gmatrices} entering the covariant realization of supercharges. Supermatrix local operators are the natural objects on which the action of the supermatrix generators introduced in the  previous section is well defined. 

Local operators on the defect organize themselves into superconformal multiplets of the $\mathfrak{su}(1,1|3)$ algebra. These are generated by the repeated action of ${\mathcal Q}^a, \bar{\mathcal Q}_a$ supercharges on the superconformal primary (SCP), the lowest dimensional operator appearing in the multiplet. A supermultiplet can be decomposed into a finite sum of conformal multiplets, which generate from the repeated application of the covariant momentum generator ${\mathcal P}$ to superconformal descendants annichilated by ${\mathcal K}$.

Supermultiplet components are labeled by quantum numbers $[\Delta, m , j_1,j_2]$, where $\Delta$ is the conformal weight, $m$ the $\mathfrak{u}(1)$ charge associated with the $M$ generator, and $(j_1, j_2)$ are the eigenvalues corresponding to two $\mathfrak{su}(3)$ Cartan generators \cite{Bianchi:2017ozk, Bianchi:2020hsz, Gorini:2020new}. For details we refer to appendix \ref{sect: su(1,1|3)}.

\vskip 5pt
The physical observables of the dSCFT are correlations functions of supermatrix local operators, defined as 
\begin{align}\label{eq:def_cor}
\frac{ 
\langle \Tr W(+ \infty,s_n) {\mathcal O}(s_n)  W(s_n, s_{n-1}) {\mathcal O} (s_{n-1}) \cdots W(s_2, s_1) {\mathcal O}(s_1) W(s_1,-\infty)\rangle}{\langle W(+\infty, - \infty) \rangle} 
\end{align}
where the insertion of ``Wilson segments'' $W(s_j, s_{j-1})$ ensures manifest gauge invariance. 
This definition can be easily understood as the expectation value on the (suitably normalized) dressed vacuum $| 0 \rangle \! \rangle \equiv W(0,-\infty)|0\rangle$ (with $| 0 \rangle \!\rangle^\dagger  \equiv \langle 0|W(+\infty,0)$) of local operators translated along the line by the covariant translation generator ${\mathcal P}= -\mathfrak{D}_3$. In fact, using the explicit expression $\mathfrak{D}_3 = \partial_3 + i [ {\mathcal L} , \cdot \}$, one can easily check that 
\begin{equation}\label{eq:covtransl}
{\mathcal {O}}(0) \; \to \;  e^{-s{\mathcal P}} {\mathcal {O}}(0) e^{s{\mathcal P}} = W(0,s) {\mathcal O}(s) W(s,0) \equiv \tilde{\mathcal {O}}(s) 
\end{equation}
where ${\mathcal O}(s)$ is the original operator evaluated at point $s$, whereas we have dubbed $\tilde{\mathcal O}(s)$ the covariantly translated operator. Correlator \eqref{eq:def_cor} can then be rewritten as
\beq\label{eq:covcorr}
\langle \! \langle \, {\rm Tr}  \, \tilde{\mathcal {O}}(s_n)  \tilde{\mathcal O} (s_{n-1}) \cdots \tilde{\mathcal O}(s_1) \,  \rangle \! \rangle 
\eeq

This suggests a systematic way for constructing local operators of the dSCFT. Originally the defect inherits local operators from the bulk theory, which are localized at the origin. Gauge covariance then requires to translate the operators at a point $s$ by acting with the covariantized momentum generator. The result is that the operators get dressed with Wilson segments as in \eqref{eq:covtransl}. 

It is interesting to investigate how the rest of the covariantized generators work on the defect correlators. To begin with, we consider the action of the covariantized SUSY supercharges defined in \eqref{covsusy}. Taking for simplicity a one-point function and assuming that $\mathbb{G}^a\to 0$ for $s\to\pm\infty$, we find that its covariantized SUSY variation works as follows\footnote{We use the shorthand notation $W_{s_1,s_2} \equiv W(s_1,s_2)$, in particular $W_{+s} \equiv W(+\infty,s)$, $W_{s-} \equiv W(s,-\infty)$. We also neglect $\Tr$. \label{foot:notation}}
\begin{equation}\label{Delta}
\begin{aligned}
& \langle\!\langle \delta_{\mathcal Q} \mathcal \mathcal{\tilde{O}}(s)  \rangle\!\rangle  \equiv \langle W_{+s}[\delta_{\mathcal Q},\mathcal{O}(s)]W_{s-}\rangle 
=\langle W_{+s} \big([\delta_Q,\mathcal{O}(s)] - [G(s),\mathcal{O}(s)] \big) W_{s-} \rangle\\
&=\langle W_{+s}(\delta_Q\mathcal{O}(s)) W_{s-} \rangle-\langle W_{+s} G(s) \mathcal{O}(s) W_{s-}\rangle+\langle W_{+s} \mathcal{O}(s) G(s) W_{s-} \rangle\\
&= \langle\delta_Q(W_{+s} \mathcal{O}(s) W_{s-}) \rangle = 0 
\end{aligned}
\end{equation}
where $\delta_Q$ is the ordinary SUSY variation defined in \eqref{matrixtransf2}, and we have used that the ordinary vacuum  is killed by $\delta_Q$. 
This result implies that the dressed vacuum defining $\langle\! \langle \cdot \rangle\! \rangle$ on the l.h.s. of \eqref{Delta} is instead killed by the covariant supercharges ${\mathcal Q}^a$. 
Correlators \eqref{eq:covcorr} are therefore invariant under the action of covariantized supersymmetry generators, while ordinary SUSY $\delta$-variations would not leave them invariant. 
This is consistent with the observation that if two supercharges were to close on an ordinary translation generated by $\partial_{3}$, gauge invariance on the defect would be broken. We need to act with supercharges that close on $\mathfrak{D}_3$ to maintain gauge invariance.

One can recursively check that the same property holds for any higher-point correlator. 
Similarly, one can check that correlators \eqref{eq:covcorr} are invariant under the action of the covariantized $\mathcal S$ generators defined in \eqref{covsuperconf}. Therefore, we conclude that the covariantized algebra built above is the correct realization of the $\mathfrak{su}(1,1|3)$  superalgebra on the defect and definition \eqref{eq:covcorr} of correlation functions is consistent with it. 

\vskip 5pt
In order to make the previous discussion more concrete and open the possibility to evaluate correlators explicitly, we now proceed to the construction of some elementary $\mathfrak{su}(1,1|3)$ supermultiplets  of the dSCFT on the Wilson line. For a systematic classification of unitary representations of the $\mathfrak{su}(1,1|3)$ algebra on the rigid line we refer to \cite{Gorini:2020new} (see also appendix \ref{sect: su(1,1|3)} for a brief review). This classification can be easily adapted to the case of the dCFT without relevant modifications. A main difference arises, instead, in the actual realization of the multiplet components in terms of ABJ(M) elementary fields. This is due to the structural difference between the algebra generators defined on the rigid line and on the Wilson line. 

As a relevant example, in the next section we construct a new long multiplet living on the Wilson line, which does not have analogue in ABJ(M) and on the rigid line. We also review the construction of the displacement multiplet using the present approach of covariant supercharges realized as supermatrices.

\subsection{The lowest dimensional supermultiplet}\label{sect:T}
We observe that the covariantized  $\mathfrak{su}(1,1|3)$ generators are not just differential operators as in the ordinary case. Rather they acquire a non-trivial dependence on local fields from the covariantizing terms. This implies that when we look for superconformal primaries (SCPs) generating supermultiplets, we should also enlarge the spectrum to include constant operators. The action of the covariant SUSY charges on constant operators may lead to non-trivial local descendants originating from the multiplication with the covariantizing term. Here, we construct an example of such a multiplet.

Constant operators can be easily constructed as linear combinations of the ${\mathcal I}, {\mathcal T}$ operators, the natural basis of even $U(N_1|N_2)$ supermatrices, given by\footnote{The particular normalization of ${\mathcal T}$ is chosen for later convenience.} 
\beq 
\quad 
{\mathcal I} = \begin{pmatrix} 
\mathds{1}_{N_1} & 0 \\
0 & \mathds{1}_{N_2}
\end{pmatrix} \; , \qquad \qquad 
{\mathcal T} = \frac12 \begin{pmatrix} 
-\mathds{1}_{N_1} & 0 \\
0 & \mathds{1}_{N_2}
\end{pmatrix} 
\eeq

Now, while covariant SUSY charges \eqref{covsusy} act on ${\mathcal I}$ trivially, this is no longer the case for ${\mathcal T}$. We obtain
\beq\label{eq:Yop}
\left[ {\mathcal Q}^a(s) , {\mathcal T} \right] = {\mathbb G}^a(s) \; ,  \quad \quad  \left[ \bar{\mathcal Q}_a(s) , {\mathcal T} \right] = \bar{\mathbb G}_a(s)  \qquad a = 1,2,3
\eeq
where ${\mathbb G}^a(s)$ and $\bar{\mathbb G}_a(s)$ are the supermatrix operators that covariantize the SUSY charges (see eq. \eqref{Gmatrices}). In addition, the constant operator ${\mathcal T}$ satisfies 
\beq
\left[ {\mathcal S}^a(s) , {\mathcal T} \right] = s \, {\mathbb G}^a \qquad 
\left[ \bar{\mathcal S}_a(s) , {\mathcal T} \right]  = s \, \bar{\mathbb G}_a
\eeq
Therefore, at the origin ($s=0$) it is a superconformal primary (SCP), with quantum numbers $[0,0,0,0]$. 
Since ${\mathcal T}$ is not annihilated by any ${\mathcal Q}, \bar{\mathcal Q}$ supercharge, it is not protected and, as we show in the next section, it acquires an anomalous dimension. The repeated application of ${\mathcal Q}^a, \bar{\mathcal Q}_a$ generates a whole  $\mathfrak{su}(1,1|3)$ long multiplet that we now construct explicitly.

We organize the resulting descendant operators in terms of conformal primaries with a specific $M$-charge and R-symmetry representation. This implies that in the derivation of the descendants, we can neglect  $\mathcal{P}$-exact terms. In other words, we set $\mathcal{P}=-\mathfrak{D}_3=0$ and treat all the Poincar\'e supercharges as anticommuting. 

The R-symmetry representation of the descendants is determined by considering that $\mathcal{T}$ is an R-symmetry singlet. In contrast, in our conventions, $\bar{\mathcal Q}_a$ belongs to the fundamental representation $\mathbf{3}$ of the $SU(3)$ R-symmetry group, and ${\mathcal Q}^a$ belongs to the antifundamental $\bar{\mathbf{3}}$.  Due to the nature of $\mathcal{T}$, some representations are forbidden by the SUSY algebra, particularly those that would correspond to the application of symmetric configurations of supercharges.  

We organize the results in terms of the \emph{level} of a conformal primary. It is defined as the number of supercharges acting on the SCP. In the present case, taking into account that $\Delta({\mathcal T}) = 0$ and $\Delta({\mathcal Q}^a) = \Delta(\bar{\mathcal Q}_a)=1/2$ (see table \ref{table1}), a conformal primary at level $p$ has dimension $p/2$.

\begin{figure}
    \centering
    \includegraphics{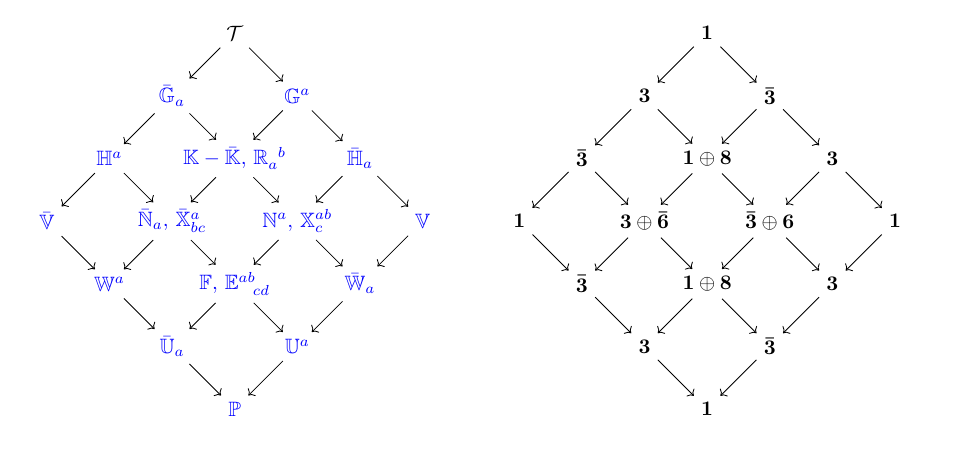}
    \caption{Diamonds corresponding to the \({\mathcal{T}}\) supermultiplet. Arrows pointing towards the left (right) mean the application of one $\bar{\mathcal Q}_a$ (${\mathcal Q}^a$). The left diagram shows how the various components have been named in the main text. The right diagram provides their decomposition in terms of $SU(3)$ irreducible representations.}
    \label{fig:supmultT}
\end{figure}

\paragraph{Level 1.} 
Referring to figure \ref{fig:supmultT}, at the first level we find $\bar{\mathbb G}_a$ and ${\mathbb G}^a$. They are superconformal descendants belonging to the fundamental and antifundamental representation of the $SU(3)$ R-symmetry group, respectively. Level 1 operators have quantum numbers $\Delta = 1/2$ and $m({\mathbb G}^a) = 1/2, m(\bar{\mathbb G}^a) = -1/2$. 

\paragraph{Level 2.}
This is obtained by acting with supercharges \eqref{covsusy} on $\bar{\mathbb G}_a$ and ${\mathbb G}^a$. According to the representation decomposition 
\beq 
\mathbf{3} \otimes \mathbf{3} =\mathbf{\Bar{3}} \oplus \mathbf{6} \nonumber
\eeq
we expect that acting with ${\mathcal Q}^a$ on ${\mathbb G}^b$, we obtain a $SU(3)$ antifundamental representation and a symmetric tensor one (similarly for its complex conjugate). However, taking into account the SUSY transformations given in appendix \ref{sect: su(1,1|3)}, we explicitly find 
\beq\label{eq:QG}
\{ \mathcal Q^a , {\mathbb G}^b \} = 2 \sqrt{\frac{\pi}{k}} \epsilon^{abc}  \begin{pmatrix}
0 & 0\\
\chi_c^2 & 0
\end{pmatrix} \equiv \epsilon^{abc}\, \bar{\mathbb{H}}_c
\qquad \quad 
\{ \bar{\mathcal Q}_a , \bar{\mathbb G}_b \}  = - 2i \sqrt{\frac{\pi}{k}}\epsilon_{abc} \begin{pmatrix}
0 & \bar\chi_2^c\\
0 & 0
\end{pmatrix} \equiv -i\epsilon_{abc}\, \mathbb{H}^c 
\eeq
Only one operator in the (anti)fundamental representation appears while the one in the $\boldmath{6}$ is missing.
The reason can be traced back to the fact that due to the anticommuting nature of the supercharges, it is impossible to realize a symmetric tensor in $(a,b)$ by applying ${\mathcal Q}^a, {\mathcal Q}^b$ to ${\mathcal T}$. In other words, a $\boldmath{6}$ symmetric tensor structure with $\Delta =m=1$  cannot be obtained from ABJ(M) elementary fields.

Similarly, according to the decomposition
\[ 
 \mathbf{3} \otimes \mathbf{\Bar{3}} = \mathbf{1} \oplus \mathbf{8} 
\]
applying ${\mathcal Q}^a$ to $\bar{\mathbb G}_b$ we should find a $SU(3)$ singlet and an adjoint. In fact, using SUSY transformations in appendix \ref{sect: su(1,1|3)}, we obtain
\begin{align}\label{eq:QY}
\{\mathcal Q^a , \bar{\mathbb G}_b \} & = 2 \sqrt{\frac{\pi}{k}}\delta_b^a \begin{pmatrix}
0 & \bar{\psi}_1\\
0 & 0
\end{pmatrix} - \frac{4\pi}{k} \begin{pmatrix}
 Y_b \bar Y^a & 0\\
0 & \bar Y^a Y_b
\end{pmatrix} \equiv  \delta_b^a \, \mathbb{K} -  \mathbb{R}\indices{_b^a} \\
\{ \bar{\mathcal Q}_a , {\mathbb G}^b \} & = 2 \sqrt{\frac{\pi}{k}}\delta_a^b \begin{pmatrix}
0 & 0\\
i \psi^1 & 0
\end{pmatrix} + \frac{4\pi}{k} \begin{pmatrix}
 Y_a \bar Y^b & 0\\
0 & \bar Y^b Y_a
\end{pmatrix} \equiv  \delta_a^b \, \bar{\mathbb{K}} + {\mathbb{R}}\indices{_a^b} \label{eq:barQbarY}
\end{align}
where we have defined 
\beq
\mathbb{K}\!=2 \sqrt{\frac{\pi}{k}}\!\begin{pmatrix}
-\frac{2}{3}\sqrt{\frac{\pi}{k}}Y_c\bar{Y}^c & \bar{\psi}_1 \\
0  &  -\frac{2}{3}\sqrt{\frac{\pi}{k}}\bar{Y}^cY_c
\end{pmatrix}\; , \qquad 
\bar{\mathbb{K}}\!=2 \sqrt{\frac{\pi}{k}}\! \begin{pmatrix}
\frac{2}{3}\sqrt{\frac{\pi}{k}}Y_c\bar{Y}^c & 0 \\
i\psi^1 &  \frac{2}{3}\sqrt{\frac{\pi}{k}}\bar{Y}^cY_c \nonumber
\end{pmatrix}
\eeq
\beq
{\mathbb{R}}\indices{_a^b}\!=\! \frac{4\pi}{k}\begin{pmatrix}
Y_a\bar{Y}^b\!-\!\frac{1}{3} \delta_a^b \, Y_c\bar{Y}^c & 0 \\
0 &  \bar{Y}^bY_a\!-\!\frac{1}{3}\delta_a^b \, \bar{Y}^cY_c
\end{pmatrix}
\eeq
Although the apparent existence of two singlets, one can easily check that 
\beq\label{eq:Bdescendant}
{\mathbb K} + \bar{\mathbb K} = [{\mathcal P}, {\mathcal T}]
\eeq
is a ${\mathcal T}$ descendant and can be removed from the spectrum. Therefore, at this level we have only one singlet $({\mathbb K} - \bar{\mathbb K})$ plus the adjoint operator ${\mathbb{R}}\indices{_a^b}$ and the two ${\mathbb H}^c, \bar{\mathbb H}_c$ (anti)fundamentals.
The adjoint operator would not be present in the non-interacting case (\(k\to \infty\)). The covariantization then acts by turning on states that are absent on the rigid line.

According to the classification of $\mathfrak{su}(1,1|3)$ representations summarised in appendix \ref{sect: su(1,1|3)}, it turns out that all the operators have $\Delta = 1$, whereas $m({\mathbb H}^c) = 1$, $m(\bar{\mathbb H}_c) = -1$ and $m(\mathbb{K}-\bar{\mathbb{K}}) = m(\mathbb{R}\indices{_a^b})=0$. 

\paragraph{Level 3.}
Using the previous arguments, acting with \({\mathcal{Q}}^a\) on \(\mathbb{{H}}_b\) (or \(\bar{\mathcal{Q}}_a\) on \(\mathbb{\bar{H}}^b\)) we should expect to produce one singlet and one adjoint, with quantum numbers $\Delta= -m = \tfrac32$.
The \(\mathbf{8}\) quantum numbers are incompatible with the gauge structure and the anticommuting nature of the supercharges. Thus this state is trivially zero. Instead, for the singlets we obtain
\begin{equation}
	\begin{aligned}
	&[\bar{Q}_a,\bar{\mathbb{H}}^{b}]=2\sqrt{\frac{\pi}{k}}\delta_a^{\;\;b}\begin{pmatrix}
		0 & \bar{D} Z \\ 0 & 0
	\end{pmatrix} \equiv \delta_a^{\;\;b}\;\bar{\mathbb{V}} \\ 
	&\left[\mathcal{Q}^a,\mathbb{H}_b\right]=-2i\sqrt{\frac{\pi}{k}}\delta^a_{\;\;b}\begin{pmatrix}
	0 & 0 \\ D \bar{Z} & 0
\end{pmatrix} \equiv -i\delta^{a}_{\;\;b} {\mathbb{V}}
\end{aligned}
	\label{}
\end{equation}
These operators appear at the two edges of the diamond in fig. \ref{fig:supmultT}. 

In order to obtain the ${\mathbb X}$ operator in the middle, we can either act with \(\bar{\mathcal{Q}}\) on the representation \(\mathbf{1} \oplus \mathbf{8}\) (operators \({\mathbb K}-\bar{\mathbb K}, {\mathbb R}\indices{_a^b}\)), or with $\mathcal{Q}$ on representation \(\mathbf{\Bar{3}}\) (\({\mathbb H}^a\) operator).
Compatibility between the two decompositions
\begin{equation}
    \mathbf{3}\otimes\left(\mathbf{1} \oplus \mathbf{8}\right)=\mathbf{3} \oplus \mathbf{3} \oplus \mathbf{\Bar{6}} \oplus \mathbf{15} \quad \qquad \quad 
    \mathbf{\Bar{3}}\otimes \mathbf{\Bar{3}}=\mathbf{3} \oplus \mathbf{\Bar{6}}
\end{equation}
implies that the additional operator in the $\mathbf{15}$ is trivially zero\footnote{We stress that the two fundamentals in the first decomposition, the one coming from \(\mathbf{3}\otimes\mathbf{1}\) and the one from \(\mathbf{3}\otimes \mathbf{8}\), are the same operator. We can write
\begin{align} 
\bar{Q}_a{\mathbb{R}_b}^c\Big|_{\mathbf{1}}\propto\delta_a^c \bar{Q}_k{\mathbb{R}_b}^k-\delta^c_b \bar{Q}_k{\mathbb{R}_a}^k \nonumber
\end{align}
and easily observe that $\bar{Q}_k{\mathbb{R}_a}^k\sim \bar{Q}_k Q^k \bar{Q}_a\mathcal{T}$, which up to descendants is the same as $\bar{Q}_a\left(\mathbb{K}-\bar{\mathbb{K}}\right)$.}. In conclusion, applying the explicit SUSY variations to the fields, at level 3 we find one extra fundamental operator ${\mathbb N}^a$ with $\Delta = \tfrac32$ and $m= -\tfrac12$, plus a $\bar{\mathbf{6}}$ operator ${\mathbb X}^{ac}_b $, with same quantum numbers, given explicitly by
\begin{align}
    &\mathbb{N}^a= 2\sqrt{\frac{\pi}{k}}\begin{pmatrix}
\frac{4}{3}\sqrt{\frac{\pi}{k}}\left(\bar\psi_1\bar Y^a\!-\!2\epsilon^{abc}Y_b\chi_c^2\right) \; & 0\\
\; -D_3\bar Y_a\!-\!\frac{2\pi}{3k}\left[\bar Y^a(3Z\bar Z\!+\!Y_k\bar Y^k)\!-\!(3\bar Z Z\!+\!\bar Y^k Y_k)\bar Y^a\right]  & -\frac{4}{3}\sqrt{\frac{\pi}{k}}\left(\bar\psi_1\bar Y^a-2\epsilon^{abc}Y_b\chi_c^2\right) 
\end{pmatrix}   \\
    & {\mathbb X}^{ac}_b = 2\sqrt{\frac{\pi}{k}}\begin{pmatrix}
2\sqrt{\frac{\pi}{k}}\epsilon^{cak}\left(Y_k\chi_b^2+Y_b\chi_k^2\right)&0\\
\frac{2\pi}{k}\left[\bar Y^c Y_b \bar Y^a\!+\!\frac{1}{2}\delta_b^c\left(\bar Y^aY_k\bar Y^k\!-\!\bar Y^kY_k\bar Y^a\right)\!-\!(c\leftrightarrow a)\right] \; 
& \; \; 2\sqrt{\frac{\pi}{k}}\epsilon^{ack}\left(\chi_b^2Y_k\!+\!\chi_k^2Y_b\right)
\end{pmatrix}
\end{align}
Similarly, the $\mathcal{Q}$ action on \(\mathbf{1} \oplus \mathbf{8}\) yields to two operators $\bar{\mathbb{N}}_a$, $\bar{\mathbb{X}}^a_{\;\;bc}$ transforming respectively in the $\bar{\mathbf{3}}$ and $\mathbf{6}$. Their field realization reads
\begin{align}
&\bar{\mathbb{N}}_a=2\sqrt{\frac{\pi}{k}}\begin{pmatrix}
\frac{4i}{3}\sqrt{\frac{\pi}{k}}\left(Y_a\psi^1\!-\!2\epsilon_{abc}\bar\chi^c_2\bar Y^b\right) \; & \; D_3 Y_a\!+\!\frac{2\pi}{3k}\left[Y_a(3\bar Z Z\!+\!Y_k\bar Y^k)\!-\!(3 Z \bar Z\!+\! Y_k \bar Y^k)Y_a\right] \\
0 & -\frac{4i}{3}\sqrt{\frac{\pi}{k}}\left(\psi^1Y_a-2\epsilon_{abc}\bar Y^b\bar\chi^c_2\right)
\end{pmatrix}\\
&\bar{\mathbb{X}}_{ac}^b=-2\sqrt{\frac{\pi}{k}}\begin{pmatrix}
-2\sqrt{\frac{\pi}{k}}\epsilon_{cak}\left(\bar\chi^k_2\bar Y^b+\bar\chi_2^b\bar Y^k\right)&
\frac{2\pi}{k}\left[ Y_c\bar Y^bY_a\!+\!\frac{1}{2}\delta_a^b\left(Y_c\bar Y^kY_k\!-\!Y_k\bar Y^kY_c\right)\!-\!(a\leftrightarrow c)\right]
\; \; \\
0 & \; 2\sqrt{\frac{\pi}{k}}\epsilon_{cak}\left(\bar Y^b\bar\chi^k_2\!+\!\bar Y^k\bar\chi_2^b\right)
\end{pmatrix}
\end{align}

\paragraph{Higher levels.} Starting from level 4, the explicit realization of the operators in terms of elementary fields becomes quite cumbersome and not very instructive. Therefore, here we simply discuss how the various structures emerge and refer to table \ref{table:Tmultiplet} for a summary of the multiplet components and their quantum numbers. 

Level 4 is obtained by acting on $\mathcal{T}$ either with three $\mathcal{Q}^a$ ($\bar{\mathcal{Q}}_a$) and one $\bar{\mathcal{Q}}_a$ ($\mathcal{Q}^a$) or with two $\mathcal{Q}^a$ and two $\bar{\mathcal{Q}}_a$. In the former case, the SUSY algebra fixes the only possible state to be of the form
$$
\mathbb{W}^a\sim \epsilon^{klm}\bar{\mathcal{Q}}_k\bar{\mathcal{Q}}_l\bar{\mathcal{Q}}_m \mathcal{Q}^a \mathcal{T}\,, \qquad
\bar{\mathbb{W}}_a \sim \epsilon_{klm}\mathcal{Q}^k\mathcal{Q}^l\mathcal{Q}^m\bar{\mathcal{Q}}_a  \mathcal{T}
$$
up to descendants. For the remaining combination of supercharges, the only non-vanishing state comes from $\epsilon_{akl}\epsilon^{bcd}\mathcal{Q}^k\mathcal{Q}^l \epsilon^{klm}\bar{\mathcal{Q}}_c\bar{\mathcal{Q}}_d\mathcal{T}$. It can be easily decomposed in $\mathbf{1}\oplus\mathbf{8}$, giving rise to a singlet ${\mathbb F}$ and a tensor $\mathbb{E}^{ab}_{\;\;\;cd}$. 

Similarly, at level 5 the states are of the form
$$
\bar{\mathbb{U}}^a \sim \epsilon^{abc}\epsilon_{klm}\mathcal{Q}^k\mathcal{Q}^l\mathcal{Q}^m\bar{\mathcal{Q}}_b\bar{\mathcal{Q}}_c  \mathcal{T}\,, \qquad
\mathbb{U}_a\sim \epsilon_{abc}\epsilon^{klm}\bar{\mathcal{Q}}_k\bar{\mathcal{Q}}_l\bar{\mathcal{Q}}_m \mathcal{Q}^b\mathcal{Q}^c \mathcal{T}
$$

Finally, the singlet at level 6 comes from the only non-vanishing contractions of the supercharges, namely that with two epsilon tensors.

\begin{table}[h!]
\begin{center}
\begin{tabular}{|c|c|c|} 
\hline
Level & Irrep & Op name \\
\hline\hline
\multirow{1}{5em}{0} & $[\mathbf{1}]_\Delta^0$ & ${\mathcal{T}}$ \\
\hline
\multirow{2}{5em}{1} & $[\mathbf{3}]_{\Delta+1/2}^{-1/2}$ & $\mathbb{G}_a$\\ 
& $[\mathbf{\bar{3}}]_{\Delta+1/2}^{1/2}$ & $\Bar{\mathbb{G}}^a$\\ 
\hline
\multirow{4}{5em}{2} & $[\mathbf{\bar{3}}]_{\Delta+1}^{-1}$ & $\bar{\mathbb{H}}^a$\\ 
& $[\mathbf{1}]_{\Delta+1}^0$ & $\mathbb{K}$\\
& $[\mathbf{8}]_{\Delta+1}^0$ & $\mathbb{R}_a^{\;\;b}$\\
& $[\mathbf{3}]_{\Delta+1}^1$ & $\mathbb{H}_a$\\ 
\hline
\multirow{4}{5em}{3} & $[\mathbf{1}]_{\Delta+\frac{3}{2}}^{-\frac{3}{2}}$ & $\mathbb{V}$\\ 
& $[\mathbf{3}]_{\Delta+3/2}^{-1/2}$ & $\mathbb{X}_a$\\
& $[\mathbf{\Bar{6}}]_{\Delta+3/2}^{-1/2}$ & $\Bar{\mathbb{Y}}_{ab}$\\
& $[\mathbf{\bar{3}}]_{\Delta+3/2}^{1/2}$ & $\bar{\mathbb{X}}^a$\\
& $[\mathbf{6}]_{\Delta+3/2}^{1/2}$ & $\mathbb{Y}_{ab}$\\
& $[\mathbf{1}]_{\Delta+3/2}^{3/2}$ & $\bar{\mathbb{V}}$\\ 
\hline
\multirow{4}{5em}{4} & $[\mathbf{\bar{3}}]_{\Delta+2}^{-1}$ & $\bar{\mathbb{W}}^a$ \\ 
& $[\mathbf{1}]_{\Delta+2}^0$ & $\mathbb{F}$ \\
& $[\mathbf{8}]_{\Delta+2}^0$ & $\mathbb{E}_a^{\,\;b}$\\
& $[\mathbf{3}]_{\Delta+2}^1$ & $\mathbb{W}_a$\\ 
\hline
\multirow{2}{5em}{5} & $[\mathbf{3}]_{\Delta+5/2}^{-1/2}$ & $\mathbb{U}_a$\\ 
& $[\mathbf{\bar{3}}]_{\Delta+5/2}^{1/2}$ & $\bar{\mathbb{U}}^a$ \\ 
\hline
\multirow{1}{5em}{6} & $[\mathbf{1}]_{\Delta+3}^0$ & $\mathbb{P}$ \\
\hline\hline
\end{tabular}
\end{center}
\caption{The list of operators in the ${\mathcal T}$ supermultiplet with their quantum numbers. We use the notation $[\mathbf{A}]_\Delta^m$, where $\mathbf{A}$ is the irrep of SU(3), $\Delta$ the scaling dimension, and $m$ the eigenvalue of the $U(1)$ generator M.}
\label{table:Tmultiplet}
\end{table}

\vskip 10pt
We close this section with a couple of further observations. 

First, we note that the SCP ${\mathcal T}$, though trivially constant, is not covariantly constant. Acting with the covariant momentum ${\mathcal P}$ according to prescription \eqref{eq:covtransl}, we find that under translation along the line, it gets mapped to
\beq\label{eq:Ttilde}
{\mathcal T} \rightarrow \tilde{\mathcal T}(s)  = W(0,s) {\mathcal T} \, W(s,0)
\eeq
However, since the covariant supercharges commute with the covariant momentum ${\mathcal P}$, identities \eqref{eq:Yop} remain true also for the covariantly translated operators. Using \eqref{eq:covtransl}, they get the form 
\beq\label{eq:easycov}
\tilde{{\mathbb G}}^a(s)  =
 \left[ {\mathcal Q}^a(s) , \tilde{\mathcal T}(s) \right] \qquad \qquad
\tilde{\bar{\mathbb G}}_a(s)  = 
 \left[ \bar{\mathcal Q}_a(s) , \tilde{\mathcal T}(s) \right]
\eeq
The further application of covariant supercharges works similarly and leads to constructing the whole supermultiplet at point $s$. 
We note that, as a consequence of \eqref{eq:Ttilde}, away from the origin the action of the superconformal charges is no longer trivial, but gives $[ {\mathcal S}^a(s) , \tilde{\mathcal T}(s) ] = {\mathcal Q}^a(s)$--exact and  $[ \bar{\mathcal S}_a(s), \tilde{\mathcal T}(s) ]  = \bar{\mathcal Q}_a(s)$-exact.

The second observation arises from comparing ABJ(M) operators localized on the rigid line and those defined on a Wilson line. There is, in fact, a highly non-trivial difference in the nature of the operators they give rise to in the two cases. 

Let's consider, for instance, the ABJ(M) elementary scalars $Y_a, \bar{Y}^a$, $a=1,2,3$. When localized on the rigid line, they give rise to $1/2$-BPS operators, killed by three of the six Poincar\`e supercharges preserved by the line\footnote{Rigorously speaking, these are not well-defined operators on the line, as they are not gauge invariant. One should rather consider combinations of the form $\Tr (Y_a\bar{Y}^a)$ as the building blocks of the local sector on the line. However, since gauge invariance does not play any role in the present discussion, we prefer to simplify the discussion by looking directly at $Y_a$.}. As such, they turn out to be the SCP of short multiplets \cite{Gorini:2020new}. For example, in the notations of appendix \ref{sect: su(1,1|3)}, $Y_1$ generates the $\mathcal{B}^{\frac{1}{3},\frac{1}{6}}_{-\frac{1}{2},1,0}$ multiplet.
Their scaling dimension is protected against quantum corrections \cite{Gorini:2020new}. 

Instead, when $Y_a, \bar{Y}^a$ are localized on the Wilson line and promoted to supermatrices, they give rise to $\bar{\mathbb G}_a$ and ${\mathbb G}^a$ operators, which are killed only by one covariant Poincar\`e supercharge. As discussed above, they are no longer SCPs. Rather they are the level 1 descendants of ${\mathcal T}$. Moreover, they belong to a long multiplet. Thus, they are expected to develop an anomalous dimension at the quantum level. We will return to this point in section \ref{sect:perturbative} where we compute their defect two-point function perturbatively. Here we provide a simple algebraic argument that explains why these operators are no longer protected on the Wilson defect. 

We consider the $\bar{\mathbb G}_1$ operator at the origin and act on it with a particular combination of covariant generators
\beq\label{eq:comb}
[-({\mathcal D}+{\mathcal M}) + {{\mathcal R}_1}^1+2{{\mathcal R}_2}^2, \bar{\mathbb G}_1 ]  \equiv [ \acomm*{\bar{\mathcal Q}_1-2{\mathcal Q}^2}{{\mathcal S}^1+\bar{\mathcal S}_2}, \bar{\mathbb G}_1]
\eeq
The l.h.s. of this expression gives $-(\Delta - 1/2) \bar{\mathbb G}_1$, whereas evaluating the r.h.s. we obtain $[\bar{\mathcal S}_2, \{ {\mathcal Q}^2 , \bar{\mathbb G}_1 \}]$ which is not vanishing, as it can be easily checked using SUSY transformations of appendix \ref{susytransfs}. Therefore, identity \eqref{eq:comb} leads to conclude that $\Delta(\bar{\mathbb G}_1) \neq 1/2$, i.e. the operator acquires non-trivial quantum dimension. The same argument holds for $\bar{\mathbb G}_2, \bar{\mathbb G}_3$ by suitably changing the linear combination of generators in \eqref{eq:comb}. We note that this result is a direct consequence of the fact that $\bar{\mathbb G}_1$ is annihilated by at most one supercharge. In particular, it is not killed by ${\mathcal Q}^2$. 
On the rigid line where instead $ \left[{\mathcal Q}^2 , Y_1 \right] = 0$, the same argument concludes that the operator is protected. 

\subsection{The displacement supermultiplet}
The displacement supermultiplet is the $su(1,1|3)$ multiplet containing the displacement operator as the top component, the operator that measures the breaking of translation invariance in the directions orthogonal to the Wilson line. 
The supermultiplet components have been worked out in \cite{Bianchi:2020hsz} by applying covariant SUSY transformations to the SCP, which in terms of the ABJ(M) elementary fields is given by\footnote{We focus on the $U(N_1|N_2)$ defect theory. A similar construction holds for its dual too.}
\begin{equation}\label{eq:Z}
{\mathbb{Z}} =2\sqrt{\frac{\pi}{k}} \begin{pmatrix}
0 & Z \\
0 & 0
\end{pmatrix} \qquad \qquad 
\bar{\mathbb{Z}}=2\sqrt{\frac{\pi}{k}}\begin{pmatrix}
0 & 0\\
\bar{Z} & 0
\end{pmatrix}
\end{equation}
where the normalization factor has been chosen for later convenience\footnote{Our definition of the SCP differs from the one in \cite{Bianchi:2020hsz} by the absence of an overall constant spinor. In fact, with our conventions on supermatrices - see appendix A - operator \eqref{eq:Z} has an automatically spinorial (odd) nature.}. These operators have quantum numbers $\Delta = 1/2$, $m = \pm 3/2$, respectively and are both R-symmetry singlets. 

Here, we quickly re-derive the whole supermultiplet by applying the supermatrix version of SUSY charges introduced in the previous sections. This helps us check the consistency of our covariant generators and, at the same time, fix notations. 

Contrary to what happens with the ${\mathbb G}^a, \bar{\mathbb G}_a$ triplets, the singlet operators maintain the same nature when they are defined on the rigid line or the Wilson line. In fact, on the 1/2-BPS line the $Z, \bar{Z}$ operators are annihilated by all the $\bar{Q}^a$ and all the $Q_a$, respectively, and therefore they generate the $\mathcal{B}^{0,\frac12}_{\frac32, 0,0}$ and $\mathcal{B}^{\frac12,0}_{-\frac32, 0,0}$ short multiplets \cite{Bianchi:2017ozk, Bianchi:2020hsz}. Studying the action of covariant supercharges \eqref{covsusy} on the ${\mathbb{Z}}, \bar{\mathbb{Z}}$ operators it is easy to realize that the same property survives on the Wilson line, that is $\{ \bar{\mathcal  Q}_a , \mathbb{Z} \}= \{ \mathcal Q^a , \bar{\mathbb{Z}} \} = 0, \, a=1,2,3$. In this case, covariantization only affects the action of non-annihilating supercharges. It follows that operators \eqref{eq:Z} are still the superprimaries of the short multiplets 
$\mathcal{B}^{0,\frac12}_{\frac 32, 0,0}$ and $\mathcal{B}^{\frac12,0}_{-\frac 32, 0,0}$. Consequently, they are expected to be protected from acquiring anomalous dimensions at the quantum level. In section \ref{sect:perturbative} we will give a perturbative confirmation of this expectation.

We now construct the whole supermultiplet by acting with supermatrix covariantized charges. For simplicity, we focus on the supermultiplet generated by ${\mathbb Z}$, but a similar procedure can be easily implemented on $\bar{\mathbb Z}$.

At level 1 we find 
\begin{equation}
	\mathbb{O}^a \equiv \{ \mathcal{Q}^a,\mathbb{Z}\}=-2\sqrt{\frac{\pi}{k}}\left(\begin{matrix}
		2\sqrt{\frac{\pi}{k}}Z\bar{Y}^a & \bar{\chi}^a_1 \\ 
		0 & 2\sqrt{\frac{\pi}{k}}\bar{Y}^a Z
	\end{matrix}\right)
	\label{eq:operOfromvariation}
\end{equation} 

Acting once more with one \(\mathcal{Q}^a\), at level 2 we obtain
\begin{equation}
	[ \mathcal{Q}^a , \mathbb{O}^b ] =\epsilon^{abc}\mathbb{\Lambda}_c
\end{equation}
with
\begin{equation}
 \mathbb{\Lambda}_c=2\sqrt{\frac{\pi}{k}}\left(\begin{matrix}
		2\sqrt{\frac{\pi}{k}}\left(\epsilon_{cde}\bar{\chi}_1^d \, \bar{Y}^e + Z\chi_c^2\right) & i D Y_c \\
		\frac{4\pi}{k}\, \epsilon_{cde}\bar{Y}^d \, Z \bar{Y}^e & 2\sqrt{\frac{\pi}{k}}\left(\epsilon_{cde}\bar{Y}^d\bar{\chi}_1^e-\chi_c^2 Z\right)
	\end{matrix}\right)
	\label{}
\end{equation}

\begin{figure}
    \centering
    \subfigure[]{\begin{tikzpicture}
        \begin{feynman}
            \vertex (a) {\(\mathbb{Z} \; [\mathbf{1}]_{1/2}^{3/2}\)};
			\vertex[below left= of a] (b) {\(\mathbb{O}^a \; [\bar{\mathbf{3}}]_{1}^{2}\)};
			\vertex[below left=of b] (c) {\(\mathbb{\Lambda}_a \; [\mathbf{3}]_{3/2}^{5/2}\)};
			\vertex[below left=of c] (d) {\(\mathbb{D} \; [\mathbf{1}]_{2}^{3} \)};
			\draw[->] (a) -- (b);
			\draw[->] (b) -- (c);
			\draw[->] (c) -- (d);
        \end{feynman}
    \end{tikzpicture}\label{fig:displa}}\qquad 
    \subfigure[]{\begin{tikzpicture}
        \begin{feynman}
            \vertex (a) {\(\bar{\mathbb{Z}} \; [\mathbf{1}]_{1/2}^{-3/2}\)};
			\vertex[below right= of a] (b) {\(\bar{\mathbb{O}}_a \; [\mathbf{3}]_{1}^{-2}\)};
			\vertex[below right=of b] (c) {\(\bar{\mathbb{\Lambda}}^a \; [\bar{\mathbf{3}}]_{3/2}^{-5/2}\)};
			\vertex[below right=of c] (d) {\(\bar{\mathbb{D}} \; [\mathbf{1}]_{2}^{-3} \)};
			\draw[->] (a) -- (b);
			\draw[->] (b) -- (c);
			\draw[->] (c) -- (d);
        \end{feynman}
    \end{tikzpicture}\label{fig:displb}}
    \caption{The displacement supermultiplet and its hermitian conjugate.}
    \label{fig:displsuperm}
\end{figure}
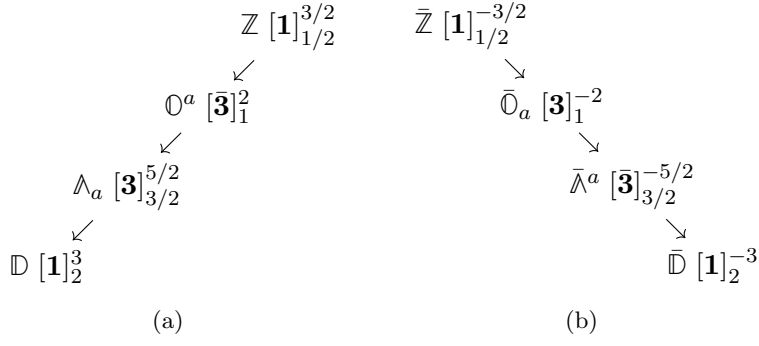

Finally, at level 3 we write
\begin{equation}
	\mathbb{D} \equiv \frac{1}{3!}\epsilon_{abc}\{\mathcal{Q}^a, [ \mathcal{Q}^b, \{\mathcal{Q}^c ,\mathbb{Z}\}]\}  
	=\frac{1}{3}\{\mathcal{Q}^a , \mathbb{\Lambda}_a\}  
\end{equation}
and the displacement operator is explicitly given by 
\begin{equation}
	\resizebox{\hsize}{!}{\(\mathbb{D}=i\left(\begin{matrix}
		\frac{4\pi}{k}\left(ZD\bar{Z}-DY_a\bar{Y}^a+i\bar{\chi}_1^a\chi_a^2\right) & 2\sqrt{\frac{\pi}{k}} D \bar{\psi}_1 \\
		8i\left(\frac{\pi}{k}\right)^{\frac{3}{2}}\left(\bar{Y}^a Z\chi_a^2-\chi_a^2Z\bar{Y}^a+\epsilon_{abc}\bar{Y}^a\bar{\chi}_1^b\bar{Y}^c\right) &\frac{4\pi}{k}\left(D\bar{Z}Z-\bar{Y}^aDY_a-i\chi_a^2\bar{\chi}_1^a\right)
	\end{matrix}\right)\)}
	\label{eq:fallingdispl}
\end{equation}
with the covariant derivative $D$ defined in \eqref{eq:derivatives}. 
The quantum numbers of these operators are reported in figure \ref{fig:displsuperm}. 

The barred operators (see fig. \ref{fig:displb}) can be obtained in a similar way acting multiple times with \(\Bar{\mathcal{Q}}_a\) on the superprimary \(\bar{\mathbb{Z}}\).

\subsection{Symmetry breaking and defect deformations}
One way to generate insertions of local primary operators on the defect is by acting with bulk symmetry generators broken by the defect's presence. This can be easily understood by observing that if we vary the Wilson line with respect to a broken symmetry, at first order in the deformation parameter, we bring down a new local operator $\delta {\cal L}$ according to
\beq\label{eq:wavyline}
\frac{\langle (\delta W) \cdots  \rangle}{\langle W \rangle} = -i \int  ds \, \langle \! \langle \delta {\cal L}(s) \cdots \rangle \! \rangle
\eeq
For a generic variation $\delta {\cal L}\equiv [\epsilon U, {\cal L} ]$ where $U$ is any of the broken generators, this identity can be more formally expressed as 
\beq
[U, W] = \int ds \, {\cal C}(s)  \, W
\eeq
where ${\cal C}(s) \equiv [U, -i{\cal L}(s) ]$ is the primary operator inserted on the defect. 

Many structural theorems follow from this set of identities, together with the algebra of (anti)commutators, which constrain the organization of these operators inside $\mathfrak{su}(1,1|3)$ supermultiplets \cite{Agmon:2020pde}. 

In particular, the conformal primary operators in the displacement supermultiplet reviewed above are associated with the action of the bulk superconformal generators broken by the Wilson line \cite{Cooke:2017qgm,Bianchi:2017ozk}. 
To be concrete, Eq. \eqref{eq:wavyline} for the broken translations $P_i$, with $i=1,2$ 
\beq
[P_i, W] = \int ds \, {\mathbb{D}_i}(s)  \, W
\eeq
provides an explicit definition for the displacement operator. The corresponding multiplet also includes the operators associated with half broken supersymmetries ${\mathbb \Lambda}_a, \bar{\mathbb \Lambda}^a$, as well as the ${\mathbb O}^a, \bar{\mathbb O}_a$ operators from the action of broken $SU(4)/SU(3)$ R-symmetry generators. 

As a consistency check of our construction, below, we review the action of transverse translations to check that we obtain precisely the displacement operator in \eqref{eq:fallingdispl} constructed by acting with the covariant generators.
Moreover, we study the action of the would-be broken $U(1)_b$ symmetry and explain why the Wilson line does not break this symmetry. 
Finally, we will use the wavy-line formalism to discuss the fate of the $\mathfrak{u}(1)_B$ symmetry. As a byproduct, we give an alternative motivation to consider $\mathcal{T}$ as a genuine defect operator. 

\subsubsection{The wavy-line}
Deforming a generic contour as \(x^\mu(s) \to x^\mu(s) +\delta x^\mu(s)\), the variation of the corresponding fermionic Wilson loop at first order in $\delta x^\mu$ leads to the insertion of the displacement operator,
whose explicit expression is given by \cite{Bianchi:2017ozk,Cooke:2017qgm} 
\begin{equation}
	\delta {\mathcal L}|_{transl} \equiv \tilde{\mathbb{D}}=\delta x^\mu \left(-i\dot{x}^\nu \mathbb{F}_{\mu\nu}+|\dot{x}| \mathcal{D}_{\mu}\mathbb{O}\right)+\frac{\dot{x}\cdot \delta \dot{x} }{|\dot{x}|}\mathbb{O}
	\label{eq:wavydisplacement}
\end{equation}
Here we have defined
\begin{equation}
	\mathbb{F}_{\mu\nu}=\left(\begin{matrix}
		F_{\mu\nu} & 0 \\ 0 & \hat{F}_{\mu\nu}
	\end{matrix}\right)=\de_\mu \mathcal{A}_\nu-\de_\nu \mathcal{A}_\mu+i[\mathcal{A}_\mu,\mathcal{A}_\nu],\qquad\mathcal{A}_\mu = \frac{1}{\sqrt{k}} \left(\begin{matrix}
		A_\mu & 0 \\ 0 & \hat{A}_\mu
	\end{matrix}\right)
	\label{opfinal}
\end{equation}
and 
\begin{equation}
	\mathbb{O}=\left(\begin{matrix}
		-\frac{2\pi}{k}\tensor{M}{^I_J}C_I\bar{C}^J & \sqrt{\frac{2\pi}{k}}\eta_I\bar{\psi}^I \\ -\sqrt{\frac{2\pi}{k}}\psi_I\bar{\eta}^I & -\frac{2\pi}{k}\tensor{M}{_J^I}\bar{C}^J C_I
	\end{matrix}\right) \, , \qquad {\rm with} \quad \mathcal{D}_{\mu}\mathbb{O}=\de_\mu\mathbb{O}+i\left[\mathcal{A}_\mu,\mathbb{O}\right] 
	\label{eq:supercovder}
\end{equation}

Specializing to the line \(x^\mu(s)=(0,0,s)\), we choose a deformation \(\delta x^\mu(s)=(\epsilon^1(s),\epsilon^2(s),0)\) orthogonal to the defect. The general expression for the displacement then reduces to 
\begin{equation}
	\tilde{\mathbb{D}}_{line}=\epsilon^k\left(-i\dot{x}^3\mathbb{F}_{k3}+\mathcal{D}_k\mathbb{O}_l\right) \equiv \epsilon^k \, \mathbb{D}_k
	\qquad k=1,2
	\label{Dline}
\end{equation}
and the operator in \eqref{eq:supercovder} reads
\begin{equation}
	\mathbb{O}_{line} =\left(\begin{matrix}
		\frac{2\pi}{k}\left(Z \bar{Z}-Y_a \bar{Y}^a \right) & 2\sqrt{\frac{\pi}{k}}\bar{\psi}_1 \\ -2i\sqrt{\frac{\pi}{k}}\psi^1 & \frac{2\pi}{k}\left(\bar{Z}Z-\bar{Y}^a Y_a\right)
	\end{matrix}\right)
	\label{eq:Ol}
\end{equation}
In particular, if we now consider the complex combination corresponding to the choice \(\epsilon^k=(1,-i)\) of the deformation parameters\footnote{The hermitian conjugate \(\bar{\mathbb{D}} \equiv \mathbb{D}_1 + i \mathbb{D}_2\) can be obtained taking the conjugate deformation parameters, i.e. \(\epsilon^k=(1,i)\).}
\begin{equation}
	\mathbb{D} \equiv \mathbb{D}_1 - i \mathbb{D}_2 =-\left(\mathbb{F}_{23}+i\mathbb{F}_{13}\right)+ D\mathbb{O}_{line}
	\label{eq:displquas}
\end{equation}
with the $D$ derivative defined in \eqref{eq:derivatives}, and use the equations of motion for the gauge fields and the fermion \(\psi^1\), we can easily prove that this operator coincides with the top component \eqref{eq:fallingdispl} of the displacement multiplet, up to the total covariant derivative
\begin{equation}
-2i\sqrt{\frac{\pi}{k}} \, \mathfrak{D}_3 \left(\begin{matrix}
		0 & 0 \\ \psi^2 & 0
	\end{matrix}\right)
	\label{eq:finaltildeD}
\end{equation}
This is the expected result. In fact, since the correlator at the r.h.s. of \eqref{eq:wavyline} is integrated along the contour, the operator insertion is always defined up to a total covariant derivative along the defect \cite{Bianchi:2020hsz}. Assuming that the correlators decay quickly enough at infinity, it is not hard to show that
\begin{equation}\label{eq:nullvar}
\int  ds \, \langle \! \langle\mathfrak{D}_3 {\cal O}(s)\cdots \rangle \! \rangle= \int ds\;  \partial_s  \langle \! \langle {\cal O}(s)\cdots \rangle \! \rangle=0  
\end{equation}
where the dots indicate possible insertions of local operators away from $s$. 

In the present framework, the identification between the operator insertion generated by the ``wavy line'' and the operator in \eqref{eq:fallingdispl}  has an even simpler explanation: Their difference \eqref{eq:finaltildeD} is a conformal descendant, but the supermultiplet construction of the previous section is blind to descendants. In conclusion, this derivation represents a non-trivial consistency check of the covariant superalgebra constructed in section \ref{sec:cov_alg} and its representations studied in this section. 

\subsubsection{The $\mathfrak{u}(1)_B$ variation}
We now consider the action of the would-be broken generator $B=M_{12}+2i {J_1}^1$ of \eqref{eq:B}. It generates the abelian factor $\mathfrak{u}(1)_B$. Being a linear combination of the transverse rotations and one broken R-symmetry generator orthogonal to the preserved $\mathfrak{u}(1)_M$ generator \eqref{eq:M}, it is supposed to be broken by the Wilson line. 

Applying $\delta_B$ to the Wilson line, the associated $\delta_B{\mathcal L}$ is non-vanishing due to a non-trivial transformation of the fermions
\beq\label{eq:rotation}
\delta_B \psi^{(1)} = -i \psi^{(1)}\; , \qquad \delta_B \bar\psi_{(1)} = i\bar\psi_{(1)} 
\eeq
According to identity \eqref{eq:wavyline}, this leads to the insertion of the defect operator 
\begin{equation}\label{eq:Bop}
\mathbb{B} = -2 \sqrt{\frac{\pi}{k}} \begin{pmatrix}
0 & \bar{\psi}_{(1)} \\
i\psi^{(1)} & 0
\end{pmatrix}
\end{equation}
However, it is easy to realize that $\mathbb{B} = -(\mathbb{K} + \bar{\mathbb{K}})$, where $\mathbb{K} + \bar{\mathbb{K}}$ is the descendant \eqref{eq:Bdescendant} appearing at level 2 of the ${\mathcal T}$ supermultiplet. Since it is a total covariant derivative, because of \eqref{eq:nullvar}, its contribution to the r.h.s. of \eqref{eq:wavyline} vanishes and we eventually obtain that $\delta_B W = 0$. It follows that $B$ is preserved, even in the presence of the Wilson line.

This proof that the Wilson line preserves the $U(1)_B$ symmetry is alternative to the argument of \cite{Agmon:2020pde} based on the fact that the non-trivial rotation \eqref{eq:rotation} of fermions can always be compensated by a gauge transformation. The relation between the two arguments relies on the fact that $\cal{T}$ is precisely the generator of the gauge transformation of \cite{Agmon:2020pde} \footnote{More generally, the $\cal{T}$ operator can be seen as a particular representative of a one-parameter family of operators 
\begin{equation}
\mathcal{T}_{\alpha} = -\frac12 \begin{pmatrix}
\mathds{1}_{N_1} & 0 \\
0 & e^{i\alpha} \mathds{1}_{N_2} 
\end{pmatrix} \qquad \quad \alpha \in \mathbb{R}    
\end{equation}
which generate $\mathbb{C}^\ast$ global gauge symmetry of the Wilson line under global gauge transformations ${\mathcal L} \to \mathcal{T}_{\alpha}{\mathcal L} \mathcal{T}_{\alpha}^{-1}$. This symmetry can be traced back to the freedom of fixing the phase of the fermionic couplings $\eta, \bar{\eta}$ defined in eqs. (\ref{eq:couplings}, \ref{eq:couplingsprop}). We thank Nadav Drukker for pointing this out.}.

As noticed in \cite{Billo:2016cpy, Agmon:2020pde}, if the transverse rotations are preserved, their action on the defect yields a descendant operator. What is relevant here is that the primary of the descendant operator is precisely $\cal{T}$. This fact provides further evidence that $\cal{T}$ is a building block of the dCFT on the Wilson line. 

\subsection{The cohomological equivalence revised}
The constant operator ${\mathcal T}$ turns out to play an interesting role also in connection with the cohomological equivalence between the bosonic 1/6 BPS and the fermionic 1/2 BPS Wilson Lines, discovered in \cite{Drukker:2009hy}. 

In fact, using the covariant supercharges, it is easy to check that the difference between the fermionic and the bosonic superconnections corresponding to line operators along direction 3, can be written as 
\begin{equation}
	\mathcal{L}_{1/2}-\mathcal{L}_{1/6}=\{ \mathcal{Q}^2+\bar{\mathcal{Q}}_2 , \Lambda \} \qquad\text{where}\qquad \Lambda= 2i\sqrt{\frac{\pi}{k}}  \begin{pmatrix}
		0 &  Y_2 \\ -\bar{Y}^2 & 0 
	\end{pmatrix} = i \left(  \bar{\mathbb G}_2 - {\mathbb G}^2 \right)
	\label{eq:cohom1}
\end{equation}
with ${\mathbb G}^2, \bar{\mathbb G}^2$ defined in \eqref{Gmatrices}. Therefore, the \(\Lambda\) operator is a combination of  \(\mathcal{T}\) descendants, precisely
\begin{equation}
	\Lambda=i[\bar{\mathcal{Q}}_2-\mathcal{Q}^2 , \mathcal{T}]
	\label{}
\end{equation}
Inserting this expression in \eqref{eq:cohom1} gives 
\begin{equation}\label{eq:alternative}
     \mathcal{L}_{1/2}-\mathcal{L}_{1/6}= 2i\left\{\mathcal{Q}^2,[\bar{\mathcal{Q}}_2,\mathcal{T}] \right\} + i\mathfrak{D}_3 {\mathcal T} = 2i\left\{\mathcal{Q}^2,[\bar{\mathcal{Q}}_2,\mathcal{T}] \right\} + i {\mathbb B}
\end{equation}
where ${\mathbb B}$ is the operator defined in \eqref{eq:Bop}. 

The appearance of ${\mathbb B}$ 
in this alternative way of writing the cohomological equivalence may be a bit suspicious. In fact, as we are going to show in section \ref{sec:andim}, at quantum level ${\mathbb B}$ acquires a positive anomalous dimension, thus apparently contradicting the general understanding that 1/6 and 1/2 BPS Wilson lines should be related by an exactly marginal deformation \cite{Correa:2019rdk}. However, we recall that ${\mathbb B}$ is a total covariant derivative and once integrated on the line it simply generates a supergauge transformation. 
Therefore, the cohomological identity in \eqref{eq:alternative} states that the difference between the two integrated superconnections is a ${\cal Q}$-exact term, up to a supergauge transformation. Once inserted into the Wilson line definition this term is completely harmless and we obtain the expected result $\langle W_{1/6} \rangle = \langle W_{1/2} \rangle$, that is the two defects differ indeed by an exactly marginal operator.  

\section{Ward Identities and perturbative analysis}\label{sect:perturbative}

This section discusses the perturbative evaluation of two-point correlation functions of local operators inserted on the Wilson line.  

Perturbation theory is in terms of the couplings $N_1/k, N_2/k$. There is no need to take any planar limit, so the calculations are trustable for any finite $N_1 N_2$, as long as $N_{1,2} \ll k$ holds. At a given order in $1/k$, the contributing Feynman diagrams arise from all possible contractions among the local operators, powers of ${\mathcal L}$ super connections coming from the expansion of the $W$'s and the action vertices. 

To begin with, we discuss a set of Ward identities that relate correlation functions of local operators belonging to the same supermultiplet. We specialize these identities to the ${\mathcal T}$ supermultiplet, obtaining useful instructions for computing its anomalous dimension perturbatively. We look at its one- and two-point functions, discovering a non-trivial mixing with the identity operator, which occurs already at the tree level. Moving at loop order, we first discuss a general prescription for the IR regularization of the infinite line, compatible with its conformal mapping on the circle. We then apply this prescription to the evaluation of the two-point functions appearing in \eqref{eq:an_dim}, thus finding the anomalous dimension of \(\mathcal{T}\) at one loop. As a by-product of the $\langle \! \langle  {\mathbb G}^a  \bar{\mathbb G}_b \rangle \! \rangle$ calculation, we easily obtain the two-point correlator of the Displacement superprimary \(\mathbb{Z}\). We discuss the technical mechanism which ensures the \(\mathbb{Z}\) protection, while 
the ${\mathbb G}^a$ protection is lost. Finally, as a consistency check, we recover the result for the Bremsstrahlung function from the \(\mathbb{Z}\) correlator up to two loops. 

\subsection{Ward Identities}

The link between primaries and descendants driven by the SUSY charges preserved by the Wilson line leads to super-Ward identities that correlators on the defect must satisfy. This is a well-known fact in any SCFT, but what makes the Ward identities special on the defect is that the covariant supercharges used to build up multiplets carry a non-trivial dependence on the $1/k$ coupling (see eq. \eqref{covsusy}). Therefore, they are responsible for mixing between loop orders, thus leading to Ward identities peculiar to the dSCFT, as we will now describe. 

In order to find the general structure of Ward identities, we consider a primary operator $P^{(n)}$ at level $n$ of a given multiplet. We can take the ${\mathcal T}$ multiplet of figure \ref{fig:supmultT}  as a reference example. $P^{(n)}$ can be a single primary or mixing of primaries if at level $n$ there is more than one primary with the same $\mathfrak{u}(1)_M$ charge. It may carry $SU(3)$ indices, but we neglect them for simplicity. Now, given the two descendants
\beq\label{eq:descendants}
D^{(n+1) \; a} = [ {\mathcal Q}^a, P^{(n)} \} \qquad , \qquad \bar{D}^{(n+1)}_a = [ \bar{\mathcal Q}_a, \bar{P}^{(n)} \} 
\eeq
we consider the two-point function $\langle \! \langle D^{(n+1) \; a}(s) \bar{D}^{(n+1)}_a(0) \rangle \! \rangle$. Expressing the operators as in \eqref{eq:descendants} and using the covariant algebra of section \ref{sec:cov_alg} we obtain the following set of Ward identities
\beq\label{eq:WI}
\langle \! \langle D^{(n+1) \; a}(s) \bar{D}^{(n+1)}_a(0) \rangle \! \rangle = - 3 \, \partial_s \langle \! \langle P^{(n)}(s) \bar{P}^{(n)}(0) \rangle \! \rangle
- \langle \! \langle D^{(n+2) \; a}_a (s) \bar{P}^{(n)}(0) \rangle \! \rangle 
\eeq
where the descendant at level $(n+2)$ is defined as $D^{(n+2) \; a}_b = [ Q^a, \bar{D}^{(n+1)}_b \}$. 

Useful information can be obtained from identity \eqref{eq:WI} when the last correlator on the r.h.s. is identically vanishing\footnote{In perturbation theory it would be enough for the correlator to vanish up to the order one is interested in.}. In this case, if we write 
\beq
\langle \! \langle P^{(n)}(s) \bar{P}^{(n)}(0) \rangle \! \rangle = \frac{C_P}{s^{2\Delta_P + 2 \gamma_P }} \qquad , \qquad 
\langle \! \langle D^{(n+1) \; a}(s) \bar{D}^{(n+1)}_a(0) \rangle \! \rangle = \frac{C_D}{s^{2\Delta_P + 1 + 2 \gamma_P}}
\eeq
the Ward identity reduces to 
\beq\label{eq:WI2}
C_D = 6(\Delta_P + \gamma_P) \, C_P 
\eeq
where $\Delta_P$ is the scaling dimension of $P^{(n)}$ and $\gamma_P$ is the corresponding anomalous dimension. Here we have already considered that the descendant has the same anomalous dimension, as follows from the covariant algebra, particularly because the supercharges have a protected dimension $1/2$.  

This identity relates the anomalous dimension of the primary to the coefficient of the correlator of the descendant. Expressing these quantities perturbatively as series in $1/k$, 
\beq
C_P(k) = \sum_{r=0}^{\infty} \frac{c_r}{k^r} \, , \qquad C_D(k) = \sum_{r=0}^{\infty} \frac{d_r}{k^r} \, , \qquad \gamma_P(k)= \sum_{r=1}^{\infty} \frac{\gamma_r}{k^r}
\eeq
at the first few orders, we read
\begin{align}\label{eq:WI3}
& {\rm Order} \; k^0 \; \; : \quad d_0 = 6 \Delta_P c_0  \nonumber \\
& {\rm Order} \; k^{-1}: \quad  \gamma_1 = \frac{d_1}{6c_0} - \Delta_P \frac{c_1}{c_0}  \nonumber \\
& {\rm Order} \; k^{-2}: \quad   \gamma_2 = \frac{d_2}{6c_0} -  \gamma_1 \frac{c_1}{c_0} -  \Delta_P \frac{c_2}{c_0}
\end{align}

These relations further simplify when applied to $P^{(n=0)} \equiv  {\mathcal T}$, the lowest dimensional superprimary on the defect with $\Delta_{\mathcal T} = 0$, introduced in section \ref{sect:T}. In this case the descendants are $D^{(1) \, a} = {\mathbb G}^a$, $\bar{D}^{(1)}_a = \bar{\mathbb G}_a$ and $D^{(2) \, a}_b = \delta_b^a {\mathbb K} - {\mathbb R}_b^a$ (see figure \ref{fig:supmultT}). It is easy to see that up to one loop (order $1/k$) one has $\langle \! \langle  D^{(2) \, a}_b(s)  {\mathcal T}(0) \rangle \! \rangle =0$. Therefore, the Ward identity reduces to (\ref{eq:WI2}, \ref{eq:WI3}) where we set $\Delta_P=0$. In particular, from the first identity in \eqref{eq:WI3}, we read 
\beq
\langle \! \langle  {\mathbb G}^a(s)  \bar{\mathbb G}_b(0) \rangle \! \rangle^{(0)} =0
\eeq
which is consistent with the fact that each operator is already of order $1/\sqrt{k}$. Moreover, the second identity in  \eqref{eq:WI3} leads to
\beq\label{eq:an_dim}
\gamma({\mathcal T})|_{1L} = \frac16 \frac{{\rm coeff }[\langle \! \langle  {\mathbb G}^a(s)  \bar{\mathbb G}_a(0) \rangle \! \rangle^{(1)}]}{\langle \! \langle  {\mathcal T}(s)  \bar{\mathcal T}(0) \rangle \! \rangle^{(0)}} 
\eeq
where the numerator means taking the overall coefficient of the two-point function at order $1/k$. We note that since the ${\mathbb G}, \bar{\mathbb G}$ operators are already of order $1/\sqrt{k}$, this means taking the overall coefficient of their two-point function at the tree level. Therefore, the tree level of the descendant measures the anomalous dimension of its superprimary. We will exploit this identity in the next subsection to infer the anomalous dimension of ${\mathcal T}$. 

\subsection{The constant operator at weak coupling}
\label{sec:andim}

Considering the constant operator ${\mathcal T}$, it is easy to see that in the ABJ theory ($N_1 \neq N_2$), its one-point function at the tree level is non-vanishing. In fact, 
\begin{equation}
	\langle \!\langle \mathcal{T}\rangle \!\rangle^{(0)}=  \frac{\langle \Tr\left[W(+\infty, -\infty) \mathcal{T}\right]\rangle}{\langle \Tr W(+\infty, -\infty) \rangle}= -\frac12 \frac{\langle \STr W(+\infty, -\infty)
	\rangle}{\langle \Tr W(+\infty, -\infty) \rangle} = -\frac{1}{2}\frac{N_1-N_2}{N_1+N_2}
	\label{}
\end{equation}
This result may signal a non-trivial mixing of ${\mathcal T}$ with the identity operator. From this consideration, it would follow that the correct operator to consider is the linear combination
\begin{equation}\label{eq:twistedyo}
	{\mathcal T}'=\mathcal{T}+\frac{N_1-N_2}{2(N_1+N_2)}\mathbb{1}
\end{equation}
that satisfies $\langle\!\langle {\mathcal T}'\rangle\!\rangle=0$. This combination does not get any correction at one-loop, as the ${\mathcal T}$ one-point function is zero at this order. However, at higher orders, there is no reason why this pattern should persist. Therefore, we cannot exclude that the linear combination coefficient in \eqref{eq:twistedyo} may get $1/k^2$ corrections. Another problematic aspect of our interpretation would arise, in any case, by observing that the odd correlation function of ${\mathcal T}'$ is non-zero already at tree-level.

Nevertheless, we observe that the new operator ${\mathcal T}'$ can safely replace ${\mathcal T}$ as the superprimary of the multiplet in figure \ref{fig:supmultT}. Adding the identity operator does not affect the descendant operators' commutation relations. Therefore, identities \eqref{eq:Yop} defining the ${\mathbb G}^a, \bar{\mathbb G}_a$ operators can be safely replaced by
\begin{equation}
	{\mathbb G}^a=\left[\mathcal{Q}^a,{\mathcal{T}}'\right] \; ,  \quad \quad  \bar{\mathbb G}_a=\left[\bar{\mathcal{Q}}_a, {\mathcal{T}}'\right] \qquad a=1,2,3
	\label{eq:defdestilde}
\end{equation}

Having identified the correct operator, we can now determine its anomalous dimension using identity \eqref{eq:an_dim}. 

First of all, at the tree level, we find 
\begin{equation}\label{eq:Ttree}
	\langle\!\langle {\mathcal{T}}'(s) {\mathcal{T}}'(0)\rangle\!\rangle^{(0)}= \frac{N_1 N_2}{(N_1+N_2)^2}
\end{equation} 
For the $\langle\!\langle {\mathbb G}^a(s) \bar{\mathbb G}_a(0)\rangle\!\rangle$ correlator at order $1/k$,  
a simple calculation leads to
\begin{equation}\label{eq:Gtree}
	\langle\!\langle {\mathbb G}^a(s)\bar{\mathbb G}_a(0)\rangle\!\rangle^{(1)}=\frac{3}{k}\frac{N_1 N_2}{N_1+N_2}\frac{1}{s}
\end{equation}
Inserting these results into \eqref{eq:an_dim}, we finally obtain
\beq\label{eq:an_dimT}
\gamma({\mathcal T}')|_{1L} = \frac{N_1 + N_2}{2k}
\eeq
A similar calculation can be done in the ABJM theory ($N_1=N_2 \equiv N$). In this case there is no apparent mixing and $\langle\!\langle {\mathcal{T}}(s) {\mathcal{T}}(0)\rangle\!\rangle^{(0)}=1/4$. Since the result in \eqref{eq:Gtree} is valid also for $N_1=N_2$, we can still use it in \eqref{eq:an_dim} and find $\gamma({\mathcal T})|_{1L} = N/k$. This is consistent with \eqref{eq:an_dimT} for $N_1=N_2$.  

We recall that the SCP of a given representation of the $\mathfrak{su}(1,1|3)$ superconformal algebra has to satisfy the unitarity bound \(\Delta\geq 0\) \cite{Bianchi:2017ozk}. The anomalous dimension \eqref{eq:an_dimT}, being always positive, is then consistent with unitarity.

Result \eqref{eq:an_dimT} is the one-loop anomalous dimension of the whole ${\mathcal T}$ multiplet in figure \ref{fig:supmultT}, in particular of the $SU(3)$ triplets $\{ {\mathbb G}^a \}, \{ \bar{\mathbb G}_a \}$, which are then non-protected operators. It is interesting to recall that these operators, together with the ${\mathbb Z}, \bar{\mathbb Z}$ (anticommuting) scalars in \eqref{eq:Z}, originate from the $SU(4)$ multiplets $C_I, \bar{C}^J, I,J=1, \dots, 4$ of the bulk theory, under decomposition \eqref{su3breaking}. In the parent theory, they concur to form protected, gauge invariant operators of the form $\Tr{(C_I \bar{C}^J)^n}$, with the trace in $I,J$ removed. Nonetheless, once localized on the line, they undergo a completely different destiny: The ${\mathbb Z}, \bar{\mathbb Z}$ scalars remain protected, being part of the displacement multiplet, whereas $ {\mathbb G}^a, \bar{\mathbb G}_a$ are no longer protected, being descendants of the non-protected constant ${\mathcal T}'$ operator. 

From a computational point of view, it would be interesting to understand the mechanism that leads on the Wilson line to finite $\langle\!\langle {\mathbb Z} \bar{\mathbb Z} \rangle\!\rangle$ correlators, but divergent $\langle\!\langle {\mathbb G} \bar{\mathbb G} \rangle\!\rangle$ ones. We devote the rest of this section to addressing this question, digging out this mechanism perturbatively, at order $1/k^2$. 

\subsection{Two-loop scalar correlators}\label{sect:corr}

We now move to evaluate the two-point functions 
\begin{equation}
    \langle\!\langle {\mathbb Z} \bar{\mathbb Z} \rangle\!\rangle\qquad \text{and}\qquad  \langle\!\langle \bar{\mathbb{G}}_a \mathbb G^b  \rangle\!\rangle
    \label{eq:twocorr}
\end{equation}
on the Wilson line. 
As already mentioned, we expect only the first correlator to be finite, as the ${\mathbb G}$ operators should acquire anomalous dimension at quantum level. 

As a by-product, we will also rederive the two-loop Bremsstrahlung function associated to the \(1/2\)-BPS Wilson loop. In fact, this is known to be captured by the coefficient of the displacement two-point function \cite{Correa:2012at}, or equivalently of its $\mathbb{Z}$ superprimary.  

We begin by evaluating the normalization factor $\langle W \rangle$ in \eqref{eq:def_cor}. At the order we are interested in, it is sufficient to evaluate the Wilson expectation value up to order $1/k$. 

In the case of a linear defect, the evaluation of $\langle {W}\rangle$ is complicated by the appearance of long distance singularities associated with the infinite domain of line integrals. Regularizing such singularities requires introducing a long distance cut-off which restricts the line integrals to integrals on a finite size segment $(-L,L)$. Moreover, short distance singularities also appear, which are suitably regularized by using dimensional regularization in $d=3-2\epsilon$ with dimensional reduction \cite{Chen:1992ee,Bianchi:2013zda,Bianchi:2013rma}. The problem of how to remove the regulators and in which order is a subtle issue that requires careful analysis. 

At one loop the Wilson line receives a non-trivial contribution coming from the exchange of a fermion propagator. Using Feynman rule \eqref{0fermion}, we obtain the following integral\footnote{We use the convention $s_{ij} \equiv s_i - s_j$ for the distance between two points on the line.}
\begin{equation}
	\int_{-L}^{L}ds_1\int_{-L}^{s_1}ds_2\;\frac{1}{s_{12}^{2-2\epsilon}}=-\frac{(2L)^{2\epsilon}}{4\epsilon\left(\frac{1}{2}-\epsilon\right)} 
	\label{eq:intLL}
\end{equation}
It follows that, including all the factors from the propagator and the traces, up to one loop the defect vacuum-to-vacuum transition amplitude is
\begin{equation}
	\langle {W}  \rangle^{(0)+(1)}=(N_1+N_2) - \frac{N_1 N_2}{k} \,  \frac{\Gamma\left(\frac{1}{2}-\epsilon\right)}{\pi^{\frac{1}{2}- \epsilon}} \, \frac{\left(2L\mu\right)^{2\epsilon}}{\epsilon}
	\label{eq:Wpert}
\end{equation}
where $\mu$ is the mass scale of dimensional regularization. We note that, although we are working in Landau gauge, this result is gauge independent (differently from what observed for the analog operator in \(\mathcal{N}=4\) SYM \cite{Griguolo:2012iq} and for amplitudes in ABJM theory \cite{Leoni:2010az}). In fact, the longitudinal part of the gauge propagator vanishes on the line, as follows from eq. \eqref{eq:xigauge}. Therefore, there is no possibility that extra gauge-dependent contributions arise from the exchange of a vector propagator. 

For finite $L$ expression \eqref{eq:Wpert} is UV divergent, against the expectations based on the BPS nature of the defect. This is due to the appearance of boundary effects induced by the IR regularization that temporarily destroy the SUSY invariance of the Wilson line. It would be interesting to better investigate how to remove these unwanted contributions for the Wilson line {\em per s\`e}, in particular which should be the correct renormalization prescription and how to safely remove the IR cut-off. However, since here we are primarily interested in evaluating defect correlators, we study how to cure this problem once we have combined this divergent term with similar terms that are expected to appear in the evaluation of the numerator in \eqref{eq:def_cor}. 

Expanding the normalization factor $\frac{1}{\langle {W}  \rangle^{(0) + (1)}}$, at the order we are interested in a generic correlator $\langle\!\langle {\mathcal O} \bar{\mathcal O} \rangle\!\rangle$ is given by
\bea\label{eq:correxp}
&& \hspace{-0.5cm} \left( \langle W {\mathcal O} W \bar{\mathcal O} W \rangle^{(1)} +
 \langle W {\mathcal O} W \bar{\mathcal O} W \rangle^{(2)} \right) 
 \times \frac{1}{N_1+N_2} \left( 1 + \frac{1}{k} \, \frac{N_1 N_2}{N_1+N_2} \,  \frac{\Gamma\left(\frac{1}{2}-\epsilon\right)}{\pi^{\frac{1}{2}- \epsilon}} \, \frac{\left(2L\mu\right)^{2\epsilon}}{\epsilon} \right) \nonumber \\
 &=& \frac{\langle W {\mathcal O} W \bar{\mathcal O} W \rangle^{(1)}}{N_1+N_2}  \nonumber \\
 &~& + \frac{\langle W {\mathcal O} W \bar{\mathcal O} W \rangle^{(2)}}{N_1+N_2} + \frac{1}{k} \, \frac{N_1 N_2}{(N_1+N_2)^2} \,  \frac{\Gamma\left(\frac{1}{2}-\epsilon\right)}{\pi^{\frac{1}{2}- \epsilon}} \, \frac{\left(2L\mu\right)^{2\epsilon}}{\epsilon} \, 
 \langle W {\mathcal O} W \bar{\mathcal O} W \rangle^{(1)}
\eea

Lowest order corresponds to the first term in this expansion. Evaluating the numerators for the two correlators \eqref{eq:twocorr}, we find that their $O(1/k)$ expression in the $\epsilon \to 0$ limit reads\footnote{We note that this should correspond to tree level, but due to the particular normalization of the operators, it is already order $1/k$.} 
\begin{equation}
\langle\!\langle  \mathbb{Z}\bar{\mathbb{Z}}  \rangle\!\rangle^{(1)}=\frac{1}{k}\frac{N_1 N_2}{N_1+N_2}\frac{1}{s} \qquad , \qquad \langle\!\langle \, \bar{\mathbb G}_a{\mathbb G}^b  \rangle \!\rangle^{(1)} =\delta_b^a  \frac{1}{k}\frac{N_1 N_2}{N_1+N_2}\frac{1}{s}
	\label{eq:tree}
\end{equation}

Now we move to order $1/k^2$, that is the last line in \eqref{eq:correxp} where the second term comes from the one-loop result for $\langle W \rangle$ multiplied by results in \eqref{eq:tree}. 

\begin{figure}[h]
	\centering
	\subfigure[]{\begin{tikzpicture} \begin{feynman}
				\vertex (a); 
				\vertex[right=0.5cm of a] (b) ;
				\vertex[right=1.5cm of b] (c) ; 
				\vertex[right=1.5cm of c] (d) ;
				\vertex[right=0.5cm of d] (e); 
				\diagram* {(b) -- [scalar,half left] (c), (b) -- [scalar, half left] (d),};
				\draw[fill=gray] (b) circle (3pt);
				\draw[blue,thick] (a) -- (e);
				\draw[fill=white] (c) circle (2pt);
				\draw[fill=white] (d) circle (2pt);
				\draw[fill=lightgray] (b) circle (3pt);
		\end{feynman} \end{tikzpicture}\label{scalaraapp}}\qquad
	\subfigure[]{\begin{tikzpicture} \begin{feynman}
				\vertex (a); 
				\vertex[right=0.5cm of a] (b) ;
				\vertex[right=1.5cm of b] (c) ; 
				\vertex[right=1.5cm of c] (d) ;
				\vertex[right=0.5cm of d] (e); 
				\diagram* {(b) -- [scalar,half left] (c), (c) -- [scalar, half left] (d),};
				\draw[blue,thick] (a) -- (e);
				\draw[fill=lightgray] (c) circle (3pt);
				\draw[fill=white] (b) circle (2pt);
				\draw[fill=white] (d) circle (2pt);
		\end{feynman} \end{tikzpicture}\label{scalarbapp}}\qquad
	\subfigure[]{\begin{tikzpicture} \begin{feynman}
				\vertex (a); 
				\vertex[right=0.5cm of a] (b) ;
				\vertex[right=1.5cm of b] (c) ; 
				\vertex[right=1.5cm of c] (d) ;
				\vertex[right=0.5cm of d] (e); 
				\diagram* {(b) -- [scalar,half left] (d), (c) -- [scalar, half left] (d),};
				\draw[thick,blue] (a) -- (e);
				\draw[fill=lightgray] (d) circle (3pt);
				\draw[fill=white] (b) circle (2pt);
				\draw[fill=white] (c) circle (2pt);
		\end{feynman} \end{tikzpicture}\label{scalarcapp}}
	\caption{Diagrams with purely bosonic contractions. White bubbles represent the two local operator insertions, whereas the grey one is the bosonic part of the ${\mathcal L}$ superconnection coming from the first order expansion of W. The diagrams take into account all possible path orderings of the operators.}
	\label{fig:bremscalar1loop}
\end{figure}
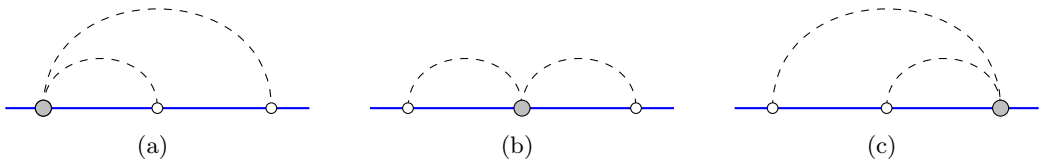
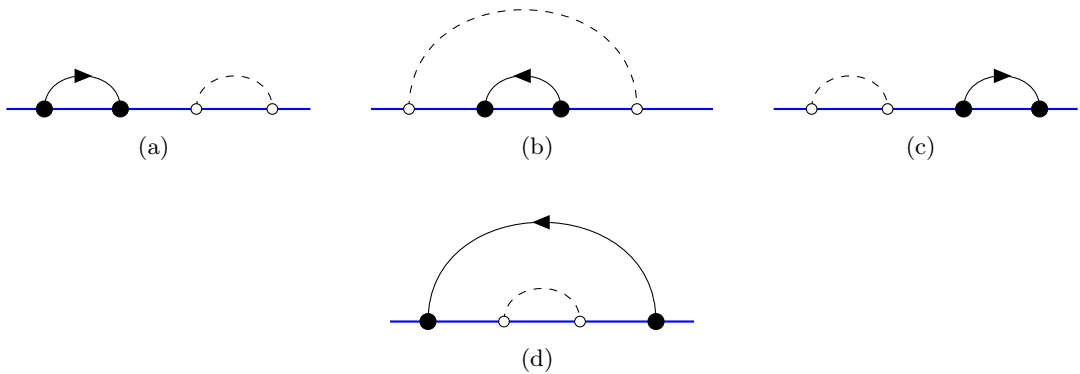
\begin{figure}[h]
	\centering
	\subfigure[]{\begin{tikzpicture} \begin{feynman}
				\vertex (a); 
				\vertex[right=0.5cm of a] (b) ;
				\vertex[right=1cm of b] (c); 
				\vertex[right=1cm of c] (d);
				\vertex[right=1cm of d] (e) ; 
				\vertex[right=0.5cm of e] (f) ;
				\diagram* {(d) -- [scalar,half left] (e),
					(b) -- [fermion, half left] (c),};
				\draw[thick,blue] (a) -- (f);
				\draw[fill=black] (b) circle (3pt);
				\draw[fill=black] (c) circle (3pt);
				\draw[fill=white] (d) circle (2pt);
				\draw[fill=white] (e) circle (2pt);
		\end{feynman} \end{tikzpicture}\label{fermionlineaapp}}\qquad
	\subfigure[]{\begin{tikzpicture} \begin{feynman}
				\vertex (a); 
				\vertex[right=0.5cm of a] (b);
				\vertex[right=1cm of b] (c); 
				\vertex[right=1cm of c] (d);
				\vertex[right=1cm of d] (e); 
				\vertex[right=1cm of e] (f);
				\diagram* {(b) -- [scalar,half left] (e),
					(d) -- [fermion, half right] (c)};
				\draw[thick,blue] (a) -- (f);
				\draw[fill=black] (d) circle (3pt);
				\draw[fill=black] (c) circle (3pt);
				\draw[fill=white] (b) circle (2pt);
				\draw[fill=white] (e) circle (2pt);
		\end{feynman} \end{tikzpicture}\label{fermionlinebapp}}\qquad
	\subfigure[]{\begin{tikzpicture} \begin{feynman}
				\vertex (a); 
				\vertex[right=0.5cm of a] (b);
				\vertex[right=1cm of b] (c); 
				\vertex[right=1cm of c] (d);
				\vertex[right=1cm of d] (e); 
				\vertex[right=.5cm of e] (f);
				\diagram* {(b) -- [scalar,half left] (c),
					(d) -- [fermion, half left] (e)};
				\draw[thick,blue] (a) -- (f);
				\draw[fill=black] (d) circle (3pt);
				\draw[fill=black] (e) circle (3pt);
				\draw[fill=white] (b) circle (2pt);
				\draw[fill=white] (c) circle (2pt);
		\end{feynman} \end{tikzpicture}\label{fermionlinecapp}}\qquad
	\subfigure[]{\begin{tikzpicture} \begin{feynman}
				\vertex (a); 
				\vertex[right=0.5cm of a] (b);
				\vertex[right=1cm of b] (c) ;
				\vertex[right=1cm of c] (d) ;
				\vertex[right=1cm of d] (e) ; 
				\vertex[right=.5cm of e] (f) ;
				\diagram* {(c) -- [scalar,half left] (d),
					(e) -- [fermion, half right] (b)};
				\draw[thick,blue] (a) -- (f);
				\draw[fill=black] (b) circle (3pt);
				\draw[fill=black] (e) circle (3pt);
				\draw[fill=white] (c) circle (2pt);
				\draw[fill=white] (d) circle (2pt);
		\end{feynman} \end{tikzpicture}\label{fermionlinedapp}}
	\caption{Diagrams with fermionic contractions (arrowed lines). White bubbles represent the two local operator insertions, whereas the black ones are the fermions from two ${\mathcal L}_F$ superconnections coming from the second order expansion of W. The diagrams take into account all possible path orderings of the operators.}
	\label{fig:bremferm1loop}
\end{figure}

The first term $\langle W {\mathcal O} W \bar{\mathcal O} W \rangle^{(2)}$ receives contributions from two sets of diagrams. Diagrams in figure \ref{fig:bremscalar1loop} come from the first order expansion of the Wilson line and involve contractions of the $\mathbb{Z}$ and $\mathbb{G}$ operators with the scalar part of the superconnection ${\mathcal L}_B$ in eq. \eqref{connecsu3}.  
The second set of diagrams are depicted in figure \ref{fig:bremferm1loop}. They come from the second order expansion of $W$ and involve self-contractions of two fermionic ${\mathcal L}_F$ terms in eq. \eqref{connecsu3}, times the free propagators $\langle Z \bar{Z} \rangle$ and $\langle \bar{Y}^a Y_b \rangle$, respectively. 

Since the tree level propagators for $Z$ and $Y$'s are the same (they all come from propagator \eqref{scalartree} evaluated on the line), it is clear that the diagramatic contributions in figure \ref{fig:bremferm1loop} are the same for both the correlators \eqref{eq:twocorr}. Instead, due to the sign difference between the two biscalars appearing in ${\mathcal L}_B$, the diagrams in figure \ref{fig:bremscalar1loop} contribute to the two correlators with an opposite sign. Therefore, if we call ${\mathcal B}^{(2)}$ the contributions from diagrams \ref{fig:bremscalar1loop} and 
${\mathcal F}^{(2)}$ the ones from diagrams \ref{fig:bremferm1loop}, we can write
\bea\label{eq:BandF}
&& \langle W \, {\mathbb Z}(s) \,  W \, \bar{\mathbb Z}(0) \,  W\rangle^{(2)} = {\mathcal F}^{(2)} + {\mathcal B}^{(2)} \\ 
&& \langle W \, \bar{{\mathbb G}}_a(s) \, W \, {\mathbb G}^b(0) \, W \rangle^{(2)}  = \delta_a^b \, \left( {\mathcal F}^{(2)} - {\mathcal B}^{(2)} \right) \nonumber
\eea

We now evaluate ${\mathcal B}^{(2)}$ and ${\mathcal F}^{(2)}$, explicitly. We compute the Feynman integrals corresponding to the diagrams in figures \ref{fig:bremscalar1loop} and \ref{fig:bremferm1loop} by using the IR regulator discussed above, plus dimensional regularization for short distance divergences. The necessary Feynman rules are listed in appendix \ref{ABJ(M)}. We evaluate one of the two correlators in the expressions \eqref{eq:BandF}.

From the diagrams in fig. \ref{fig:bremscalar1loop} we obtain
\begin{align}
    \ref{scalaraapp}&=	\frac{N_1^2 N_2}{k^2} \;  \frac{\Gamma^2\left(\frac{1}{2}-\epsilon\right)}{2\pi^{1-2\epsilon}} \,\left(\frac{L}{L+s}\right)^{2\epsilon} \, \frac{s^{4\epsilon-1}}{2\epsilon} \\
    \ref{scalarbapp}&=\frac{N_1 N_2^2 }{k^2} \; \frac{\Gamma^2\left(\frac{1}{2}-\epsilon\right)}{2\pi^{1-2\epsilon}} \, \frac{\Gamma^2(2\epsilon)}{\Gamma(4\epsilon)} \, s^{4\epsilon-1}\\
    \ref{scalarcapp}&=\frac{N_1^2 N_2}{k^2} \; \frac{\Gamma^2\left(\frac{1}{2}-\epsilon\right)}{2\pi^{1-2\epsilon}}  \,  \left(\frac{L-s}{L}\right)^{2\epsilon} \, \frac{s^{4\epsilon-1}}{2\epsilon}
\end{align}
We see that these contributions are regular in the limit \(L\to \infty\). Therefore, removing the IR cut-off they eventually
sum up to the following UV divergent contribution
\bea
{\mathcal B}^{(2)}&=& \frac{1}{\epsilon} \; \frac{N_1 N_2}{2k^2} \, \frac{\Gamma^2\left(\frac{1}{2}-\epsilon\right)}{\pi^{1-2\epsilon}} \left(N_1 + N_2\frac{\Gamma^2(1+2\epsilon)}{\Gamma(1 + 4 \epsilon)}\right) \, \frac{1}{s^{1-4\epsilon}} \nonumber \\
&\sim& \frac{1}{\epsilon} \; \frac{N_1 N_2(N_1+N_2)}{2k^2} \, \frac{1}{s} + O(\epsilon) 
	\label{eq:bosonic1loop}
\eea

Now we move to the fermionic contributions. The double integrals coming from diagrams in figure \ref{fig:bremferm1loop} evaluate to
\begin{align}
	\ref{fermionlineaapp}&=\frac{N_1 N_2^2}{k^2} \, \frac{\Gamma^2\left(\frac{1}{2}-\epsilon\right)}{2\pi^{1-2\epsilon}} \, \left[-L^{2\epsilon}\right] \, \frac{s^{2\epsilon-1}}{\epsilon}\\
	\ref{fermionlinebapp}&=\frac{N_1^2 N_2 }{k^2}  \, \frac{\Gamma^2\left(\frac{1}{2}-\epsilon\right)}{2\pi^{1-2\epsilon}} \, \left[-s^{2\epsilon}\right] \, \frac{s^{2\epsilon-1}}{\epsilon}\\
	\ref{fermionlinecapp}&=\frac{N_1 N_2^2}{k^2}  \, \frac{\Gamma^2\left(\frac{1}{2}-\epsilon\right)}{2\pi^{1-2\epsilon}} \, \left[-(L-s)^{2\epsilon}\right] \, \frac{s^{2\epsilon-1}}{\epsilon} \\
	\ref{fermionlinedapp}&=\frac{N_1 N_2^2}{k^2} \, \frac{\Gamma^2\left(\frac{1}{2}-\epsilon\right)}{2\pi^{1-2\epsilon}} \, \left[L^{2\epsilon}-s^{2\epsilon}-(2L)^{2\epsilon}+(L+s)^{2\epsilon}\right] \, \frac{s^{2\epsilon-1}}{\epsilon}
\end{align}
and sum up to
\begin{equation}\label{eq:calF}
    \begin{aligned}
    {\mathcal F}^{(2)} = - \frac{1}{\epsilon} \; \frac{N_1 N_2}{2k^2} \, & \frac{\Gamma^2\left(\frac{1}{2}-\epsilon\right)}{\pi^{1-2\epsilon}} \, \frac{1}{s^{1-2\epsilon}}\\
    \times &\left( (N_1+N_2) s^{2\epsilon} + N_2\left( (L-s)^{2\epsilon} - (L+s)^{2\epsilon} + (2L)^{2\epsilon} \right)\right) 
    \end{aligned}
\end{equation}
In this case the $L \to \infty$ limit is not totally safe as long as $\epsilon \neq 0$. In fact, while the second and the third terms cancel each other in this limit, we are left with a divergent contribution proportional to $(2L)^{2\epsilon}$ which is problematic. However, this is exactly of the same form of the last term in \eqref{eq:correxp} coming from the expansion of the denominator $\langle W \rangle$. Therefore, we have to sum up all the contributions before discussing how to remove the IR regulator.

Focusing for the time being only on the problematic terms, for both correlators we have the following contribution (reiserting the mass scale $\mu$)
\begin{equation}\label{eq:sick}
    - \frac{1}{2k^2}\,  \frac{N_1 N_2^2 (N_1-N_2) }{(N_1+N_2)^2} \, \frac{(2L\mu)^{2\epsilon}}{\epsilon} \, \frac{1}{s} +  c \,  (2L\mu)^{2\epsilon}  + O(\epsilon)
\end{equation}
where $c$ is an UV finite function of the couplings and the position $s$. We see that the problematic term is eventually proportional to $(N_1-N_2)$, and it vanishes for $N_1=N_2$. It is therefore convenient to split the discussion of the ABJM and ABJ cases. 

\paragraph{The {\bf $N_1=N_2 \equiv N$} case.} When the defect lives in the ABJM theory, the divergent term in \eqref{eq:sick} vanishes identically. This means that, at least at order $1/k^2$, the bad divergent one-loop contribution to the Wilson expectation value is needed to cancel exactly a similar term which arises in the evaluation of the correlators in \eqref{eq:BandF}. The rest of expression \eqref{eq:sick} does not present any problem and can be safely removed by sending for instance $\epsilon \to 0$ and then $L \to \infty$. In the ABJM case it can be actually checked that the result is independent of the order of limits. A similar pattern was already encountered in \cite{Bianchi:2017ozk}. 

Having removed the $(2L)^{2\epsilon}$ terms, from \eqref{eq:bosonic1loop} and \eqref{eq:calF} it is now easy to realize that 
\beq
{\mathcal F}^{(2)} = -{\mathcal B}^{(2)} + O(\epsilon)
\eeq
Therefore, from eqs. \eqref{eq:BandF} it follows that 
\bea
&& \langle \! \langle \, {\mathbb Z}(s) \bar{\mathbb Z}(0) \, \rangle\!\rangle^{(2)} = O(\epsilon) \\
&& \langle\!\langle \, \bar{{\mathbb G}}_a(s) {\mathbb G}^b(0) \, \rangle\!\rangle^{(2)} = -\delta_a^b \, \frac{N^2}{k^2} \left( \frac{1}{\epsilon } + 4 \log{s}+ 2\gamma +2\log{(4 \pi )} \right) \, \frac{1}{s} + O(\epsilon) \nonumber 
\eea

The first line is perfectly consistent with the expectations: not only the $\langle \! \langle {\mathbb Z} \bar{\mathbb Z} \rangle\!\rangle $ correlator is finite, in addition its one-loop coefficient is zero, in agreement with the fact that the Bremsstrahlung function is known to get no corrections at order $1/k^2$ \cite{Lewkowycz:2013laa, Bianchi:2014laa, Bianchi:2017svd}. 

More interesting is the second line. The appearance of the $1/\epsilon$ divergence signals the necessity of renormalizing the $\mathbb{G}^a$ operators, which consequently acquire an anomalous dimension. It is easy to show that renormalizing the operators as $\mathbb{G}_R^a = Z_{\mathbb G}^{-1} \mathbb{G}^a$ (the same for $\bar{\mathbb{G}}_a$) and applying the usual procedure which in minimal subtraction scheme allows to read the anomalous dimension from the $1/\epsilon$ pole of $Z_{\mathbb G}$, one finds $\gamma({\mathbb G}^a)|_{1L} = \tfrac{N}{k}$, in agreement with \eqref{eq:an_dimT} for $N_1=N_2 \equiv N$.

\paragraph{The {\bf $N_1 \neq N_2$} case.} In the ABJ theory the previous calculations reveal that the $1/\epsilon$ pole in \eqref{eq:sick} proportional to the IR regulator is not vanishing. This term, mixing UV and IR divergences, renders the regularization prescriptions ambiguous. In fact, this term is divergent for $L \to \infty$, as long as $\epsilon \neq 0$. On the other hand, if we keep $L$ finite and choose an UV renormalization prescription which removes completely the first term in \eqref{eq:sick}, the dependence on the IR cut-off disappears and one can safely take the $L \to \infty$ limit afterwards. It follows that the perturbative corrections to the correlators on the line can be anything, depending on the order of the $L \to \infty$ and $\epsilon \to 0$ limits and the renormalization prescription that we adopt.  

We fix this ambiguity by choosing a different prescription to regularize the IR divergences in the ABJ case. This regularization is analysed in details in appendix \ref{sect:cut-off} and basically amounts to conformally mapping the cut-off line onto the cut-off circle to avoid long distance bad behavior. As discussed in the appendix, this new prescription simply amounts to discard the terms $(L-s)^{2\epsilon}, (L+s)^{2\epsilon}$ and $(2L)^{2\epsilon}$, as they were to be cancelled by extra degrees of freedom placed at the two edges of the cut-off line\footnote{A different regularization scheme that one might try is the gauge averaging proposed in \cite{Leoni:2010az}.}. 

Using this prescription, and still using dimensional regularization to keep UV divergences under control, the result in \eqref{eq:calF} reads 
\begin{equation}
{\mathcal F}^{(2)} = -	\frac{N_1 N_2(N_1+N_2)}{2k^2} \, \frac{\Gamma^2\left(\frac{1}{2}-\epsilon\right)}{\pi^{1-2\epsilon}} \, \frac{1}{\epsilon} \, \frac{1}{s^{1 - 4\epsilon}} = - {\mathcal B}^{(2)} + O(\epsilon)
\end{equation}
Therefore, expanding around $\epsilon = 0$, from eqs. \eqref{eq:BandF} we finally obtain
\bea
&& \langle\!\langle {\mathbb Z}(s) \bar{\mathbb Z}(0) \rangle\!\rangle^{(2)} = O(\epsilon) \\
&& \langle\!\langle \bar{{\mathbb G}}_a(s) {\mathbb G}^b(0) \rangle\!\rangle^{(2)} = -\delta_a^b \, \frac{N_1 N_2}{k^2} \left( \frac{1}{\epsilon } + 4 \log{s}+ 2\gamma +2\log{(4 \pi )} \right) \, \frac{1}{s} + O(\epsilon) \nonumber 
\eea
Once more, the first correlator is consistent with the absence of $1/k^2$ corrections to the Bremsstrahlung function of the 1/2-BPS Wilson loop, whereas renormalizing the second correlator we obtain the one-loop anomalous dimension of ${\mathbb G}^a$ which agrees with the expression in \eqref{eq:an_dimT}. 


\section{The constant operator at strong coupling} \label{sect:strong}
In this section, we propose a holographic interpretation of the $\mathcal{T}$ multiplet. Our conjecture relies on a similar situation in 4d $\mathcal{N}=4$ SYM. Therefore, we begin by briefly recalling what happens in four dimensions. 

To this end, we focus on the one-dimensional dCFT defined on the $\frac{1}{2}$-BPS Wilson loop \cite{Rey:1998ik, Maldacena:1998im} of the $\mathcal{N}=4$ SYM theory. The lightest local operators one can consider are the scalars $\Phi^I$, $I=1,\dots, 6$. When localized on the defect, these are the SCPs of two supermultiplets. Precisely, one can choose $\Phi^a$ $a=1,\dots,5$ to be the lowest operators of the displacement multiplet, which is a short multiplet, while $\Phi^6$ generates a long multiplet.

In \cite{Giombi:2017cqn}, a holographic description of the dCFT on the Wilson loop has been proposed. Given the minimal surface dual to the straight Wilson line, which defines an $AdS_2$ metric inside the $AdS_5\times S^5$ background \cite{Maldacena:1998im}, the holographic dual of the dCFT is the $AdS_2$ QFT for the transverse fluctuations around the minimal surface, obtained by expanding the worldsheet superstring action in the static gauge. According to the holographic dictionary, the $\Phi^a$ operators with $a=1, \dots , 5$ are mapped to the fluctuations $y^a$ in the $S^5$ directions, whereas the unprotected $\Phi^6$ scalar is conjectured to be dual to the lightest bound state $y^a y_a$. Since this is the lightest operator exchanged in the OPE $y^a\times y^a$, one can use bootstrap methods to compute the anomalous dimension of the bound state.

Here, we generalize this proposal to the ABJM theory.
In this case, the $\frac{1}{2}$-BPS Wilson line admits a holographic description in terms of a minimal area superstring worldsheet on $AdS_4\times \mathbb{C}{\rm P}^3$ \footnote{In the ABJM theory, the duals of $\frac{1}{2}$-BPS Wilson operators can be more generally obtained in terms of minimal M2-brane configurations in M-theory on $AdS_4 \times S^7/Z_k$ \cite{Lietti:2017gtc}. They reduce to $AdS_4\times \mathbb{C}{\rm P}^3$ type IIA string solutions in the regime $k \ll N \ll k^5$.}. Following the 4d counterpart, one can consider the $AdS_2$ QFT, which arises from expanding the superstring action on $AdS_4\times \mathbb{C}{\rm P}^3$ in the static gauge around the Wilson line solution. We interpret it as the gravitational dual of the dCFT defined on the $\frac{1}{2}$-BPS Wilson line. The fluctuations transverse to the $AdS_2$ solution are in one-to-one correspondence with the operators in the displacement multiplet \cite{Bianchi:2020hsz}. An important difference with respect to the 4d case is that the SCP $\mathbb{Z}$, being an anticommuting supermatrix operator, corresponds to a {\em fermionic} fluctuation $z$ in the worldsheet theory.

At weak coupling, it is tempting to make an analogy between the lightest non-protected operator $\mathcal{T}$ of the dCFT and the non-protected scalar $\Phi^6$ in 4d. It is then natural to take inspiration from the 4d duality $\Phi^6 \sim y^a y_a$ to conjecture a duality between the $\mathcal{T}$ excitations and the lightest bound state built from the fluctuations dual to the displacement multiplet, that is 
\beq
\mathcal{T} \sim z\bar{z}
\eeq
The quantum numbers of the bound state $z\bar{z}$ are $[1,0,0,0]$. While the $\mathfrak{u}(1)_M$ and R-symmetry quantum numbers match those of $\mathcal{T} $, scaling dimensions are different. However, this may not be a problem since $\mathcal{T}$ is not protected. It is, in fact, conceivable that its quantum dimension, being a function of $k$, $N_1$ and $N_2$, interpolates between the dimension at weak coupling (zero at lowest order) and the one at strong coupling captured by $z\bar{z}$. Indeed, the same pattern occurs in the four-dimensional case.

A couple of qualitative arguments can be used to support our conjecture. First of all, the fact that the  $\mathcal{T}$ dimension flows in the IR to a larger value is consistent with our perturbative findings. In fact, at weak coupling, we have found a positive anomalous dimension (see \eqref{eq:an_dimT}), which signals an increasing flow towards the IR. Second, for $N_1=N_2 \equiv N$, the anomalous dimension of the $\mathbb{Z}\bar{\mathbb{Z}}$ operator dual to the bound state $z\bar{z}$ has been computed at strong coupling in \cite{Bianchi:2020hsz}, and reads
\begin{equation}
\Delta_{\mathbb{Z}\bar{\mathbb{Z}}}=1-3\epsilon
\end{equation}
where $\epsilon \sim {(N/k)}^{-\frac12}$ is the coupling constant. Again, the negative sign of the correction, signaling a decreasing flow towards the UV, agrees with our proposal.

\section{Conclusions and perspectives}\label{sect:conclusions}
 The study of dCFT's defined through supersymmetric Wilson lines in ABJ(M) theory is still on its infancy. Already the maximal 1/2 BPS case presents peculiarities and unexpected properties, due to the fermionic couplings appearing in its field theoretical definition. In this paper we have observed the existence of a long multiplet whose highest weight state is obtained by inserting into the Wilson line a constant supermatrix operator ${\mathcal T}$. We have derived the full supermultiplet exploiting an explicit covariant representation of the preserved supercharges. While the relation between ${\mathcal T}$ and the honest local operators ${\mathbb G}^a(x)$ might seem an artifact of taking the covariant version of the supercharges, perturbation theory supports our interpretation. In fact ${\mathbb G}^a(x)$ is not protected and acquires at quantum level the same anomalous dimension as ${\mathcal T}$, suggesting that ${\mathbb G}^a(x)$ is truly a descendant of ${\mathcal T}$. Another piece of evidence for the consistency of our construction comes from strong coupling considerations. If we were not to assume that ${\mathbb G}^a(x), \bar{\mathbb G}_a(x)$ are ${\mathcal T}$ descendants, we could not find any obvious operator corresponding to the $z\bar{z}$ bound state appearing in this regime. It turns out that the quantum dimension of the constant operator ${\mathcal T}$ is compatible with an interpolating function between weak and strong coupling. In this respect, it would be certainly interesting to apply bootstrap techniques to verify our intuition, mimicking the 4d analog \cite{Grabner:2020nis, Ferrero:2021bsb, Cavaglia:2021bnz, Cavaglia:2022qpg}. We have also noted that, even more mysteriously, ${\mathcal T}$ enters the cohomological equivalence between 1/2 BPS and 1/6 BPS Wilson loops. We remark that "constant" local operators inserted into Wilson lines were previously considered in the literature. For example “defect changing operators”, which change the scalar coupled to the Wilson loop have been studied in \cite{Kim:2017sju}, while in \cite{Gabai:2022vri} it has been shown that the holonomy itself gets contribution from the constant part.

While in the case of ABJM the situation seems quite clear at weak coupling, for $N_1\neq N_2$ we found some subtle and somehow unexpected effect. Bad terms, proportional to $(N_1-N_2)$, arise in our computations, inducing a strong dependence on the regularization procedure. We adopted a regularization consistent with the same calculation on a circular Wilson loop, finding a reasonable result for the anomalous dimensions. It is certainly worth to explore more deeply this last feature, maybe in connection with the parity properties of ABJ theory.
More generally, it would be important to have a more clear picture on the correct way to define the 1/2 BPS Wilson line at perturbative level, maybe resorting to a well-defined limiting procedure that involves boundary operators connected by the line.

As a final remark, we stress that no computation of four-point functions has been attempted so far for defect operators in the ABJ(M) theory, at perturbative level. It could be useful to have some results in this direction, also to understand the behaviour of ${\mathcal T}$ in the OPE expansion.

\vskip 35pt 

\acknowledgments
We thank Lorenzo Bianchi, Diego Correa, Shota Komatsu, Carlo Meneghelli and Guillermo Silva for interesting discussions and useful insights. The work of Luigi Guerrini and Paolo Soresina is supported by Della Riccia Foundation. This work has been supported in part by Italian Ministero dell'Universit\`a e Ricerca (MUR), and by Istituto Nazionale di Fisica Nucleare (INFN) through the ``Gauge Theories, Strings, Supergravity'' (GSS) and ``Gauge and String Theory'' (GAST) research projects.

\newpage
\appendix

\section{Supermatrix identities}\label{supermatrices}

In this appendix we shortly review the main rules concerning supermatrices which have been used along the text. We refer to volume III of \cite{Cornwell:1989bx} for a more complete introduction. 

Given a block supermatrix 
\begin{equation}
X = \begin{pmatrix}
X_{00} &  X_{01} \\
X_{10}  & X_{11}
\end{pmatrix} 
\end{equation}
the matrix is called {\em even} if the $ X_{00}, X_{11}$ entries are bosonic and $X_{01}, X_{10}$ are fermionic. It is called {\em odd} in the opposite case.
We define even supermatrices to have grade $|X| =0$ and odd ones to have grade $|X| =1$.

The (anti)commutator of two supermatrices is given by
\begin{equation}\label{supercomm}
[ X , Y \} = X Y - (-1)^{|X||Y|} Y X
\end{equation}

Given a scalar $\alpha$ with grade $|\alpha| = 0$ (grassmann even) or $|\alpha| = 1$ (grassmann odd), the left product of $X$ by $\alpha$ is defined as
\begin{equation}\label{left}
\alpha \cdot X = \begin{pmatrix}
\alpha X_{00} &  \, \alpha X_{01} \\
\hat{\alpha} X_{10}  & \, \hat{\alpha} X_{11}
\end{pmatrix}  \qquad {\rm where} \quad  \hat{\alpha} = (-1)^{|\alpha|} \alpha
\end{equation}
Similarly the right product is given by
\begin{equation}\label{right}
 X \cdot \alpha = \begin{pmatrix}
 X_{00} \alpha &  \, X_{01} \hat\alpha  \\
 X_{10} \alpha & \,  X_{11} \hat{\alpha}
\end{pmatrix}  
\end{equation}
Note that $\alpha \cdot X = (-1)^{|\alpha||X|} X \cdot \alpha$.

\vskip 25pt

\section{ABJ(M) action and Feynman rules}\label{ABJ(M)}

Here we shortly summarize the basic notions about ABJ(M) theory needed to perform the perturbative calculations of section \ref{sect:perturbative}. We stick to conventions of \cite{Bianchi:2018bke,Gorini:2020new}, to which we refer for more details. 

We work in euclidean space with coordinates $x^\mu=(x^1,x^2,x^3)$ and metric $\delta_{\mu\nu}$. 
Gamma matrices satisfying the usual Clifford algebra $\{\gamma^\mu,\gamma^\nu\}=2\delta^{\mu\nu}\mathbb 1$, are chosen to be the Pauli matrices 
\beq
(\gamma^\mu)_\alpha^{\; \beta} \equiv (\sigma^\mu)_\alpha^{\; \beta} \qquad \quad \mu = 1,2,3
\eeq

Spinorial indices are raised and lowered according to
\[
\psi^\alpha=\varepsilon^{\alpha\beta}\psi_\beta,\qquad \psi_\alpha=\varepsilon_{\alpha\beta}\psi^\beta \qquad {\rm with } \qquad \varepsilon^{12}=-\varepsilon_{12}=1
\]
Therefore, we also define the symmetric matrices 
\beq
(\g^\mu)_{\a \b } \equiv \varepsilon_{\b \g} (\g^\mu)_\a^{\; \g} = (-\sigma^3, i {\rm I}, \sigma^1) \qquad \qquad (\g^\mu)^{\a \b } \equiv \varepsilon^{\a \g} (\g^\mu)_\g^{\; \b} = (\sigma^3, i {\rm I}, -\sigma^1)
\eeq

\vskip 10pt

The field content of the $U(N_1)_k \times U(N_2)_{-k}$ ABJ(M) theory includes two gauge fields $(A_\mu)_i^j$, $(\hat A_\mu)_{\hat i}^{\hat j}$ belonging to the adjoint representation of $U(N_1)$ and $U(N_2)$ respectively, minimally coupled to four matter multiplets $(C_I, \bar{\psi}^I)_{I=1, \dots , 4}$ in the $(N_1, \bar{N}_2)$ representation of the gauge group and their conjugates $(\bar{C}^I, \psi_I)_{I=1, \dots , 4}$ in the 
$(\bar{N}_1, N_2)$. 

Introducing bulk covariant derivatives  
\begin{equation}
\begin{aligned}
&D_\mu C_I =\de_\mu C_I + i  A_\mu C_I- i  C_I \hat{A}_\mu ,\qquad D_\mu \bar{C}^I =\de_\mu \bar{C}^I + i  \hat{A}_\mu \bar{C}^I-i \bar{C}^I A_\mu \\
&D_\mu \bar{\psi}^I =\de_\mu \bar{\psi}^I + i  A_\mu \bar{\psi}^I- i \bar{\psi}^I \hat{A}_\mu ,\qquad D_\mu \psi_I =\de_\mu \psi_I +i  \hat{A}_\mu \psi_I-i \psi_I A_\mu 
\end{aligned}
\label{covd}
\end{equation}
the Euclidean gauge-fixed action is given by 
\begin{equation}\label{action}
S=S_{\rm CS}+S_{\rm mat} + S_{\rm pot}+S_{\rm gf}
\end{equation}
where
\bea
S_{\rm CS} &=&  -\frac{ik}{4\pi}\int d^3x \ \varepsilon^{\mu\nu\rho}\left[
\Tr\left(A_\mu\de_\nu A_\rho+\frac23 i A_\mu A_\nu A_\rho\right)-\Tr\left(\hat{A}_\mu\de_\nu\hat{A}_\rho+\frac23 i \hat{A}_\mu\hat{A}_\nu\hat{A}_\rho\right)\right] \non \\
&& \label{CS} \\
S_{\rm mat} &=& \int d^3x \ \Tr\left[D_\mu C_I D^\mu\bar{C}^I - i\bar{\psi}^I\gamma^\mu D_\mu\psi_I\right] \nonumber \\
&= & \int d^3x \ \Tr\left[\de_\mu C_I\de^\mu \bar{C}^I - i\bar{\psi}^I\gamma^\mu\de_\mu\psi_I+ \left(\bar{\psi}^I\gamma^\mu\hat{A}_\mu\psi_I 
- \bar{\psi}^I\gamma^\mu\psi_I A_\mu\right) \right. \notag \\
&&\left.\qquad\qquad+ i \left(A_\mu C_I\de^\mu\bar{C}^I-C_I\hat{A}_\mu\de^\mu \bar{C}^I-\de_\mu C_I\bar{C}^I A^\mu+\de_\mu C_I\hat{A}^\mu\bar{C}^I\right) \right. \notag \\
&&\left.\qquad\qquad+\left(A_\mu C_I\bar{C}^I A^\mu-A_\mu C_I \hat{A}^\mu\bar{C}^I-C_I \hat{A}_\mu\bar{C}^I A^\mu+C_I\hat{A}_\mu\hat{A}^\mu\bar{C}^I\right)\right] \label{Smat}
\eea
\bea 
&& \hspace{-1.8cm} S_{\rm pot} \equiv  S_{\rm 6pt} + S_{\rm 4pt} \non \\
&&  \hspace{-1.8cm} \qquad  = -\frac{4\pi^2}{3 k^2} \int d^3x \, \Tr \Big[ C_I \bar{C}^I C_J \bar{C}^J C_K \bar{C}^K + \bar{C}^I C_I \bar{C}^J C_J \bar{C}^K C_K
\non \\
&&  \hspace{-1.8cm} ~  \qquad \qquad \qquad \qquad  \quad + 4 \, C_I \bar{C}^J C_K \bar{C}^I C_J \bar{C}^K - 6 \, C_I \bar{C}^J C_J \bar{C}^I C_K \bar{C}^K \Big] 
\nonumber \\
&&  \hspace{-0.3cm} -\frac{2\pi i}{k} \int d^3x \, \Tr \Big[ \bar{C}^I C_I \Psi_J \bar{\Psi}^J - C_I \bar{C}^I \bar{\Psi}^J \Psi_J
+2 \, C_I \bar{C}^J \bar{\Psi}^I \Psi_J 
\non \\
&&  \qquad \qquad \quad    - 2 \, \bar{C}^I C_J \Psi_I \bar{\Psi}^J - \epsilon_{IJKL} \bar{C}^I\bar{\Psi}^J \bar{C}^K \bar{\Psi}^L + \epsilon^{IJKL} C_I \Psi_J C_K \Psi_L \Big]
\label{S4pt}
\eea
with $\epsilon_{1234}=\epsilon^{1234} =1$, and the gauge-fixing plus ghost terms read
\beq
S_{\rm gf} = \frac{k}{4\pi}\int d^3x \ \Tr\left[\frac{1}{\alpha}\left(\de_\mu A^\mu\right)^2+\de_\mu\bar{c}D^\mu c-\frac{1}{\alpha}\left(\de_\mu\hat{A}^\mu\right)^2-\de_\mu\bar{\hat{c}}D^\mu\hat{c}\right] 
\eeq
For the group generators we use the following relations
\beq
\Tr (T^A T^B) = \delta^{AB} \; , \qquad [ T^A , T^B ] = i f^{AB}_{\; \; \; \; \, C} \, T^C
\eeq

\vskip 15pt

In doing perturbative calculations it is convenient to rescale the gauge fields in the action as 
\beq\label{eq:rescaling}
A_\mu \to \frac{1}{\sqrt{k}} A_\mu \qquad ,  \qquad \hat{A}_\mu \to \frac{1}{\sqrt{k}} \hat{A}_\mu
\eeq
Having performed this rescaling, the tree-level propagators read:
\begin{itemize}
	\item Scalar propagator
	\begin{align}
	\langle {(C_I)_i}^{\hat{j}} (x)\ {(\bar C^J)_{\hat k}}^l (y)  \rangle &=\delta^J_I\delta^l_i\delta^{\hat j}_{\hat k} \ \frac{\Gamma(\frac{1}{2}-\epsilon)}{{4\pi}^{\frac{3}{2}-\epsilon}}\frac{1}{{|x-y|}^{1-2\epsilon}} \label{scalartree}\\
	\end{align}  

	\item Fermion propagator
	\begin{equation}
	\label{0fermion}
	\langle {(\psi_{\alpha I})_{\hat i}}^j (x)\ {(\bar \psi^{J\beta})^{\hat l}}_k (y)  \rangle= \delta^J_I\delta^{\hat l}_{\hat i}\delta^j_k \ i \, \frac{\Gamma(\frac{3}{2}-\epsilon)}{{2\pi}^{\frac{3}{2}-\epsilon}} \ {(\gamma^\mu)_\alpha}^\beta\ \frac{(x-y)_\mu}{{|x-y|}^{3-2\epsilon}}
	\end{equation}

	\item Vector propagators in Landau gauge ($\alpha =0$)
\bea\label{prop:vector}
	&& \langle {(A_\mu)_i}^j (x)\ {(A_\nu)_k}^l (y)  \rangle =  \delta^l_i\delta^j_k \   i  \, \frac{\Gamma(\frac{3}{2}-\epsilon)}{{\pi}^{\frac{1}{2}-\epsilon}}\ \varepsilon_{\mu\nu\rho}\ \frac{(x-y)^\rho}{{|x-y|}^{3-2\epsilon}} \non \\
	&&	\langle {(\hat A_\mu)_{\hat i}}^{\hat j} (x)\ {(\hat A_\nu)_{\hat k}}^{\hat l} (y)  \rangle =  - \, \delta^{\hat l}_{\hat i}\delta^{\hat j}_{\hat k} \   i  \, \frac{\Gamma(\frac{3}{2}-\epsilon)}{{\pi}^{\frac{1}{2}-\epsilon}}\ \varepsilon_{\mu\nu\rho}\ \frac{(x-y)^\rho}{{|x-y|}^{3-2\epsilon}}
\eea
In a generic $\alpha$-gauge the propagators would acquire an extra term, precisely
\beq\label{eq:xigauge}
\langle A_\mu (x)\ A_\nu (y)  \rangle =\frac{i}{2} \varepsilon_{\mu\nu\rho}\ \frac{(x-y)^\rho}{|x-y|^{3}} + \frac{\alpha}{4}\left[\frac{\delta_{\mu\nu}}{|x-y|}-\frac{(x-y)_\mu(x-y)_\nu}{|x-y|^3}\right] 
\eeq
and similarly for $\hat{A}_\mu$. We note that, independently of the value of $\alpha$, the $\alpha$-term is identically zero for the propagator $\langle A_3 (s)\ A_3 (0)  \rangle$ evaluated on the line placed along the third direction.  

\end{itemize}

\noindent
At the order we are working the ghost propagators do not enter, whereas the vertices can be easily read from terms \eqref{CS}, \eqref{Smat} and \eqref{S4pt} of the action after performing rescaling \eqref{eq:rescaling}. 

\vskip 10pt

We choose the Wilson line along direction 3. Therefore, it is convenient to relabel gauge fields and covariant derivatives (see their definition in \eqref{covd}) localized on the defect, as
\begin{equation}\label{gauge}
A_\mu \, \to \, (A \equiv A_1-iA_2,\ \bar A \equiv A_1+iA_2,\ A_3)\qquad \hat A_\mu \, \to \, (\hat A \equiv \hat A_1-i\hat A_2,\ \hat{\bar A} \equiv \hat A_1+i\hat A_2,\ \hat A_3)\nonumber  
\end{equation}
\begin{equation}\label{eq:derivatives}
D_\mu \, \to \, ( D \equiv D_1-iD_2,\ \bar{ D} \equiv D_1+iD_2,\ {D}_3)
\end{equation}

Similarly, matter fields localized on the Wilson line are conveniently split according to their $\mathfrak{su}(3)$ representation. Precisely, we rename
\begin{equation}\label{su3breaking}
 C_I=(Z,Y_a) \ \qquad \bar C^I=(\bar Z, \bar Y^a) \ \qquad \psi_{I}=(\psi, \chi_{a}) \ \qquad \bar\psi^I=(\bar\psi, \bar\chi^a) \qquad a = 1,2,3
\end{equation}
where $ Y_a (\bar Y^a), \chi_a (\bar\chi^a)$ belong to the ${\mathbf 3} (\bar{\mathbf 3})$ of $\mathfrak{su}(3)$, while $Z,\bar Z, \psi, \bar\psi$ are $SU(3)$-singlets.  

\vskip40pt

\section{The $\mathfrak{su}(1,1|3)$ superalgebra}\label{sect: su(1,1|3)}

In this appendix we describe the superalgebra preserved by the maximally supersymmetric Wilson line in ABJ(M) theory. We also review some useful details of its representation theory.

The insertion of the $\frac{1}{2}$-BPS Wilson line in ABJ(M) theory breaks the bulk $\mathfrak{osp}(6|4)$ superalgebra to the one-dimensional $\mathfrak{su}(1,1|3)\oplus \mathfrak{u}(1)_B$ superconformal algebra.
We are not going to describe the bulk superalgebra in detail\footnote{We refer to \cite{Gorini:2020new} for a complete presentation in our notations, including the explicit embedding of $\mathfrak{su}(1,1|3)\oplus \mathfrak{u}(1)_B$ into $\mathfrak{osp}(6|4)$.}.
Here we limit to recall that the bosonic part of $\mathfrak{osp}(6|4)$ contains the three-dimensional conformal algebra generated by translations $P_\mu$, rotations $M_{\mu\nu}$, dilatations $D$, and special conformal transformations $K_\mu$. The $SU(4)$ R-symmetry group is generated by ${J_I}^J$, $I,J=1,\dots,4$, with ${J_I}^I=0$. 

The $\mathfrak{su}(1,1|3)$ superalgebra on the Wilson line contains the 1d $\mathfrak{su}(1,1)$ conformal algebra, spanned by $P$, $K$, $D$. These are the generators of translations and special conformal transformations along the direction of the Wilson line, and dilatations, respectively. 

The Wilson line preserves a residual $\mathfrak{su}(3)$ R-symmetry, whose generators are denoted by ${R_a}^b$, with $a,b=1,2,3$ and ${R_a}^a\equiv0$. 

Finally, the bosonic sector of the superalgebra includes the $\mathfrak{u}(1)_M$ factor generated by 
\beq\label{eq:M}
M=3i M_{12}-2{J_1}^1 
\eeq
namely the combination of the generator of the rotation in the transverse direction and broken R-symmetry preserving the fermionic part of the superconnection. There is a second preserved abelian factor $\mathfrak{u}(1)_B$ generated by 
\beq \label{eq:B}
B=M_{12}+2i {J_1}^1 
\eeq
It is the sum of the orthogonal rotation and the broken R-symmetry commuting with $\mathfrak{su}(1,1|3)$. 

Looking at the fermionic sector, the $\mathfrak{su}(1,1|3)$ superalgebra contains twelve odd generators: six Poincaré supercharges $Q^a$, $\bar Q_a$ and six superconformal charges $S^a$, $\bar S_a$. The upper index $a=1,2,3$ defines the fundamental representation of the residual $\mathfrak{su}(3)$ R-symmetry algebra, while the lower index indicates the anti-fundamental one.
The fermionic supercharges close on the 1d conformal algebra, spanned by $P$, $K$, $D$.

The complete set of non-vanishing (anti)commutation relations is the following
\begin{equation}\label{su1,1}
[D,P]=P \ \qquad [D,K]=-K \ \qquad [P,K]=-2D    
\end{equation}
\begin{equation}\label{su3}
[{R_a}^b, {R_c}^d]=\delta_a^d {R_c}^b- \delta_c^b {R_a}^d     
\end{equation}
\begin{equation}\label{anticomm}
\begin{aligned}
\{Q^a, \bar Q_b\}&= \delta^a_b \, P\qquad \quad &\{S^a, \bar S_b\} &=\delta^a_b \, K \\
\{Q^a, \bar S_b\}&= \delta^a_b\bigg(D+\frac{1}{3}M\bigg)- {R_b}^a &
\{\bar Q_a, S^b\}&= \delta_a^b\bigg(D-\frac{1}{3}M\bigg)+ {R_a}^b
\end{aligned}
\end{equation}
together with the mixed commutation rules
\begin{equation}\label{comm}
\begin{aligned} 
[D,Q^a]&=\frac{1}{2}Q^a\qquad &[K,Q^a]&=S^a\qquad &[{R_a}^b, Q^c]&=\delta^c_a Q^b-\frac{1}{3}\delta_a^b Q^c\qquad & [M, Q^a]&=\frac{1}{2}Q^a\\ 
[D,\bar Q_a]&=\frac{1}{2}\bar Q_a \qquad
 &[K,\bar Q_a]&=\bar S_a &[{R_a}^b, \bar Q_c]&=-\delta^b_c \bar Q_a+\frac{1}{3}\delta_a^b \bar Q_c &[M, \bar Q_a]&=-\frac{1}{2}\bar Q_a    \\ 
[D, S^a]&=-\frac{1}{2}S^a\qquad 
 &[P, S^a]&=-Q^a &[{R_a}^b, S^c]&=\delta^c_a S^b-\frac{1}{3}\delta_a^b S^c &[M,S^a]&=\frac{1}{2}S^a  \\
[D, \bar S_a]&=-\frac{1}{2}\bar S_a &[P, \bar 
S_a]&=-\bar Q_a &[{R_a}^b, \bar S_c]&=-\delta^b_c \bar S_b+\frac{1}{3}\delta_a^b \bar S_c & [M, \bar S_a]&=-\frac{1}{2}\bar S_a \\
& \ 
\end{aligned}
\end{equation}
It is convenient to recall that the action of the $\mathfrak{su}(3)$ generators on fields in the fundamental and anti-fundamental representations reads
\begin{equation}\label{fund3}
[{R_a}^b, \Phi_c]=\frac{1}{3}\delta^b_a \Phi_c-\delta^b_c \Phi_a \qquad  [{R_a}^b, \bar{\Phi}^c]=\delta^c_a \bar{\Phi}^b - \frac{1}{3}\delta^b_a \bar{\Phi}^c
\end{equation}
 
\vskip 15pt 
A brief classification of the $\mathfrak{su}(1,1|3)$ multiplets goes as follows \cite{Bianchi:2017ozk}.

Multiplet components are classified in terms of the four Dynkin labels $[\Delta, m, j_1, j_2]$ associated to the bosonic subalgebra $\mathfrak{su}(1,1)\oplus \mathfrak{u}(1)_M \oplus  \mathfrak{su}(3)$. $\Delta$ is the conformal weight, $m$ the $\mathfrak{u}(1)_M$ charge, whereas $(j_1, j_2)$ are the eigenvalues corresponding to two $\mathfrak{su}(3)$ Cartan generators $J_1$ and $J_2$ that we choose to be
\begin{equation}\label{su(3)cartan}
\begin{aligned}
& J_1\equiv \frac{{R_2}^2-{R_1}^1}{2} =  - \frac{2{R_1}^1 + {R_3}^3}{2} \\
&J_2\equiv \frac{{R_3}^3-{R_2}^2}{2} =  \frac{{R_1}^1 + 2{R_3}^3}{2} 
\end{aligned}
\end{equation}
Here we have exploited the traceless property ${R_a}^a=0$ to remove the dependence on ${R_2}^2$. 

With this choice of the basis, the supercharges possess well-defined Dynkin labels, whose values are displayed in Table \ref{table1}.
\begin{table}
\begin{center}
\begin{tabular}{ |c|c| } 
 \hline
 {\rm Generators} & $[\Delta, m, j_1, j_2]$  \\ 
 \hline\hline
 $Q^1$ $\; \bar Q_1$ & $\left[\frac{1}{2},\frac{1}{2},-1,0\right]$ \quad $\left[\frac{1}{2},-\frac{1}{2},1,0\right]$ \\  
 $Q^2$ $\; \bar Q_2$ & $\left[ \frac{1}{2},\frac{1}{2},1,-1\right] $  \quad $\left[\frac{1}{2},-\frac{1}{2},-1,1\right]$ \\ 
 $Q^3$ $\; \bar Q_3$ & $\left[\frac{1}{2},\frac{1}{2},0,1\right]$  \quad $\left[\frac{1}{2},-\frac{1}{2},0,-1\right]$ \\
\hline\hline
$S^1$ $\; \bar S_1$ &  $\left[-\frac{1}{2},\frac{1}{2},-1,0\right]$ \quad  $\left[-\frac{1}{2},-\frac{1}{2},1,0\right]$ \\
$S^2$ $\; \bar S_2$ & $\left[- \frac{1}{2},\frac{1}{2},1,-1\right]$ \quad $\left[-\frac{1}{2},-\frac{1}{2},-1,1\right]$ \\
$S^3$ $\; \bar S_3$ & $\left[-\frac{1}{2},\frac{1}{2},0,1\right]$ \quad $\left[-\frac{1}{2},-\frac{1}{2},0,-1\right]$ \\
 \hline
\end{tabular}
\caption{Table of Dynkin labels of fermionic generators. For a generic element $v_\mu$ transforming in a weight-$\mu$ representation, the Dynkin label corresponding to a generator  $H_{i}$ of the Cartan subalgebra is defined as $j_i (v_\mu) \equiv 2[H_{i}, v_\mu]$. }
\label{table1}
\end{center}
\end{table}
When  localized on the line, also the ABJ(M) fundamental fields  have definite  quantum numbers. Their values are listed in Table \ref{table2} for the scalar fields and in Table \ref{table3} for the fermionic ones.
\begin{table}
\begin{center}
\begin{tabular}{ |c|c| } 
 \hline
 {\rm Scalar fields} & $[\Delta, m, j_1, j_2]$  \\ 
 \hline\hline
 $Z$ $, \; \bar Z$ & $\left[\frac{1}{2},\frac{3}{2},0,0\right]$ \quad $\left[\frac{1}{2},-\frac{3}{2},0,0\right]$ \\  
 $Y_1$ $,\; \bar Y^1$ & $\left[ \frac{1}{2},-\frac{1}{2},1,0\right] $  \quad $\left[\frac{1}{2},\frac{1}{2},-1,0\right]$ \\ 
 $Y_2$ $,\; \bar Y^2$ & $\left[\frac{1}{2},-\frac{1}{2},-1,1\right]$  \quad $\left[\frac{1}{2},\frac{1}{2},1,-1\right]$ \\
  $Y_3$ $,\; \bar Y^3$ & $\left[\frac{1}{2},-\frac{1}{2},0,-1\right]$  \quad $\left[\frac{1}{2},\frac{1}{2},0,1\right]$ \\
\hline
 \hline
\end{tabular}
\caption{Quantum number assignments to scalar matter fields of the ABJ(M) theory defined in eq. \eqref{su3breaking}.} 
\label{table2}
\end{center}
\end{table}
\begin{table}
\begin{center}
\begin{tabular}{ |c|c| } 
 \hline
 {\rm Fermionic fields} & $[\Delta, m, j_1, j_2]$  \\ 
 \hline\hline
 $(\psi)_1$ $, \; (\psi)_2$ & $\left[1,3,0,0\right]$ \quad $\left[1,0,0,0\right]$ \\  
 $(\bar \psi)_1$ $,\; (\bar \psi)_2$ & $\left[1,0,0,0\right]$  \quad $\left[1,-3,0,0\right]$ \\ 
 $ (\chi_1)_1$ $,\; (\chi_1)_2$ & $\left[ 1,1,1,0\right]$  \quad $\left[1,-2,1,0\right]$ \\
 $ (\bar \chi^1)_1$ $,\; (\bar \chi^1)_2$ & $\left[ 1,2,-1,0\right]$  \quad $\left[1,-1,-1,0\right]$ \\
 $ (\chi_2)_1$ $,\; (\chi_2)_2$ & $\left[1,1,-1,1\right]$  \quad $\left[1,-2,-1,1\right]$ \\
 $ (\bar \chi^2)_1$ $,\; (\bar \chi^2)_2$ & $\left[1,2,1,-1\right]$  \quad $\left[1,-1,1,-1\right]$ \\
 $ (\chi_3)_1$ $,\; (\chi_3)_2$ & $\left[1,1,0,-1\right]$  \quad $\left[1,-2,0,-1\right]$ \\
 $ (\bar \chi^3)_1$ $,\; (\bar \chi^3)_2$ & $\left[1,2,0,1\right]$  \quad $\left[1,-1,0,1\right]$ \\
\hline
 \hline
\end{tabular}
\caption{Quantum number assignments to fermionic matter fields of the ABJ(M) theory defined in eq. \eqref{su3breaking}.} 
\label{table3}
\end{center}
\end{table}
Finally, the Dynkin labels of the covariant derivatives defined in \eqref{eq:derivatives} are given by
\begin{equation}
{ D}\ [1,3,0,0]\qquad \bar{ D} \ [1,-3,0,0]\qquad { D}_3\ [1,0,0,0]    
\end{equation}

The relevant superconformal multiplets constructed in \cite{Bianchi:2017ozk} are the following (for a systematic classification, see \cite{Agmon:2020pde}):
\vskip .5cm

\paragraph{The $\mathcal A$ Multiplets}

These are long multiplets, denoted by $\mathcal{A}^{\Delta}_{m;j_1,j_2}$.  Their highest weight, namely their super-conformal primary (SCP), is identified by requiring that
\begin{align}
 S^a\ket{\Delta,m,j_1,j_2}^{\text{hw}}&=0 & \bar S_a\ket{\Delta,m,j_1,j_2}^{\text{hw}}&=0 & R_{a+1}^{\quad \; \,  a}\ket{\Delta,m,j_1,j_2}^{\text{hw}}&=0
\end{align}
where we have exploited the state-operator correspondence. 
The entire multiplet is then built by acting with the supercharges $Q^a$ and $\bar Q_a$. 
For unitary representations, the Dynkin labels of the highest weight are constrained by the following inequalities \cite{Agmon:2020pde}
\begin{equation}\label{Amultiplet}
\Delta\ge\begin{cases}
\frac{1}{3}(2j_2+j_1-m), \qquad m<\frac{j_2-j_1}{2}\\
\frac{1}{3}(j_2+2j_1+m), \qquad m\ge\frac{j_2-j_1}{2}
\end{cases}
\end{equation}
The constant operator $\mathcal{T}$ is the SCP of the long multiplet $\mathcal{A}^{\Delta}_{0;0,0}$ constructed explicitly in section 
\ref{sect:T}. Here $\Delta$ is a function of the coupling constants of the theory. 
 
\paragraph{The $\mathcal B$ Multiplets}
These are obtained by imposing that the highest weight is annihilated by some of the $Q$ or $\bar Q$ charges, ({\it shortening condition}). 
We may also have mixed multiplets where the highest weight is annihilated both by some $Q^a$ and some $\bar Q_a$. We denote these multiplets as $\mathcal{B}^{\frac{1}{N}\frac{1}{M}}_{m;j_1,j_2}$, where $\frac{1}{N}$ and $\frac{1}{M}$ denote the fraction of $Q$ and $\bar Q$ annihilating the states, respectively. For instance, the displacement operator sits in the $\mathcal{B}^{0\frac{1}{2}}_{\frac{3}{2};0,0}\oplus\mathcal{B}^{\frac{1}{2}0}_{-\frac{3}{2};0,0}$ multiplet.
Each $\mathcal{B}$ multiplet has its specific unitarity bounds, which are detailed in \cite{Agmon:2020pde}.

\vskip 35pt

 \section{Supersymmetry and superconformal transformations}\label{susytransfs}

The ABJ(M) $\frac{1}{2}$-BPS Wilson line is invariant under the following supersymmetry transformations 
\begin{itemize}
\item Scalars
\begin{align} 
Q^aZ&=-\bar{\chi}_1^a 		& \bar{Q}_aZ&=0 		&Q^a\bar{Z}&=0			& \bar{Q}_a\bar{Z}&=i\chi_a^1 \non \\
Q^a Y_b&= \delta^a_b\bar{\psi}_1		&\bar{Q}_aY_b&=- i\epsilon_{abc}\bar{\chi}_2^c		&Q^a\bar{Y}^b&=-\epsilon^{abc}\chi^2_c		&	\bar{Q}_a\bar{Y}^b&=-i\delta_a^b\psi^1
\end{align}

\item Fermions
\begin{subequations}
\begin{align} 
\bar{Q}_a\psi^1&=0			&Q^a\psi^1&= - iD_3\bar{Y}^a - \frac{2\pi i}{k}\left( \bar{Y}^al_B-\hat{l}_B\bar{Y}^a \right) 	\\
Q^a\psi^2&=-i D \bar{Y}^a			&\bar{Q}_a\psi^2&=-\frac{4\pi }{k}\epsilon_{abc}\bar{Y}^bZ\bar{Y}^c	\\
\bar{Q}_a\chi_b^1&=	 \epsilon_{abc} \, \bar{D} \bar{Y}^c		&Q^a\chi_b^1&= i\delta_b^aD_3\bar{Z} + \frac{4\pi i}{k}\left(\bar{Z}\Lambda^a_b-\hat{\Lambda}^a_b\bar{Z}\right)	
\\
Q^a\chi_b^2&=i\delta^a_b \, D \bar{Z}			&\bar{Q}_a\chi_b^2&= -\epsilon_{abc}D_3\bar{Y}^c - \frac{2\pi }{k}\epsilon_{acd}\left( \bar{Y}^c\Theta_b^d-\hat{\Theta}_b^d\bar{Y}^c \right) 	\\
Q^a\bar{\psi}_1&=0			&\bar{Q}_a\bar{\psi}_1&=- D_3 Y_a - \frac{2\pi }{k}\left( Y_a\hat{l}_B-l_B Y_a \right)	\\
\bar{Q}_a\bar{\psi}_2&=- \bar{D}Y_a			&Q^a\bar{\psi}_2&= \frac{4\pi i}{k}\epsilon^{abc}Y_b\bar{Z}Y_c	\\
Q^a\bar{\chi}^b_1&=	- i\epsilon^{abc} \, D Y_c		&\bar{Q}_a\bar{\chi}^b_1&=\delta_b^aD_3Z + \frac{4\pi}{k}\left(Z\hat{\Lambda}^a_b-\Lambda^a_bZ\right)	\\
\bar{Q}_a\bar{\chi}^b_2&= \delta_a^b \, \bar{D}Z			&Q^a\bar{\chi}^b_2&=i\epsilon^{abc}D_3Y_c + \frac{2\pi i}{k}\epsilon^{acd}\left(Y_c\hat{\Theta}_d^b-\Theta_d^b Y_c \right)
\end{align}
\end{subequations}

\item Gauge fields
\begin{subequations}
\begin{align} 
Q^aA_3&=- \frac{2\pi i}{k}\left(\bar{\psi}_1\bar{Y}^a-\bar{\chi}_1^a\bZ+\epsilon^{abc}Y_b\chi_c^2 \right)			
& \quad \bar{Q}_aA_3&=\frac{2\pi }{k}  \left(Z\chi^1_a-Y_a\psi^1 -\epsilon_{abc}\bar{\chi}_2^b\bar{Y}^c \right)	\non \\
Q^aA&=0		& \quad	\bar{Q}_a A &=-\frac{4\pi}{k}\left(Y_a\psi^2-Z\chi_a^2 - \epsilon_{abc}\bar{\chi}_1^b\bar{Y}^c\right) \non	\\
Q^a \bar{A}&=- \frac{4\pi i}{k}	\left( \bar{\psi}_2\bar{Y}^a-\bar{\chi}^a_2\bZ -\epsilon^{abc}Y_b\chi_c^1\right)		
& \quad \bar{Q}_a \bar{A} &=0	\non \\
Q^a\hat{A}_3&=-\frac{2\pi i}{k}\left(\bar{Y}^a\bar{\psi}_1-\bZ\bar{\chi}_1^a + \epsilon^{abc}\chi_c^2Y_b \right) & \quad \bar{Q}_a\hat{A}_3&=\frac{2\pi }{k} \left(\chi^1_aZ-\psi^1Y_a - \epsilon_{abc}\bar{Y}^c\bar{\chi}_2^b \right)	\non  \\
Q^a \hat{A} &=0		
& \quad \bar{Q}_a \hat{A} &=\frac{4\pi}{k}\left(\psi^2Y_a-\chi_a^2Z - \epsilon_{abc}\bar{Y}^c\bar{\chi}_1^b\right)	 \non	\\
Q^a \hat{\bar A} &= - \frac{4\pi i}{k}	\left( \bar{Y}^a\bar{\psi}_2-\bZ\bar{\chi}^a_2 - \epsilon^{abc}\chi_c^1Y_b\right)	& \quad	
\bar{Q}_a \hat{\bar A} &=0
\end{align}
\end{subequations}
\end{itemize}

\noindent
For the superconformal charges acting on the fields we find

\begin{itemize}
\item Scalars
\begin{align} 
S^aZ&=-s\bar{\chi}_1^a 		& \bar{S}_aZ&=0 		&S^a\bar{Z}&=0			& \bar{S}_a\bar{Z}&=is\chi_a^1 \non \\
S^a Y_b&=s \delta^a_b\bar{\psi}_1		&\bar{S}_aY_b&=-is\epsilon_{abc}\bar{\chi}_2^c		&S^a\bar{Y}^b&=-s\epsilon^{abc}\chi^2_c		&	\bar{S}_a\bar{Y}^b&=-is\delta_a^b\psi^1
\end{align}

\item Fermions
\begin{subequations}
\begin{align} 
\bar{S}_a\psi^1&=0			&S^a\psi^1&= -isD_3\bar{Y}^a - \frac{2\pi i s}{k}\left( \bar{Y}^al_B-\hat{l}_B\bar{Y}^a \right)+i \bar Y^a 	\\
S^a\psi^2&=-i s D \bar{Y}^a			&\bar{S}_a\psi^2&=-\frac{4\pi s }{k}\epsilon_{abc}\bar{Y}^bZ\bar{Y}^c	\\
\bar{S}_a\chi_b^1&=	s\epsilon_{abc} \, \bar{D} \bar{Y}^c		&S^a\chi_b^1&= is\delta_b^aD_3\bar{Z} + \frac{4\pi i s}{k}\left(\bar{Z}\Lambda^a_b-\hat{\Lambda}^a_b\bar{Z}\right)-i\bar Z\\
S^a\chi_b^2&=is\delta^a_b \, D \bar{Z}	&\bar{S}_a\chi_b^2&= -s\epsilon_{abc}D_3\bar{Y}^c - \frac{2\pi s}{k}\epsilon_{acd}\left( \bar{Y}^c\Theta_b^d-\hat{\Theta}_b^d\bar{Y}^c \right)+\epsilon_{abc}\bar Y^c\\
S^a\bar{\psi}_1&=0			&\bar{S}_a\bar{\psi}_1&=-sD_3 Y_a - \frac{2\pi s}{k}\left( Y_a\hat{l}_B-l_B Y_a \right)+Y_a	\\
\bar{S}_a\bar{\psi}_2&=-s \bar{D}Y_a			&S^a\bar{\psi}_2&= \frac{4\pi i s}{k}\epsilon^{abc}Y_b\bar{Z}Y_c	\\
S^a\bar{\chi}^b_1&=-is\epsilon^{abc} \, D Y_c		&\bar{S}_a\bar{\chi}^b_1&=s\delta_b^aD_3Z +\frac{4\pi s}{k}\left(Z\hat{\Lambda}^b_a-\Lambda^b_a Z\right)-\delta_a^b Z\\
\bar{S}_a\bar{\chi}^b_2&= s\delta_a^b \, \bar{D}Z			&S^a\bar{\chi}^b_2&=is\epsilon^{abc}D_3Y_c + \frac{2\pi is}{k}\epsilon^{acd}\left(Y_c\hat{\Theta}_d^b-\Theta_d^b Y_c \right)-i\epsilon^{abc}Y_c 
\end{align}
\end{subequations}

\item Gauge fields
\begin{subequations}
\begin{align} 
S^aA_3&=- \frac{2\pi i}{k}s\left(\bar{\psi}_1\bar{Y}^a-\bar{\chi}_1^a\bZ+\epsilon^{abc}Y_b\chi_c^2 \right)			
& \quad \bar{S}_aA_3&=\frac{2\pi }{k} s \left(Z\chi^1_a-Y_a\psi^1 -\epsilon_{abc}\bar{\chi}_2^b\bar{Y}^c \right)	\non \\
\bar{S}_a A &=-\frac{4\pi}{k}s\left(Y_a\psi^2-Z\chi_a^2 - \epsilon_{abc}\bar{\chi}_1^b\bar{Y}^c\right) & S^aA&=0\non	\\
S^a \bar{A}&=- \frac{4\pi i}{k}	s\left( \bar{\psi}_2\bar{Y}^a-\bar{\chi}^a_2\bZ -\epsilon^{abc}Y_b\chi_c^1\right)		
& \quad \bar{S}_a \bar{A} &=0	\non \\
S^a\hat{A}_3&=-\frac{2\pi i}{k}s\left(\bar{Y}^a\bar{\psi}_1-\bZ\bar{\chi}_1^a + \epsilon^{abc}\chi_c^2Y_b \right) & \quad \bar{S}_a\hat{A}_3&=\frac{2\pi }{k}s \left(\chi^1_aZ-\psi^1Y_a - \epsilon_{abc}\bar{Y}^c\bar{\chi}_2^b \right)	\non  \\
S^a \hat{A} &=0		
& \quad \bar{S}_a \hat{A} &=\frac{4\pi}{k}s\left(\psi^2Y_a-\chi_a^2Z - \epsilon_{abc}\bar{Y}^c\bar{\chi}_1^b\right)	 \non	\\
S^a \hat{\bar A} &= - \frac{4\pi i}{k}s	\left( \bar{Y}^a\bar{\psi}_2-\bZ\bar{\chi}^a_2 - \epsilon^{abc}\chi_c^1Y_b\right)	& \quad	
\bar{S}_a \hat{\bar A} &=0
\end{align}
\end{subequations}

\end{itemize}
where we have defined the bilinear scalar fields
\begin{align}
\label{eq:bilinears}
\begin{pmatrix}
\Lambda_a^b & 0\\
0 & \hat{\Lambda}_a^b
\end{pmatrix}
&=
\begin{pmatrix}
Y_a\bar{Y}^b+\frac{1}{2}\delta_a^bl_B & 0\\
0 & \bar{Y}^b Y_a +\frac{1}{2}\delta_a^b\hat{l}_B 
\end{pmatrix} \non \\
\begin{pmatrix}
\Theta_a^b & 0\\
0 & \hat{\Theta}_a^b
\end{pmatrix}
&=
\begin{pmatrix}
Y_a\bar{Y}^b-\delta_a^b (Z\bZ+Y_c\bY^c) & 0\\
0 & \bar{Y}^b Y_a -\delta_a^b(\bZ Z+\bY^cY_c)
\end{pmatrix} \non \\
\begin{pmatrix}
l_B & 0\\
0 & \hat{l}_B
\end{pmatrix}
&=
\begin{pmatrix}
Z\bZ-Y_c\bY^c & 0\\
0 & \bZ Z-\bY^cY_c
\end{pmatrix}
\end{align}
It is easy to show that these transformations match the $\mathfrak{su}(1,1|3)$ superalgebra described in appendix \ref{sect: su(1,1|3)}.

\vskip25pt

\section{Details on the closure of the covariant algebra}\label{app:closure}

In this appendix we show explicitly that the covariantized supersymmetry transformations generated by the covariant supercharges in \eqref{covsusy} provide a representation of the $\mathfrak{su}(1,1|3)$ algebra. 
To prove this statement we study the action of the anticommutators $\acomm*{\bar{\mathcal{Q}}_a}{\mathcal{Q}^b}$, $\acomm*{\bar{\mathcal{Q}}_a}{\mathcal{S}^b}$, $\acomm*{\mathcal{Q}^a}{\bar{\mathcal{S}}_b}$ and $\acomm*{\bar{\mathcal{S}}_a}{\mathcal{S}^b}$ on the local operators introduced in the main text, namely $\mathbb{Z}$ and $\mathbb{G}^a$. 

To begin with, we evaluate
\begin{equation}
\acomm*{\bar{\mathcal{Q}}_a}{\mathcal{Q}^b}\mathbb{Z}=\bar{\mathcal{Q}}_a \mathcal{Q}^b\mathbb{Z}
\end{equation}
where we used that $\bar{\mathcal{Q}}_a\mathbb{Z}=0$. Exploiting the explicit variations of the fields listed in the previous appendix, we first compute
\begin{equation}
\mathcal{Q}^b\mathbb{Z}=\begin{pmatrix}0&Q^b Z\\0&0\end{pmatrix}-\acomm{\mathbb{G}^b}{\mathbb{Z}}=\begin{pmatrix}-2\sqrt{\frac{\pi}{k}}Z\bar Y^b&-\bar{\chi}_1^b\\0&-2\sqrt{\frac{\pi}{k}}\bar Y^bZ\end{pmatrix}
\end{equation}
A second variation yields
\begin{equation}
\acomm*{\bar{\mathcal{Q}}_a}{\mathcal{Q}^b}\mathbb{Z}=
\begin{pmatrix}
-2\sqrt{\frac{\pi}{k}}i Z\psi^1&&\delta_a^b\left(D_3 Z+\frac{2\pi}{k}(Z\hat\ell_B-\ell_B Z)\right)\\
0 && 2\frac{\pi}{k}i\psi^1Z\delta_b^a
\end{pmatrix}
\end{equation}
where we have used the definitions in \eqref{eq:bilinears}.

It is not hard to recast this term in the following form
\begin{equation}
\acomm*{\bar{\mathcal{Q}}_a}{\mathcal{Q}^b}\mathbb{Z}=-\delta_a^b  \left(\partial_3 \mathbb{Z}+i\comm*{\mathcal{L}}{\mathbb{Z}}\right)\equiv\delta_a^b \, \mathcal{P}\,\mathbb{Z}\,,
\end{equation}
from which we read the \emph{covariantized translation} $\mathcal{P}=-\partial_3- i\comm*{\mathcal{L}}{\cdot} \equiv - \mathfrak{D}_3$.
It is straightforward to evaluate the other anticommutators 
\begin{align}
\acomm*{\bar{\mathcal{Q}}_a}{\mathcal{S}^b}\mathbb{Z}&=\delta_a^b\left[\left(-s\mathfrak{D}_3+\frac{1}{2}\right)\mathbb{Z}-\frac{1}{2}\mathbb{Z} \right]\equiv\delta_a^b\left(\mathcal{D}-\frac{1}{3}M\right)\mathbb{Z}\,,\\
\acomm*{\mathcal{Q}^b}{\bar{\mathcal{S}}_a}\mathbb{Z}&=\delta_a^b\left[\left(-s\mathfrak{D}_3+\frac{1}{2}\right)\mathbb{Z}+\frac{1}{2}\mathbb{Z} \right]\equiv\delta_a^b\left(\mathcal{D}+\frac{1}{3}M\right)\mathbb{Z}\,.
\end{align}
Comparing with the abstract algebra \eqref{anticomm}, and recalling that $\mathbb{Z}$ is a $\mathfrak{su}(3)$ singlet with $\Delta= 1/2$ and $M$-charge 3/2, we find perfect agreement. The first contribution is the action of the dilation generator, which acts as $\mathcal{D}=-s\mathfrak{D}_3+\Delta$. The other piece corresponds to the action of $M$. Finally, applying the $\{\mathcal{S}, \bar{\mathcal S} \}$ anticommutator we find
\begin{align}
\acomm*{\bar{\mathcal{S}}_a}{\mathcal{S}^b}\mathbb{Z}&=\delta_a^b\left(-s^2\mathfrak{D}_3+s\right)\mathbb{Z}\equiv\delta_a^b\, \mathcal{K} \, \mathbb{Z}
\end{align}
from which we read the covariantized action of $K$, that is $\mathcal{K}=-s^2\mathfrak{D}_3+2s\Delta $.

The same computation can be repeated for all the operators of the theory. For instance, for the $\mathbb{G}^a$ operators we obtain
\begin{align}
\acomm*{\bar{\mathcal{Q}}_a}{\mathcal{Q}^b}\mathbb{G}^c&=\delta_a^b \, \mathcal{P}\, \mathbb{G}^c\,,\\
\acomm*{\bar{\mathcal{Q}}_a}{\mathcal{S}^b}\mathbb{G}^c&=\left(\delta^b_a\mathcal{D}-\frac{1}{3}M+{R_a}^b\right)\mathbb{G}^c\,,\\
\acomm*{\mathcal{Q}^b}{\bar{\mathcal{S}}_a}\mathbb{G}^c&=\left(\delta^b_a\mathcal{D}+\frac{1}{3}M-{R_a}^b\right)\mathbb{G}^c\,,\\
\acomm*{\bar{\mathcal{S}}_a}{\mathcal{S}^b}\mathbb{G}^c&=\delta_a^b \, \mathcal{K} \, \mathbb{G}^c
\end{align}
where ${R_a}^b$ acts on $\mathbb{G}^c$ according to the rule in \eqref{fund3} and we used that $\mathbb{G}^c$ has $M$-charge 1/2. 

This provides a derivation of \eqref{covariant_gen} and proves that the covariantized algebra is a representation of the $\mathfrak{su}(1,1|3)$ superalgebra on the space of supermatrix operators.

\section{IR regulator: The cut-off line vs the cut-off circle }\label{sect:cut-off} 

In this appendix we discuss in details the IR regularization prescription that has been adopted in the main text for the ABJ theory. To this end, we focus the discussion on the perturbative evaluation of the Wilson line itself. The same prescription then applies to the evaluation of defect correlators. 

As already discussed, the evaluation of $\langle {W}\rangle$ for a line contour is complicated by the appearance of long distance singularities associated with the infinite domain of line integrals. Such singularities, if regularized by introducing a long distance cut-off $L$, lead to unwanted terms like the one in \eqref{eq:Wpert}. These terms, mixing short and long distance divergences, render the order of the two operations - UV renormalization and removal of the IR cut-off - ambiguous. 

On the other hand, the perturbative evaluation of $\langle {W}\rangle$ on a circular contour does not present any particular problem,  since long distance divergences are obviously absent. Regularizing short distance singularities by using dimensional regularization with dimensional reduction, the one-loop correction is known to vanish, while the two-loop correction \cite{Bianchi:2013rma, Bianchi:2013zda, Griguolo:2013sma} turns out to be finite  as expected, given the BPS nature of the defect. 

\begin{figure}[]
	\centering
	\subfigure[]{\begin{tikzpicture}
			\begin{feynman}
				\vertex (a);
				\vertex[above left=of a] (e1);
				\vertex[above right=of a] (e2);
				\vertex[below right=of a] (e4);
				\vertex[below=1.6cm of a](e5) {\(-\pi+\eta\)};
				\vertex[above=1.6cm of a] (e6) {\(\pi-\eta\)};
				\diagram*{ (e2) -- [fermion, quarter right] (e4)
				};
				\draw[thick,blue] (a) circle (1.5);
				\draw[blue] (-1.4,0) -- (-1.6,0);
				\draw[red] (0,1.4) -- (0,1.6);
				\draw[red] (0,-1.4) -- (0,-1.6);
				\draw[fill=black] (e4) circle (3pt);
				\draw[fill=black] (e2) circle (3pt);
			\end{feynman}
		\end{tikzpicture}\label{fig:cutcircle1}}\qquad
	\subfigure[]{\begin{tikzpicture}
			\begin{feynman}
				\vertex (a);
				\vertex[above left=of a] (e1);
				\vertex (b1) at (-0.3,1.45);
				\vertex (b2) at (-1.45,0.3);
				\vertex[above right=of a] (e2);
				\vertex[below right=of a] (e4);
				\vertex[below left=of a] (e3);
				\vertex[below=1.6cm of a](e5) {\(-\pi+\eta\)};
				\vertex[above=1.6cm of a] (e6) {\(\pi-\eta\)};
				\diagram*{ (b1) -- [fermion, quarter left] (b2)
				};
				\draw[thick,blue] (a) circle (1.5);
				\draw[blue] (-1.4,0) -- (-1.6,0);
				\draw[red] (0,1.4) -- (0,1.6);
				\draw[red] (0,-1.4) -- (0,-1.6);
				\draw[fill=black] (b1) circle (3pt);
				\draw[fill=black] (b2) circle (3pt);
			\end{feynman}
		\end{tikzpicture}\label{fig:cutcircle5}}\qquad
	\subfigure[]{\begin{tikzpicture}
			\begin{feynman}
				\vertex (a);
				\vertex[above left=of a] (e1);
				\vertex[above right=of a] (e2);
				\vertex[below right=of a] (e4);
				\vertex[below=1.6cm of a](e5) {\(-\pi+\eta\)};
				\vertex[above=1.6cm of a] (e6) {\(\pi-\eta\)};
				\diagram*{ (e1) -- [fermion, quarter right] (e2)
				};
				\draw[thick,blue] (a) circle (1.5);
				\draw[blue] (-1.4,0) -- (-1.6,0);
				\draw[red] (0,1.4) -- (0,1.6);
				\draw[red] (0,-1.4) -- (0,-1.6);
				\draw[fill=black] (e2) circle (3pt);
				\draw[fill=black] (e1) circle (3pt);
			\end{feynman}
		\end{tikzpicture}\label{fig:cutcircle2}}\qquad
	\subfigure[]{\begin{tikzpicture}
			\begin{feynman}
				\vertex (a);
				\vertex[above left=of a] (e1);
				\vertex[above right=of a] (e2);
				\vertex[below right=of a] (e4);
				\vertex[below left=of a] (e3);
				\vertex[below=1.6cm of a](e5) {\(-\pi+\eta\)};
				\vertex[above=1.6cm of a] (e6) {\(\pi-\eta\)};
				\diagram*{ (e1) -- [fermion, quarter left] (e3)
				};
				\draw[thick,blue] (a) circle (1.5);
				\draw[blue] (-1.4,0) -- (-1.6,0);
				\draw[red] (0,1.4) -- (0,1.6);
				\draw[red] (0,-1.4) -- (0,-1.6);
				\draw[fill=black] (e1) circle (3pt);
				\draw[fill=black] (e3) circle (3pt);
			\end{feynman}
		\end{tikzpicture}\label{fig:cutcircle3}}\qquad
	\subfigure[]{\begin{tikzpicture}
			\begin{feynman}
				\vertex (a);
				\vertex[above left=of a] (e1);
				\vertex[above right=of a] (e2);
				\vertex[below right=of a] (e4);
				\vertex[below left=of a] (e3);
				\vertex[below=1.6cm of a](e5) {\(-\pi+\eta\)};
				\vertex[above=1.6cm of a] (e6) {\(\pi-\eta\)};
				\diagram*{ (e4) -- [fermion, quarter right] (e3)
				};
				\draw[thick,blue] (a) circle (1.5);
				\draw[blue] (-1.4,0) -- (-1.6,0);
				\draw[red] (0,1.4) -- (0,1.6);
				\draw[red] (0,-1.4) -- (0,-1.6);
				\draw[fill=black] (e4) circle (3pt);
				\draw[fill=black] (e3) circle (3pt);
			\end{feynman}
		\end{tikzpicture}\label{fig:cutcircle4}}\qquad
	\subfigure[]{\begin{tikzpicture}
			\begin{feynman}
				\vertex (a);
				\vertex[above left=of a] (e1);
				\vertex (b1) at (-0.3,-1.45);
				\vertex (b2) at (-1.45,-0.3);
				\vertex[above right=of a] (e2);
				\vertex[below right=of a] (e4);
				\vertex[below left=of a] (e3);
				\vertex[below=1.6cm of a](e5) {\(-\pi+\eta\)};
				\vertex[above=1.6cm of a] (e6) {\(\pi-\eta\)};
				\diagram*{ (b1) -- [fermion, quarter right] (b2)
				};
				\draw[thick,blue] (a) circle (1.5);
				\draw[blue] (-1.4,0) -- (-1.6,0);
				\draw[red] (0,1.4) -- (0,1.6);
				\draw[red] (0,-1.4) -- (0,-1.6);
				\draw[fill=black] (b1) circle (3pt);
				\draw[fill=black] (b2) circle (3pt);
			\end{feynman}
		\end{tikzpicture}\label{fig:cutcircle6}}\qquad
	\caption{One-loop diagrams for the cut-off Wilson circle.}
	\label{fig:cutcircle}
\end{figure}
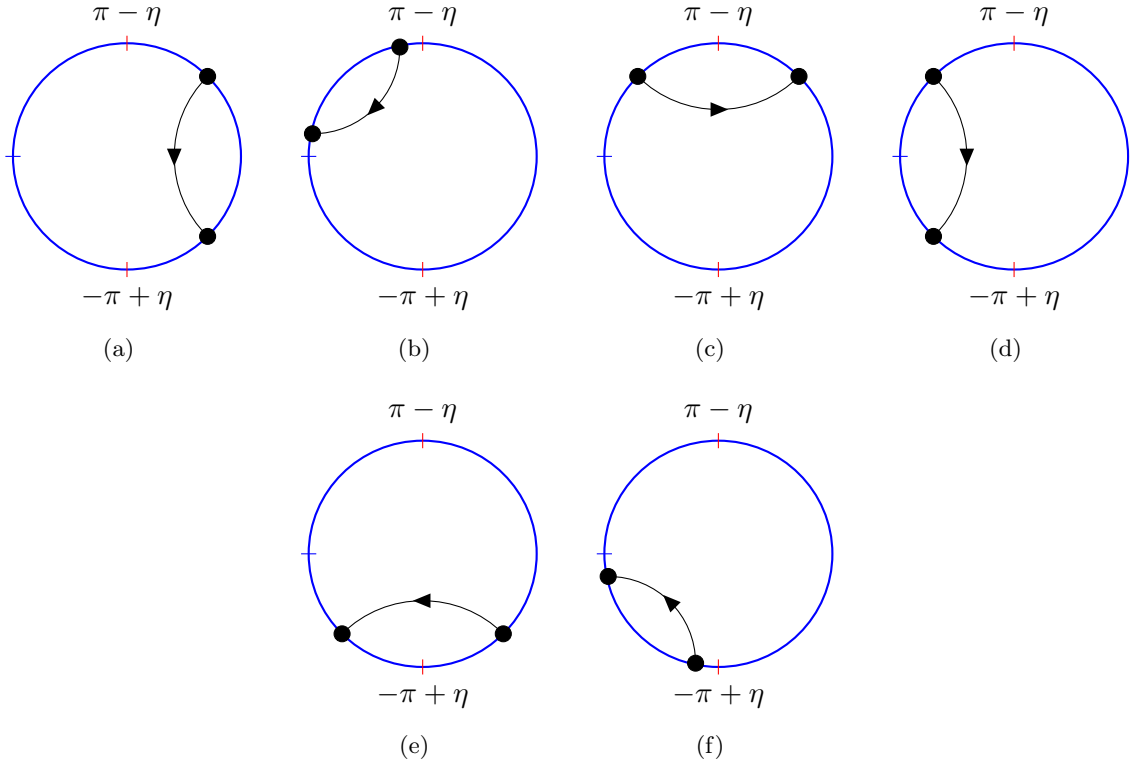

Therefore, it is convenient to regularize long distance singularities on the line by conformally mapping the line onto the circle. More precisely, in order to better understand the origin of the unwanted terms arising in the linear case and how one should interpret this regularization, we consider mapping the cut-off line onto a cut-off circle. This will help understanding that the apparent ambiguity in dealing with these terms can be traced back to the non-complete control of the contributions from degrees of freedom placed at the edges of the cut-off line. 

In order to better understand this point, we map the segment $(-L,L)$ onto a cut-off circle defined for $\tau \in [-\pi+\eta,\pi-\eta]$, where $\eta = 2 \arccot{(2L)}$. The limit \( L \rightarrow\infty\) corresponds to \(\eta\rightarrow 0\) on the circle.

Consequently, we rewrite the VEV of the circular Wilson loop as (we use the notation of footnote \ref{foot:notation})
\begin{equation}
	\langle {W}  \rangle \equiv \langle {W}_{\pi, -\pi}  \rangle =\langle \underbrace{{W}_{\pi,\pi-\eta}}_{line\;[+\infty,L]}  \underbrace{{W}_{\pi-\eta,-\pi+\eta}}_{line\;[L,-L]}\underbrace{{W}_{-\pi+\eta,-\pi}}_{line\;[-L, -\infty]} \rangle
	\label{VEVcircle}
\end{equation}
where we have explicitly indicated to which portion of the straight line each single term corresponds to under conformal mapping. 

On the circle the integrals corresponding to the three pieces can be computed exactly and are convergent for any value of $\eta$.
At one loop, the cut-off line integral in \eqref{eq:intLL} is equivalent to the following circular integral  
\begin{equation}
	(\ref{fig:cutcircle1})=\int_{-\pi+\eta}^{\pi-\eta}d\tau_1\int_{-\pi+\eta}^{\tau_1}d\tau_2 \de_{\tau_1}\de_{\tau_2}\sin^{2\epsilon}{\left(\frac{\tau_{12}}{2}\right)}=-\sin^{2\epsilon}\left(\pi-\eta\right)
	\label{eq:intetaeta}
\end{equation}
which, contrary to the line integral, is finite independently of the order of the $\epsilon \to 0$ and $\eta \to 0$ limits, though  umbiguous. However, in order to reproduce the circle VEV in \eqref{VEVcircle}, this result has to be completed with the contributions from the ``external'' regions $(\pi, \pi - \eta)$ and $(-\pi + \eta, -\pi)$. Performing fermion-fermion contractions in all possible orders, these contributions are explicitly given by (see figure \ref{fig:cutcircle})
\begin{align}
	(\ref{fig:cutcircle5})&=\int_{\pi-\eta}^{\pi}d\tau_1 \int_{\pi-\eta}^{\tau_1}d\tau_2 \partial_{\tau_2}\partial_{\tau_1}\sin^{2\epsilon}\left(\frac{\tau_{12}}{2}\right)=-\sin^{2\epsilon}\left(\frac{\eta}{2}\right)
	\label{c1} \\
	(\ref{fig:cutcircle2})&=\int_{\pi-\eta}^{\pi}d\tau_1 \int_{-\pi+\eta}^{\pi-\eta}d\tau_2 \partial_{\tau_2}\partial_{\tau_1}\sin^{2\epsilon}\left(\frac{\tau_{12}}{2}\right)
	=\sin^{2\epsilon}\left(\frac{\eta}{2}\right)-\sin^{2\epsilon}\left(\pi-\frac{\eta}{2}\right)+\sin^{2\epsilon}\left(\pi-\eta\right)
	\label{c2}\\
	(\ref{fig:cutcircle3})&=\int_{\pi-\eta}^{\pi}d\tau_1 \int_{-\pi}^{-\pi+\eta}d\tau_2 \partial_{\tau_2}\partial_{\tau_1}\sin^{2\epsilon}\left(\frac{\tau_{12}}{2}\right)=2\sin^{2\epsilon}\left(\pi-\frac{\eta}{2}\right)-\sin^{2\epsilon}\left(\pi-\eta\right)
	\label{c3} \\
	(\ref{fig:cutcircle4})&=\int_{-\pi+\eta}^{\pi-\eta}d\tau_1 \int_{-\pi}^{-\pi+\eta}d\tau_2 \partial_{\tau_2}\partial_{\tau_1}\sin^{2\epsilon}\left(\frac{\tau_{12}}{2}\right)=\sin^{2\epsilon}\left(\frac{\eta}{2}\right)\!-\!\sin^{2\epsilon}\left(\pi-\frac{\eta}{2}\right)\!+\!\sin^{2\epsilon}\left(\pi-\eta\right) \\
	(\ref{fig:cutcircle6})&=\int_{-\pi}^{-\pi+\eta}d\tau_1 \int_{-\pi}^{\tau_1}d\tau_2 \partial_{\tau_2}\partial_{\tau_1}\sin^{2\epsilon}\left(\frac{\tau_{12}}{2}\right)=-\sin^{2\epsilon}\left(\frac{\eta}{2}\right)
	\label{c4}
\end{align}
It is easy to see that they sum up to $\sin^{2\epsilon}\left(\pi-\eta\right)$ and cancel exactly the ``line'' contribution \eqref{eq:intetaeta}. 

This cancellation is expected in order to reproduce the one loop result $\langle {W}_{circle} \rangle^{(1)} = 0$. However, revisited from the line perspective, it is quite instructive. In fact, contributions (\ref{c1}-\ref{c4}) from the pieces external to the cut-off circle can be interpreted as coming from extra degrees of freedom that one should place at the boundaries of the cut-off line. Neglecting them causes the aforementioned ambiguities, while taking them into account would lead to a vanishing unambiguous result. 
Mapping the line to the circle is then a correct prescription to regularize the line, since it automatically captures the extra degrees of freedom at finite $L$. 
Operationally, this is equivalent to computing correlation functions directly on the line, neglecting extra terms such as 
$(L-s)^{2\epsilon}, (L+s)^{2\epsilon}$ and $(2L)^{2\epsilon}$ in the results of section \ref{sect:corr}.

\newpage
\bibliography{biblio}

\providecommand{\href}[2]{#2}\begingroup\raggedright\begin{thebibliography}{10}

\bibitem{Billo:2016cpy}
M.~Bill\`o, V.~Gon\c{c}alves, E.~Lauria and M.~Meineri, \emph{{Defects in
  conformal field theory}},
  \href{https://doi.org/10.1007/JHEP04(2016)091}{\emph{JHEP} {\bfseries 04}
  (2016) 091} [\href{https://arxiv.org/abs/1601.02883}{{\ttfamily
  1601.02883}}].

\bibitem{Liendo:2012hy}
P.~Liendo, L.~Rastelli and B.C.~van Rees, \emph{{The Bootstrap Program for
  Boundary CFT$_d$}},
  \href{https://doi.org/10.1007/JHEP07(2013)113}{\emph{JHEP} {\bfseries 07}
  (2013) 113} [\href{https://arxiv.org/abs/1210.4258}{{\ttfamily 1210.4258}}].

\bibitem{McAvity:1995zd}
D.M.~McAvity and H.~Osborn, \emph{{Conformal field theories near a boundary in
  general dimensions}},
  \href{https://doi.org/10.1016/0550-3213(95)00476-9}{\emph{Nucl. Phys. B}
  {\bfseries 455} (1995) 522}
  [\href{https://arxiv.org/abs/cond-mat/9505127}{{\ttfamily
  cond-mat/9505127}}].

\bibitem{Affleck:1995ge}
I.~Affleck, \emph{{Conformal field theory approach to the Kondo effect}},
  {\emph{Acta Phys. Polon. B} {\bfseries 26} (1995) 1869}
  [\href{https://arxiv.org/abs/cond-mat/9512099}{{\ttfamily
  cond-mat/9512099}}].

\bibitem{Wilson:1974sk}
K.G.~Wilson, \emph{{Confinement of Quarks}},
  \href{https://doi.org/10.1103/PhysRevD.10.2445}{\emph{Phys. Rev. D}
  {\bfseries 10} (1974) 2445}.

\bibitem{Aharony:2013hda}
O.~Aharony, N.~Seiberg and Y.~Tachikawa, \emph{{Reading between the lines of
  four-dimensional gauge theories}},
  \href{https://doi.org/10.1007/JHEP08(2013)115}{\emph{JHEP} {\bfseries 08}
  (2013) 115} [\href{https://arxiv.org/abs/1305.0318}{{\ttfamily 1305.0318}}].

\bibitem{Gaiotto:2014kfa}
D.~Gaiotto, A.~Kapustin, N.~Seiberg and B.~Willett, \emph{{Generalized Global
  Symmetries}}, \href{https://doi.org/10.1007/JHEP02(2015)172}{\emph{JHEP}
  {\bfseries 02} (2015) 172} [\href{https://arxiv.org/abs/1412.5148}{{\ttfamily
  1412.5148}}].

\bibitem{Zarembo:2002an}
K.~Zarembo, \emph{{Supersymmetric Wilson loops}},
  \href{https://doi.org/10.1016/S0550-3213(02)00693-4}{\emph{Nucl. Phys.}
  {\bfseries B643} (2002) 157}
  [\href{https://arxiv.org/abs/hep-th/0205160}{{\ttfamily hep-th/0205160}}].

\bibitem{Maldacena:1998im}
J.M.~Maldacena, \emph{{Wilson loops in large N field theories}},
  \href{https://doi.org/10.1103/PhysRevLett.80.4859}{\emph{Phys. Rev. Lett.}
  {\bfseries 80} (1998) 4859}
  [\href{https://arxiv.org/abs/hep-th/9803002}{{\ttfamily hep-th/9803002}}].

\bibitem{Rey:1998ik}
S.-J.~Rey and J.-T.~Yee, \emph{{Macroscopic strings as heavy quarks in large N
  gauge theory and anti-de Sitter supergravity}},
  \href{https://doi.org/10.1007/s100520100799}{\emph{Eur. Phys. J. C}
  {\bfseries 22} (2001) 379}
  [\href{https://arxiv.org/abs/hep-th/9803001}{{\ttfamily hep-th/9803001}}].

\bibitem{Giombi:2017cqn}
S.~Giombi, R.~Roiban and A.A.~Tseytlin, \emph{{Half-BPS Wilson loop and
  AdS$_2$/CFT$_1$}},
  \href{https://doi.org/10.1016/j.nuclphysb.2017.07.004}{\emph{Nucl. Phys. B}
  {\bfseries 922} (2017) 499}
  [\href{https://arxiv.org/abs/1706.00756}{{\ttfamily 1706.00756}}].

\bibitem{Cooke:2017qgm}
M.~Cooke, A.~Dekel and N.~Drukker, \emph{{The Wilson loop CFT: Insertion
  dimensions and structure constants from wavy lines}},
  \href{https://doi.org/10.1088/1751-8121/aa7db4}{\emph{J. Phys. A} {\bfseries
  50} (2017) 335401} [\href{https://arxiv.org/abs/1703.03812}{{\ttfamily
  1703.03812}}].

\bibitem{Giombi:2018qox}
S.~Giombi and S.~Komatsu, \emph{{Exact Correlators on the Wilson Loop in
  $\mathcal{N}=4$ SYM: Localization, Defect CFT, and Integrability}},
  \href{https://doi.org/10.1007/JHEP05(2018)109}{\emph{JHEP} {\bfseries 05}
  (2018) 109} [\href{https://arxiv.org/abs/1802.05201}{{\ttfamily
  1802.05201}}].

\bibitem{Pestun:2007rz}
V.~Pestun, \emph{{Localization of gauge theory on a four-sphere and
  supersymmetric Wilson loops}},
  \href{https://doi.org/10.1007/s00220-012-1485-0}{\emph{Commun. Math. Phys.}
  {\bfseries 313} (2012) 71} [\href{https://arxiv.org/abs/0712.2824}{{\ttfamily
  0712.2824}}].

\bibitem{Correa:2012at}
D.~Correa, J.~Henn, J.~Maldacena and A.~Sever, \emph{{An exact formula for the
  radiation of a moving quark in N=4 super Yang Mills}},
  \href{https://doi.org/10.1007/JHEP06(2012)048}{\emph{JHEP} {\bfseries 06}
  (2012) 048} [\href{https://arxiv.org/abs/1202.4455}{{\ttfamily 1202.4455}}].

\bibitem{Correa:2012hh}
D.~Correa, J.~Maldacena and A.~Sever, \emph{{The quark anti-quark potential and
  the cusp anomalous dimension from a TBA equation}},
  \href{https://doi.org/10.1007/JHEP08(2012)134}{\emph{JHEP} {\bfseries 08}
  (2012) 134} [\href{https://arxiv.org/abs/1203.1913}{{\ttfamily 1203.1913}}].

\bibitem{Aharony:2008ug}
O.~Aharony, O.~Bergman, D.L.~Jafferis and J.~Maldacena, \emph{{N=6
  superconformal Chern-Simons-matter theories, M2-branes and their gravity
  duals}}, \href{https://doi.org/10.1088/1126-6708/2008/10/091}{\emph{JHEP}
  {\bfseries 10} (2008) 091} [\href{https://arxiv.org/abs/0806.1218}{{\ttfamily
  0806.1218}}].

\bibitem{Aharony:2008gk}
O.~Aharony, O.~Bergman and D.L.~Jafferis, \emph{{Fractional M2-branes}},
  \href{https://doi.org/10.1088/1126-6708/2008/11/043}{\emph{JHEP} {\bfseries
  11} (2008) 043} [\href{https://arxiv.org/abs/0807.4924}{{\ttfamily
  0807.4924}}].

\bibitem{Drukker:2009hy}
N.~Drukker and D.~Trancanelli, \emph{{A Supermatrix model for N=6 super
  Chern-Simons-matter theory}},
  \href{https://doi.org/10.1007/JHEP02(2010)058}{\emph{JHEP} {\bfseries 02}
  (2010) 058} [\href{https://arxiv.org/abs/0912.3006}{{\ttfamily 0912.3006}}].

\bibitem{Ouyang:2015iza}
H.~Ouyang, J.-B.~Wu and J.-j.~Zhang, \emph{{Novel BPS Wilson loops in
  three-dimensional quiver Chern\textendash{}Simons-matter theories}},
  \href{https://doi.org/10.1016/j.physletb.2015.12.021}{\emph{Phys. Lett. B}
  {\bfseries 753} (2016) 215}
  [\href{https://arxiv.org/abs/1510.05475}{{\ttfamily 1510.05475}}].

\bibitem{Ouyang:2015bmy}
H.~Ouyang, J.-B.~Wu and J.-j.~Zhang, \emph{{Construction and classification of
  novel BPS Wilson loops in quiver Chern\textendash{}Simons-matter theories}},
  \href{https://doi.org/10.1016/j.nuclphysb.2016.07.018}{\emph{Nucl. Phys. B}
  {\bfseries 910} (2016) 496}
  [\href{https://arxiv.org/abs/1511.02967}{{\ttfamily 1511.02967}}].

\bibitem{Mauri:2017whf}
A.~Mauri, S.~Penati and J.-j.~Zhang, \emph{{New BPS Wilson loops in $
  \mathcal{N}=4 $ circular quiver Chern-Simons-matter theories}},
  \href{https://doi.org/10.1007/JHEP11(2017)174}{\emph{JHEP} {\bfseries 11}
  (2017) 174} [\href{https://arxiv.org/abs/1709.03972}{{\ttfamily
  1709.03972}}].

\bibitem{Drukker:2008zx}
N.~Drukker, J.~Plefka and D.~Young, \emph{{Wilson loops in 3-dimensional N=6
  supersymmetric Chern-Simons Theory and their string theory duals}},
  \href{https://doi.org/10.1088/1126-6708/2008/11/019}{\emph{JHEP} {\bfseries
  11} (2008) 019} [\href{https://arxiv.org/abs/0809.2787}{{\ttfamily
  0809.2787}}].

\bibitem{Berenstein:2008dc}
D.~Berenstein and D.~Trancanelli, \emph{{Three-dimensional N=6 SCFT's and their
  membrane dynamics}},
  \href{https://doi.org/10.1103/PhysRevD.78.106009}{\emph{Phys. Rev. D}
  {\bfseries 78} (2008) 106009}
  [\href{https://arxiv.org/abs/0808.2503}{{\ttfamily 0808.2503}}].

\bibitem{Chen:2008bp}
B.~Chen and J.-B.~Wu, \emph{{Supersymmetric Wilson Loops in N=6 Super
  Chern-Simons-matter theory}},
  \href{https://doi.org/10.1016/j.nuclphysb.2009.09.015}{\emph{Nucl. Phys.}
  {\bfseries B825} (2010) 38}
  [\href{https://arxiv.org/abs/0809.2863}{{\ttfamily 0809.2863}}].

\bibitem{Kapustin:2009kz}
A.~Kapustin, B.~Willett and I.~Yaakov, \emph{{Exact Results for Wilson Loops in
  Superconformal Chern-Simons Theories with Matter}},
  \href{https://doi.org/10.1007/JHEP03(2010)089}{\emph{JHEP} {\bfseries 03}
  (2010) 089} [\href{https://arxiv.org/abs/0909.4559}{{\ttfamily 0909.4559}}].

\bibitem{Drukker:2019bev}
N.~Drukker et~al., \emph{{Roadmap on Wilson loops in 3d
  Chern\textendash{}Simons-matter theories}},
  \href{https://doi.org/10.1088/1751-8121/ab5d50}{\emph{J. Phys. A} {\bfseries
  53} (2020) 173001} [\href{https://arxiv.org/abs/1910.00588}{{\ttfamily
  1910.00588}}].

\bibitem{Drukker:2020opf}
N.~Drukker, \emph{{BPS Wilson loops and quiver varieties}},
  \href{https://doi.org/10.1088/1751-8121/aba5bd}{\emph{J. Phys. A} {\bfseries
  53} (2020) 385402} [\href{https://arxiv.org/abs/2004.11393}{{\ttfamily
  2004.11393}}].

\bibitem{Bianchi:2017ozk}
L.~Bianchi, L.~Griguolo, M.~Preti and D.~Seminara, \emph{{Wilson lines as
  superconformal defects in ABJM theory: a formula for the emitted radiation}},
  \href{https://doi.org/10.1007/JHEP10(2017)050}{\emph{JHEP} {\bfseries 10}
  (2017) 050} [\href{https://arxiv.org/abs/1706.06590}{{\ttfamily
  1706.06590}}].

\bibitem{Bianchi:2020hsz}
L.~Bianchi, G.~Bliard, V.~Forini, L.~Griguolo and D.~Seminara, \emph{{Analytic
  bootstrap and Witten diagrams for the ABJM Wilson line as defect CFT$_1$}},
  \href{https://arxiv.org/abs/2004.07849}{{\ttfamily 2004.07849}}.

\bibitem{Bianchi:2018scb}
L.~Bianchi, M.~Preti and E.~Vescovi, \emph{{Exact Bremsstrahlung functions in
  ABJM theory}}, \href{https://doi.org/10.1007/JHEP07(2018)060}{\emph{JHEP}
  {\bfseries 07} (2018) 060}
  [\href{https://arxiv.org/abs/1802.07726}{{\ttfamily 1802.07726}}].

\bibitem{Agmon:2020pde}
N.B.~Agmon and Y.~Wang, \emph{{Classifying Superconformal Defects in Diverse
  Dimensions Part I: Superconformal Lines}},
  \href{https://arxiv.org/abs/2009.06650}{{\ttfamily 2009.06650}}.

\bibitem{Lewkowycz:2013laa}
A.~Lewkowycz and J.~Maldacena, \emph{{Exact results for the entanglement
  entropy and the energy radiated by a quark}},
  \href{https://doi.org/10.1007/JHEP05(2014)025}{\emph{JHEP} {\bfseries 05}
  (2014) 025} [\href{https://arxiv.org/abs/1312.5682}{{\ttfamily 1312.5682}}].

\bibitem{Bianchi:2014laa}
M.S.~Bianchi, L.~Griguolo, M.~Leoni, S.~Penati and D.~Seminara, \emph{{BPS
  Wilson loops and Bremsstrahlung function in ABJ(M): a two loop analysis}},
  \href{https://doi.org/10.1007/JHEP06(2014)123}{\emph{JHEP} {\bfseries 06}
  (2014) 123} [\href{https://arxiv.org/abs/1402.4128}{{\ttfamily 1402.4128}}].

\bibitem{Bianchi:2017svd}
M.S.~Bianchi, L.~Griguolo, A.~Mauri, S.~Penati, M.~Preti and D.~Seminara,
  \emph{{Towards the exact Bremsstrahlung function of ABJM theory}},
  \href{https://doi.org/10.1007/JHEP08(2017)022}{\emph{JHEP} {\bfseries 08}
  (2017) 022} [\href{https://arxiv.org/abs/1705.10780}{{\ttfamily
  1705.10780}}].

\bibitem{Bianchi:2018bke}
M.S.~Bianchi, L.~Griguolo, A.~Mauri, S.~Penati and D.~Seminara, \emph{{A matrix
  model for the latitude Wilson loop in ABJM theory}},
  \href{https://doi.org/10.1007/JHEP08(2018)060}{\emph{JHEP} {\bfseries 08}
  (2018) 060} [\href{https://arxiv.org/abs/1802.07742}{{\ttfamily
  1802.07742}}].

\bibitem{Griguolo:2021rke}
L.~Griguolo, L.~Guerrini and I.~Yaakov, \emph{{Localization and duality for
  ABJM latitude Wilson loops}},
  \href{https://doi.org/10.1007/JHEP08(2021)001}{\emph{JHEP} {\bfseries 08}
  (2021) 001} [\href{https://arxiv.org/abs/2104.04533}{{\ttfamily
  2104.04533}}].

\bibitem{Cardinali:2012ru}
V.~Cardinali, L.~Griguolo, G.~Martelloni and D.~Seminara, \emph{{New
  supersymmetric Wilson loops in ABJ(M) theories}},
  \href{https://doi.org/10.1016/j.physletb.2012.10.051}{\emph{Phys. Lett.}
  {\bfseries B718} (2012) 615}
  [\href{https://arxiv.org/abs/1209.4032}{{\ttfamily 1209.4032}}].

\bibitem{Lietti:2017gtc}
M.~Lietti, A.~Mauri, S.~Penati and J.-j.~Zhang, \emph{{String theory duals of
  Wilson loops from Higgsing}},
  \href{https://doi.org/10.1007/JHEP08(2017)030}{\emph{JHEP} {\bfseries 08}
  (2017) 030} [\href{https://arxiv.org/abs/1705.02322}{{\ttfamily
  1705.02322}}].

\bibitem{Griguolo:2012iq}
L.~Griguolo, D.~Marmiroli, G.~Martelloni and D.~Seminara, \emph{{The
  generalized cusp in ABJ(M) N = 6 Super Chern-Simons theories}},
  \href{https://doi.org/10.1007/JHEP05(2013)113}{\emph{JHEP} {\bfseries 05}
  (2013) 113} [\href{https://arxiv.org/abs/1208.5766}{{\ttfamily 1208.5766}}].

\bibitem{forthcoming}
L.~Griguolo, L.~Guerrini, S.~Penati, D.~Seminara and P.~Soresina, ``In
  progress.''.

\bibitem{Gorini:2020new}
N.~Gorini, L.~Griguolo, L.~Guerrini, S.~Penati, D.~Seminara and P.~Soresina,
  \emph{{The topological line of ABJ(M) theory}},
  \href{https://doi.org/10.1007/JHEP06(2021)091}{\emph{JHEP} {\bfseries 06}
  (2021) 091} [\href{https://arxiv.org/abs/2012.11613}{{\ttfamily
  2012.11613}}].

\bibitem{Correa:2019rdk}
D.H.~Correa, V.I.~Giraldo-Rivera and G.A.~Silva, \emph{{Supersymmetric mixed
  boundary conditions in AdS$_{2}$ and DCFT$_{1}$ marginal deformations}},
  \href{https://doi.org/10.1007/JHEP03(2020)010}{\emph{JHEP} {\bfseries 03}
  (2020) 010} [\href{https://arxiv.org/abs/1910.04225}{{\ttfamily
  1910.04225}}].

\bibitem{Chen:1992ee}
W.~Chen, G.W.~Semenoff and Y.-S.~Wu, \emph{{Two loop analysis of nonAbelian
  Chern-Simons theory}},
  \href{https://doi.org/10.1103/PhysRevD.46.5521}{\emph{Phys. Rev. D}
  {\bfseries 46} (1992) 5521}
  [\href{https://arxiv.org/abs/hep-th/9209005}{{\ttfamily hep-th/9209005}}].

\bibitem{Bianchi:2013zda}
M.S.~Bianchi, G.~Giribet, M.~Leoni and S.~Penati, \emph{{1/2 BPS Wilson loop in
  N=6 superconformal Chern-Simons theory at two loops}},
  \href{https://doi.org/10.1103/PhysRevD.88.026009}{\emph{Phys. Rev. D}
  {\bfseries 88} (2013) 026009}
  [\href{https://arxiv.org/abs/1303.6939}{{\ttfamily 1303.6939}}].

\bibitem{Bianchi:2013rma}
M.S.~Bianchi, G.~Giribet, M.~Leoni and S.~Penati, \emph{{The 1/2 BPS Wilson
  loop in ABJ(M) at two loops: The details}},
  \href{https://doi.org/10.1007/JHEP10(2013)085}{\emph{JHEP} {\bfseries 10}
  (2013) 085} [\href{https://arxiv.org/abs/1307.0786}{{\ttfamily 1307.0786}}].

\bibitem{Leoni:2010az}
M.~Leoni and A.~Mauri, \emph{{On the infrared behaviour of 3d Chern-Simons
  theories in N=2 superspace}},
  \href{https://doi.org/10.1007/JHEP11(2010)128}{\emph{JHEP} {\bfseries 11}
  (2010) 128} [\href{https://arxiv.org/abs/1006.2341}{{\ttfamily 1006.2341}}].

\bibitem{Grabner:2020nis}
D.~Grabner, N.~Gromov and J.~Julius, \emph{{Excited States of One-Dimensional
  Defect CFTs from the Quantum Spectral Curve}},
  \href{https://doi.org/10.1007/JHEP07(2020)042}{\emph{JHEP} {\bfseries 07}
  (2020) 042} [\href{https://arxiv.org/abs/2001.11039}{{\ttfamily
  2001.11039}}].

\bibitem{Ferrero:2021bsb}
P.~Ferrero and C.~Meneghelli, \emph{{Bootstrapping the half-BPS line defect CFT
  in N=4 supersymmetric Yang-Mills theory at strong coupling}},
  \href{https://doi.org/10.1103/PhysRevD.104.L081703}{\emph{Phys. Rev. D}
  {\bfseries 104} (2021) L081703}
  [\href{https://arxiv.org/abs/2103.10440}{{\ttfamily 2103.10440}}].

\bibitem{Cavaglia:2021bnz}
A.~Cavagli\`a, N.~Gromov, J.~Julius and M.~Preti, \emph{{Integrability and
  conformal bootstrap: One dimensional defect conformal field theory}},
  \href{https://doi.org/10.1103/PhysRevD.105.L021902}{\emph{Phys. Rev. D}
  {\bfseries 105} (2022) L021902}
  [\href{https://arxiv.org/abs/2107.08510}{{\ttfamily 2107.08510}}].

\bibitem{Cavaglia:2022qpg}
A.~Cavagli\`a, N.~Gromov, J.~Julius and M.~Preti, \emph{{Bootstrability in
  defect CFT: integrated correlators and sharper bounds}},
  \href{https://doi.org/10.1007/JHEP05(2022)164}{\emph{JHEP} {\bfseries 05}
  (2022) 164} [\href{https://arxiv.org/abs/2203.09556}{{\ttfamily
  2203.09556}}].

\bibitem{Kim:2017sju}
M.~Kim, N.~Kiryu, S.~Komatsu and T.~Nishimura, \emph{{Structure Constants of
  Defect Changing Operators on the 1/2 BPS Wilson Loop}},
  \href{https://doi.org/10.1007/JHEP12(2017)055}{\emph{JHEP} {\bfseries 12}
  (2017) 055} [\href{https://arxiv.org/abs/1710.07325}{{\ttfamily
  1710.07325}}].

\bibitem{Gabai:2022vri}
B.~Gabai, A.~Sever and D.-l.~Zhong, \emph{{Line Operators in
  Chern-Simons-Matter Theories and Bosonization in Three Dimensions}},
  \href{https://arxiv.org/abs/2204.05262}{{\ttfamily 2204.05262}}.

\bibitem{Cornwell:1989bx}
J.F.~Cornwell, \emph{{Group Theory in Physics. Volume III: Supersymmetries and
  Infinite-Dimensional Algebras}}, vol.~10 of \emph{Techniques of Physics},
  Academic Press, London (1989).

\bibitem{Griguolo:2013sma}
L.~Griguolo, G.~Martelloni, M.~Poggi and D.~Seminara, \emph{{Perturbative
  evaluation of circular 1/2 BPS Wilson loops in N = 6 Super Chern-Simons
  theories}}, \href{https://doi.org/10.1007/JHEP09(2013)157}{\emph{JHEP}
  {\bfseries 09} (2013) 157} [\href{https://arxiv.org/abs/1307.0787}{{\ttfamily
  1307.0787}}].

\end{thebibliography}\endgroup

\end{document}